\documentclass[12pt, review]{elsarticle}

\pdfoutput=1

\usepackage[hidelinks]{hyperref}
\usepackage[section]{placeins} 
\usepackage{amsmath, amsthm, amssymb, calrsfs, wasysym, verbatim, bbm, color, graphics, geometry, setspace, caption, float,siunitx,hyperref}
\usepackage{pdfpages}
\usepackage{soul}
\usepackage[nomarkers,nofiglist]{endfloat}
\newfloat{Video}{H}{lop}
\floatname{Video}{Video}

\journal{Acta Materialia}

\begin{document}

\begin{frontmatter}

\title{From irregular to regular eutectic growth in the Al\mbox{-}Al\textsubscript{3}Ni system:\\ in~situ observations during directional solidification}

\author[ummseaffil]{Paul Chao}
\ead{pchao@umich.edu}

\author[ummseaffil]{Shanmukha Kiran Aramanda}
\ead{askiran@umich.edu}

\author[bnlaffil]{Xianghui Xiao\corref{mycorrespondingauthor}}
\ead{xiao@bnl.gov}

\author[inspaffil]{Sabine Bottin\mbox{-}Rousseau}
\ead{sabine.bottin@insp.jussieu.fr}

\author[inspaffil]{Silvère Akamatsu}
\ead{silvere.akamatsu@insp.jussieu.fr}

\author[ummseaffil,umcheaffil]{Ashwin J. Shahani\corref{mycorrespondingauthor}}
\cortext[mycorrespondingauthor]{Corresponding authors.}
\ead{shahani@umich.edu}

\address[ummseaffil]{Department of Materials Science and Engineering, University of Michigan, 2300 Hayward St., Ann Arbor, MI 48109, USA}
\address[umcheaffil]{Department of Chemical Engineering, University of Michigan, 2300 Hayward St., Ann Arbor, MI 48109, USA}
\address[bnlaffil]{Brookhaven National Laboratory, Upton, NY 11973, USA}
\address[inspaffil]{Sorbonne Université, Institut des NanoSciences de Paris, Paris Cedex 05, France} 

\begin{abstract}
We investigate the irregular eutectic growth dynamics of the Al\mbox{-}Al\textsubscript{3} alloy, in which one of the solid phases (Al\textsubscript{3}Ni) grows faceted from the liquid. Leveraging in~situ optical microscopy and synchrotron transmission x\mbox{-}ray microscopy, we address the question of the degree of coupling between Al and Al\textsubscript{3}Ni at the growth front and that of the shape of the microstructures left behind in the bulk solid during directional solidification. Real\mbox{-}time optical observations bring evidence for a morphological transition from a eutectic\mbox{-}grain dependent, irregular eutectic growth at low solidification velocity $V$ (typically $1~{\rm \mu ms^{-1}}$), to a weakly anisotropic, regular growth at higher $V$ (reaching $10~{\rm \mu ms^{-1}}$). Unprecedented x\mbox{-}ray nano\mbox{-}imaging of the solid\mbox{-}liquid interface, and 3D characterization of the growth patterns, were made possible by a new DS setup at Brookhaven National Laboratory’s NSLS\mbox{-}II. At low $V$, the leading tips of partly faceted Al\textsubscript{3}Ni crystals are observed to grow not far ahead of the Al growth front. Correlating in~situ images and postmortem 3D tomographic reconstructions reveals that the presence of faceted and non\mbox{-}faceted regions of Al\textsubscript{3}Ni crystals in the solid is a direct consequence of coupling and decoupling during DS, respectively. Upon increasing $V$, the lead distance of Al\textsubscript{3}Ni vanishes, and the shape of Al\textsubscript{3}Ni ceases to be governed by faceted growth. These observations cast light onto the basic mechanisms (faceted growth, diffusive coupling, and the dynamics of trijunctions) governing a faceted to rod\mbox{-}like transition upon increasing $V$ in the Al\textsubscript{3}Ni system, with broad implications to a large class of irregular eutectics.
\end{abstract}

\begin{keyword}
Eutectic solidification, in~situ experiments, x\mbox{-}ray synchrotron radiation, solidification microstructures, crystalline anisotropy
\end{keyword}

\end{frontmatter}


\section{Introduction}\label{sec_introduction}

The solidification of a binary eutectic alloy leads to the direct transformation of the liquid into a two-phase solid.  The resulting composite microstructures are a frozen trace left in the bulk solid by growth patterns that self\mbox{-}organize at the solidification front, depending on alloy characteristics and the experimental conditions~\cite{kurz_fundamentals_1992, dantzig2016solidification}. During directional solidification (DS) of nonfaceted eutectics, the formation of ``regular", e.g., lamellar or rod\mbox{-}like, microstructures is primarily governed, during growth, by the coupling between solute diffusion in the liquid and capillary effects at the involved interfaces~\cite{Jackson_Hunt_1966, akamatsu_eutectic_2016}. However, in many metallic eutectics, the composite solid is made of crystals of a stiff intermetallic phase dispersed in a ductile, solid-solution matrix. This contributes to an increase of the mechanical strength of the bulk material (see, e.g., Refs.~\cite{decker1973alloy} and \cite{varin2002intermetallic}), in a way that depends on the shape, size, and spatial distribution of the intermetallic crystals~\cite{aikin_mechanical_1997, kaufman2004aluminum}. In a majority of alloys belonging to this class, the eutectic growth mechanism is made complex by the faceted growth of the intermetallic phase from the liquid. The interplay of solute redistribution by diffusion and a faceted growth kinetics during faceted/nonfaceted eutectic solidification remains poorly understood. 

In this paper, we present an in~situ experimental study of the formation of eutectic growth patterns in the Al\mbox{-}Al\textsubscript{3}Ni system. This was made possible thanks to recent developments in real\mbox{-}time observation of thin-sample DS using optical microscopy (OM) and new capabilities in full\mbox{-}field imaging via synchrotron transmission x\mbox{-}ray microscopy (TXM). The former technique allows one to monitor the solidification dynamics over large space and time scales with a micrometer resolution, and for growth velocities ranging from 0.1 to $10~{\rm \mu ms^{-1}}$. The second technique facilitates for the first time and at higher (nanoscale) resolution the direct observation of the trijunction geometry between liquid, Al, and Al\textsubscript{3}Ni, as solidification proceeds, relying on a two-zone furnace that permits studies of DS with good control of the thermal field. This \textcolor{black}{allows} us to establish the relationship between the morphology of the growth front patterns during solidification and the 3D shape of the Al\textsubscript{3}Ni crystals frozen in the solid. The insights gained can be applied to a wide array of eutectic systems that contain a solid with a fully or partially faceted growth shape from the melt. 

In DS, the binary Al\mbox{-}Al\textsubscript{3}Ni eutectic alloy is often regarded as a prototypical regular, rod\mbox{-}like eutectic made of Al\textsubscript{3}Ni rods with an apparent circular cross\mbox{-}section dispersed in an Al matrix~\cite{chadwick_eutectic_1963,jaffrey_nucleation_1970,hertzberg_microstructure_1965,ratke_ordering_2000,kaya_unidirectional_2010}. This holds true at a relatively high solidification rate --- typically larger than $10~{\rm \mu ms^{-1}}$. There is, however, experimental evidence that the shape of the intermetallic crystals produced during DS in this system qualitatively depend on the growth velocity $V$.  Indeed, ``irregular"  Al-Al\textsubscript{3}Ni microstructures with faceted Al\textsubscript{3}Ni crystals have been reported to form at lower solidification rates in the $1~{\rm \mu ms^{-1}}$ range~\cite{cantor_crystallography_1975,lemkey_microstructure_1965,knowles_structure_1983,knowles_structure_1983-1, zhuang_eutectic_2001, smartt_rod_1972}, while ``broken lamellar"  shapes were observed at intermediate $V$ values  \cite{jaffrey_lamellar_1969,knowles_structure_1983} (at comparable thermal gradients). This ensemble of observations suggests the existence of a transition from irregular to rod\mbox{-}like for average growth rates in an unusually low growth velocity range as compared to, e.g., the Al\mbox{-}Si system (see, e.g., Ref.~\cite{hosch_analysis_2009}).  

Let us recall that at the atomic\mbox{-}scale, the interface between a crystal and the liquid can be rough (typical of metals) or smooth (typical of semiconductors and intermetallics)~\cite{woodruff1973solid}, thus leading to macroscopically rounded or faceted shapes, respectively. Typically, regular eutectics are composed of two fully nonfaceted crystal phases, whereas, in an irregular eutectic, at least one of the eutectic solids is faceted. In the latter case, the irregular, non\mbox{-}periodic arrangement of crystals in the microstructure is basically attributed to the low mobility of the faceted interfaces, and their nonlinear growth behavior that depends on undercooling, crystallographic orientation, and defects~\cite{beatty2000monte, jackson2002interface}. The dimensionless factor $\alpha = \xi \frac{\Delta S_f}{R}$, with $\Delta S_f$ the entropy of fusion, $R$ the gas constant, and $\xi$ a crystallographic parameter, was introduced by Jackson for categorizing solid-liquid interfaces into the two classes, faceted and nonfaceted, depending on $\alpha$ being larger or smaller than 2, respectively. The parameter $\xi$ is bounded between \textcolor{black}{one\mbox{-}fourth} and one~\textcolor{black}{\cite{glicksman_principles_2011}}, and is highest for interfaces parallel to close-packed crystal lattice planes. Interestingly, based on the experimentally measured entropy of fusion ($17.5~{\rm J~molK^{-1}}$) of the Al\textsubscript{3}Ni intermetallic~\cite{maracsli1996solid}, the corresponding Jackson factor  $\alpha$ is found to be at most 2.10, which hardly exceeds the faceted~vs.~nonfaceted threshold. It can thus be expected that, at a given growth rate, some parts of the Al\textsubscript{3}Ni-liquid interfaces are rough while others are smooth, and that the faceted character of Al\textsubscript{3}Ni crystals evolves upon varying the solidification rate.


The dynamics of Al-Al\textsubscript{3}Ni growth patterns actually remains debated. During eutectic growth at low velocity, conflicting accounts exist on whether one phase necessarily leads or not at the growth front. We define the \textit{leading phase} as the one that extends deeper into the melt during directional eutectic solidification. For example, in one of the earliest reports, Jaffrey~et~al.~\cite{jaffrey_lamellar_1969} described a coupled-growth pattern (Fig.~9 of that reference), with pointed Al\textsubscript{3}Ni crystal tips exposed to the liquid. An alternate scenario, envisaged by Fan and Makhlouf \cite{fan_al3ni_2015}, is that faceted Al\textsubscript{3}Ni crystals grow ahead of the Al solid solution, with a substantial lead distance between the Al\textsubscript{3}Ni tips and the Al solid, and thus weak or no diffusive coupling between the two growing solids. In both cases, many questions still remain open, particularly concerning the establishment (or not) of equilibrated triple-junction (trijunction) lines, and the extent that the Al\textsubscript{3}Ni crystals frozen behind the solidification front keep the same faceted shape as during growth. 

One of the main reasons why the above phenomena (coupled~vs.~decoupled growth; irregular~vs.~regular growth) have not yet been addressed with clear\mbox{-}cut conclusions is the dearth of 3D space- and time\mbox{-}resolved experimental data. In metallic systems, attempts to examine the growth front morphology have traditionally relied on ``quench-and-look" experiments. Characterizing the solidification front with postmortem, 2D metallography provides some initial idea of the growth behavior, but it is unknown how much the quenching process distorts the solid\mbox{-}liquid interface shape, in particular in the case of non\mbox{-}isothermal growth~\cite{maracsli1996solid, peng_competitive_2020}. It is thus of key importance to perform in~situ experimentation, a domain that experienced considerable progress during the last two decades~\cite{akamatsu_situ_2016, shahani_characterization_2020}. Few studies of that kind have been dealing with irregular eutectics. Recent efforts shine new light on the degree of coupling between the two eutectic solids during thin-sample DS of a transparent faceted/non-faceted system investigated by real\mbox{-}time, optical microscopy~\cite{mohagheghi2020decoupled}. In that study, it was observed that the faceted phase was leading ahead of the non\mbox{-}faceted phase into the undercooled liquid, thereby resulting in a decoupled and non\mbox{-}isothermal growth front at low velocities ($V<0.5~{\rm \mu ms^{-1}}$). The striking resemblance of that growth dynamics with observations made in metallic eutectics using in~situ x\mbox{-}ray imaging is worth highlighting \cite{hou2018competition,shahani_mechanism_2016}. A similar investigation, combining large\mbox{-}scale optical observations with high\mbox{-}resolution and 3D x\mbox{-}ray imaging techniques during DS of a metallic system, and aiming at bringing clearer evidence of general and/or specific features associated with irregular eutectic growth, is still missing. The Al\mbox{-}Al\textsubscript{3}Ni eutectic presents itself as a good model system for such an investigation.

The paper is structured as follows.  In Sec.~\ref{sec:methods}, we outline our experimental methods for optical and x\mbox{-}ray microscopy.  In Sec.~\ref{sec_results}, we describe the results of real\mbox{-}time, DS experiments performed under various conditions.  In Sec.~\ref{sec_disc}, we discuss separately the influence of (i) solidification front morphology and (ii) crystal orientation on the shapes of the intermetallic crystals.  Conclusions and perspectives are presented in Sec.~\ref{sec_conc}. 

\section{Methods} \label{sec:methods}
\subsection{The Al\mbox{-}Al\textsubscript{3}Ni eutectic}
The Al-Ni phase diagram presents a eutectic equilibrium at a temperature $T_E={\rm 913~K}$, involving the liquid at eutectic concentration $C_E=2.9~{\rm at\%Ni}=6.1~{\rm wt\%Ni}$, the aluminum rich terminal solid solution (Al), also noted as $\alpha$ ($C_\alpha=0.11~{\rm at\%Ni}$), and the intermetallic phase Al\textsubscript{3}Ni ($C_{\rm Al\textsubscript{3}Ni}=25~{\rm at\%Ni}$). The uncertainty in $C_E$ is about 0.5~wt\%Ni, determined from the standard deviation of reported values in Refs.~\cite{lemkey_microstructure_1965, barclay_off-eutectic_1971, gwyer1908alloys, fink1934correlation, fink1934equilibrium,okamoto2000phase,du1996thermodynamic,ansara1997thermodynamic,huang1998thermodynamic,chen2011thermodynamics,yang2022thermodynamic}, which encompass several thermodynamic assessments.  The equilibrium volume fraction of Al\textsubscript{3}Ni phase in the two-phase solid is about 0.11. The $\alpha$ solid solution is a face centered cubic (fcc) crystal phase. The Al\textsubscript{3}Ni solid is orthorhombic (a=0.6598~nm; b=0.7352~nm; c=0.4802~nm)~\cite{bradley1937crystal,pearson2013handbook}. The most common growth direction of Al\textsubscript{3}Ni crystals is [010] with solid\mbox{-}liquid facets parallel to the ${\rm (101)}$ and  ${\rm (10\bar 1)}$ lattice planes~\cite{cantor_crystallography_1975,jaffrey_lamellar_1969,wang_alignment_2008, li_effect_2010, jardine_effect_1986,farag_effect_1976}.

\subsection{Thin-sample directional solidification: optical microscopy}

In~situ experiments using optical microscopy (OM) during thin\mbox{-}sample DS were first carried out with model transparent alloys~\cite{hunt1966binary}. This OM\mbox{-}based technique has been recently adapted to metallic alloys, making it possible to image the microstructure of the contact surface of the thin metallic film with a transparent wall in reflected\mbox{-}light mode~\cite{witusiewicz2013situ,akamatsu2011determination,bottinrousseau_lockedlamellar_2021}. It is applied here for the first time to the Al\mbox{-}Al\textsubscript{3}Ni system, and will be complemented with some postmortem observations (see below). 

\begin{figure}[ht!]
\centering \includegraphics[width=\textwidth,height=\textheight,keepaspectratio]{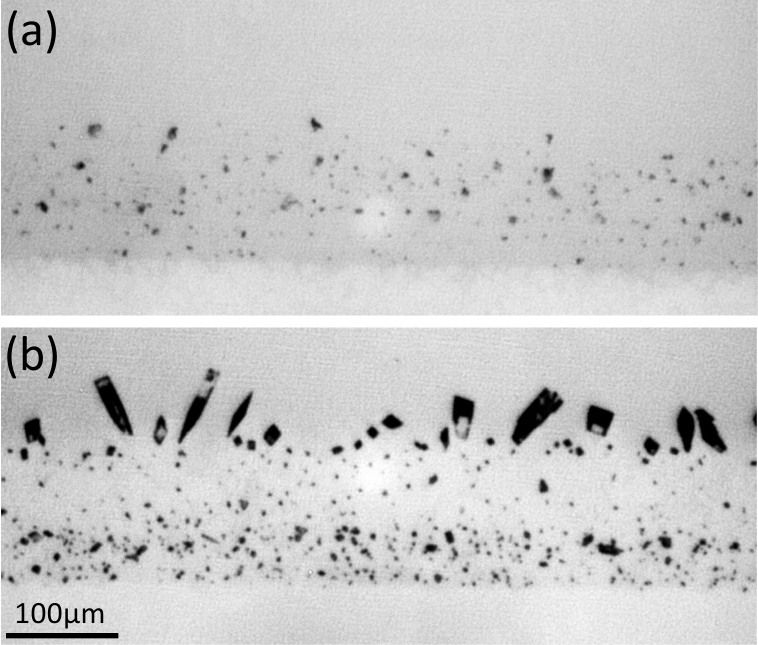}
\caption{(a) In~situ optical image of the initial stages of growth during DS ($V = 0.5~{\rm \mu ms^{-1}}$) of the same thin  Al\mbox{-}Al\textsubscript{3}Ni sample as in Fig.~\ref{Ech169_InSitu}. The growth direction is upward. Top (light grey): liquid. Bottom (white): unmelted solid. Some Al\textsubscript{3}Ni crystals (black) that appeared previously in the liquid (as shown in b), are encapsulated by the solid Al; some of them grow with faceted shapes. (b) End of the directional melting (about 5 minutes before the image in (a)) with small Al\textsubscript{3}Ni crystals dispersed in the liquid. This brings evidence that the sample was of slightly hypereutectic concentration, with Al\textsubscript{3}Ni being the primary phase.}
\label{SI_initial}
\end{figure}

Details pertaining to the preparation of thin metallic samples, DS, and real-time optical imaging can be found in Ref.~\cite{bottinrousseau_lockedlamellar_2021} and references therein. Thin Al\mbox{-}Ni films are prepared by plasma sputtering on 1\mbox{-}mm thick sapphire plates of lateral dimensions $60\times10~{\rm mm^2}$ (Situs GmbH). The magnetron sputtering machine (low\mbox{-}pressure Ar plasma) was equipped with pure Al and Ni sources ($99.99~{\rm mol\%}$; Neyco). A metallic film of total nominal thickness of $10~{\rm \mu m}$ is obtained by depositing first an Al layer on the sapphire\mbox{-}plate substrate, and then a Ni layer on top of it. The ratio between the thicknesses of the two layers is calculated so that the nominal concentration $C_0$ of the film (after melting) is equal to $C_E$.  Before solidification, a sapphire plate (top plate) is pressed on top of the free surface of the bimetallic film, and glued (ResbondTM 908) at room temperature under protective argon pressure. Some in~situ observations during the early stages of an experiment (see Fig.~\ref{SI_initial}) indicate that the samples were actually slightly hypereutectic, that is, with a slight excess of Ni as compared to the eutectic concentration ($C_0>C_E$). 

\textcolor{black}{The DS setup that we used in this study permits the optical observation of eutectic growth patterns in Al-based alloys in thin samples \cite{bottinrousseau_lockedlamellar_2021}. The design is similar to that of standard thin-sample DS apparatus utilized for low-melting alloys \cite{witusiewicz2013situ,akamatsu2011determination,akamatsu2007real}.} 
In the DS setup, a fixed, uniaxial temperature gradient ($G \approx$ 70~${\rm Kcm^{-1}}$ \textcolor{black}{in the region of interest}) is realized between two metallic blocks with individual thermal regulations, separated from each other by a gap of $1~{\rm cm}$. \textcolor{black}{The directional solidification} is performed by translating the sample at an imposed velocity $V$ ($0.3{-}10~{\rm \mu ms^{-1}}$) along the main axis {\bf z} of the temperature gradient \textcolor{black}{(linear DC motor; Physik Instrumente)}. The surface of contact between the substrate plate and the metallic film is observed in real-time with a reflected\mbox{-}light optics and a camera. The images (typically $600\times 450~{\rm \mu m^{2}}$) are recorded and analyzed on a \textcolor{black}{personal} computer. \textcolor{black}{Subsequently, \textit{in-situ} images undergo numerical filtering for global noise reduction and contrast enhancement. In the processed images, the liquid phase appears as dark grey, the Al\textsubscript{3}Ni intermetallic phase as almost black, and the solid Al phase as light grey, exhibiting a faint contrast with the liquid.} The large\mbox{-}scale microstructure (panorama) of the solidified film can be reconstructed numerically.

At the beginning of an experiment, a thin sample is placed in the DS setup at room temperature, and the two metallic blocks are progressively heated to the desired temperatures (773~K on the colder side; 973~K on the hotter side). The metallic film then undergoes a partial directional melting. \textcolor{black}{At the end of the partial-melting stage, a dispersion of small Al\textsubscript{3}Ni crystals can be seen in a band of finite extension along the temperature gradient  (Fig.~\ref{SI_initial}(a)).  This, as stated above, brings clear evidence that the sample is slightly hypereutectic, with Al\textsubscript{3}Ni being the primary phase \cite{{akamatsu2001formation}}. The solidification is started after a short annealing time. During the first stages, most of the preexisting Al\textsubscript{3}Ni crystals are encapsulated by the solid Al, but some of them grow with faceted shapes, thus giving rise to decoupled-growth patterns at low solidification velocity (Fig.~\ref{SI_initial}(b)). Individual intermetallic crystals grow with an inclination that is merely determined, without a clear selective process, by the orientation of the primary crystals. A similar sequence has been recently observed in a different irregular-eutectic system (a model transparent alloy) in thin samples  \cite{mohagheghi2024pinning}.} Long solidification runs at constant velocity, separated by stepwise changes of $V$ are performed. The samples could be used for solidification over several hours, before partial dewetting occurred in the hottest part of the solidification bench.

For ex~situ investigation after complete solidification, the top plate of the sample was removed without substantial damage. Cross-sectional (transverse) views of the eutectic microstructure (parallel to the plane of an isotherm during solidification) were acquired by using Focused Ion Beam (FIB) with Xe ions for surface milling, in a Thermo Fisher Helios G4 PFIB UXe dual-beam, scanning electron microscope (SEM), so as to unveil subsurface features (Fig.~\ref{SI_fig06_fib}). In the operation mode used, the ions travel perpendicularly to the sample surface, thereby enabling sectioning perpendicular to the surface of the metallic film, and to the DS axis as well (the surface of the cross-sections thus nominally corresponds to the plane of an isotherm during solidification). Transverse sections are observed at a 38\textdegree~tilt to the SEM, while the (polished) surface of the film is viewed at a 52\textdegree~tilt.

\subsection{Synchrotron x-ray microscopy and tomography}
\begin{figure}[ht!]
\centering \includegraphics[width=\textwidth,height=\textheight,keepaspectratio]{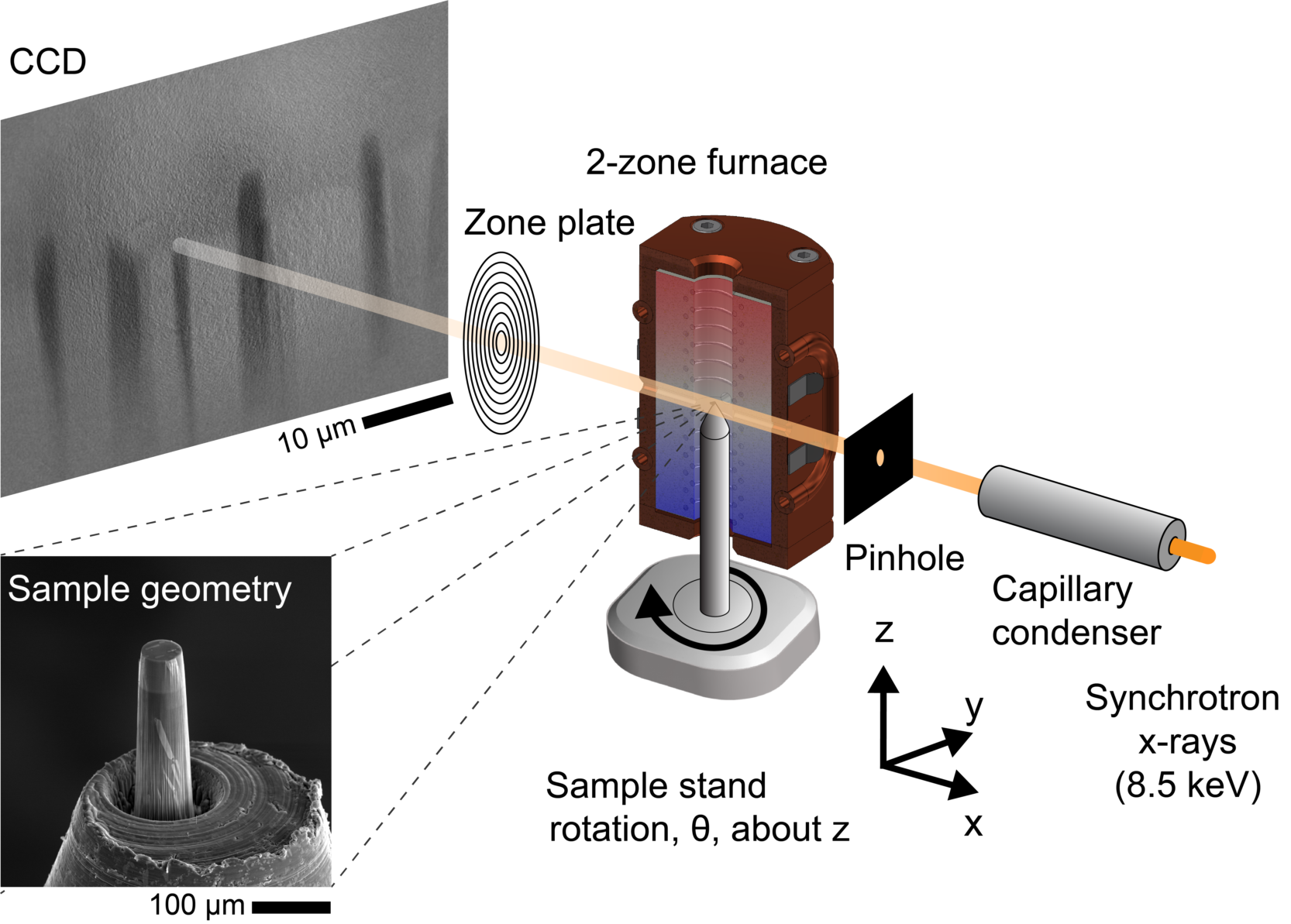}
\caption{Schematic representation of the experimental setup for in~situ transmission x\mbox{-}ray microscopy (TXM) during DS at the FXI beamline 18\mbox{-}ID at NSLS-II. The condenser focuses the x\mbox{-}rays onto the sample, which is located inside the two\mbox{-}zone furnace.  Behind the furnace, the zone plate magnifies and images the sample onto a charge\mbox{-}coupled device based detector.  Samples were constructed with various geometries, including a cylindrical pillar, shown inset. Other geometries are depicted in Fig.~\ref{SI_fig03_TXM_prep}. \textcolor{black}{The axial temperature distribution around the center of the two zones is shown in Fig.~\ref{SI_fig02_temp}.}}
\label{fig06_TXM_method}
\end{figure}

To peer inside the bulk of the material as it solidifies, we combined two techniques using synchrotron\mbox{-}based TXM, namely, in~situ x\mbox{-}radiography and postmortem tomography, following a multimodal imaging strategy initiated in Ref.~\cite{chao2022pseudo}.  We investigated near\mbox{-}eutectic and hypereutectic alloys of composition Al-6wt\%Ni and Al-10wt\%Ni, respectively. High\mbox{-}purity ingots were cast by vacuum arc remelting (Materials Preparation Center at Ames Laboratory, Ames, IA, USA). Prior to the synchrotron campaign, alloy samples were machined into rods (1~mm in diameter; 4~cm in length) via Electrical Discharge Machining (Fig.~\ref{SI_fig01_template}). For preparation, these rods were processed in a regular DS setup at a velocity $V = 1.4~{\rm \mu ms^{-1}}$. The microstructures, examined by SEM, exhibit elongated Al\textsubscript{3}Ni crystals in a solid Al matrix  (Fig.~\ref{SI_fig02_comp_series}), in overall agreement with reference observations \cite{scheil1957untersuchungen,barclay_off-eutectic_1971,kurz1979dendrite}. We also observed blocky primary Al\textsubscript{3}Ni crystals in hypereutectic samples  \cite{juarez-hernandez_growth_1998,sun_situ_2018,ding_intergrowth_2020, ding_new_2020,yu_effect_2021}. The samples used for in~situ DS experiments at the synchrotron were prepared from those rods. We shaped rectangular or circular pillars with their largest dimension not exceeding $50~{\rm \mu m}$  --- the size of the sample is primarily limited by the x\mbox{-}ray attenuation length and the field\mbox{-}of\mbox{-}view (FOV) of the camera. For seeding purposes, we carefully selected two\mbox{-}phase microstructures with Al\textsubscript{3}Ni crystals aligned perpendicular to the x\mbox{-}ray beam, see Fig.~\ref{SI_fig03_TXM_prep}. That is, the growth direction of Al\textsubscript{3}Ni ([010]) lies in the plane of the sample surface.  In practice, the micrometer-scale samples (see inset of Fig.~\ref{fig06_TXM_method}) were cut via micromachining on a mini\mbox{-}lathe prior to final shaping by FIB milling on a Thermo Fisher Helios G4 PFIB UXe dual-beam SEM.  

To realize DS at the synchrotron, we have designed a vertical, two\mbox{-}zone, resistively\mbox{-}heated furnace, based on a previous setup described elsewhere~\cite{antonelli2020versatile}. \st{Technical details on the new furnace will be elaborated upon in a subsequent report.} \textcolor{black}{The two heating zones are constructed using Kanthal based resistive heater coils, each capable of reaching temperatures up to 1373~K. The temperatures are controlled with a PID controller, allowing for heating and cooling rates up to $0.1~{\rm Kmin^{-1}}$. The heater assembly is mounted on a motorized stage stack for precise positioning.} The furnace has two x\mbox{-}ray transparent windows allowing the beam to pass, enabling full\mbox{-}field nano\mbox{-}imaging in transmission during the solidification process.  Independent control of the upper (hot) and lower (cold) heating zones provides a stable, tunable temperature gradient $G$ in the range of $0-20~{\rm Kmm^{-1}}$. \textcolor{black}{The axial temperature distribution around the center of the two zones is given in Fig.~\ref{SI_fig02_temp}.} The samples are kept immobile (in translation) in the setup. They can be rotated for tomographic imaging. Solidification is performed under ambient atmosphere by cooling the two blocks at a rate $\dot{T}$ in the range of 0.1 to $25~{\rm Kmin^{-1}}$. In practice, the velocity of the isotherms (and hence the solidification front) under steady\mbox{-}state conditions is approximately given by $V = \dot{T}/G$. Importantly, the new furnace is 3.1~cm by 2.3~cm by 4.0~cm in width by depth by height. The ``miniature" form factor is ideal for the limited space of the synchrotron beamline (sensitive x\mbox{-}ray optics are positioned near the sample). We identified minimal cross\mbox{-}talk between the two zones during calibration and testing. Measurements of the thermal profiles (not shown) indicated that the furnace could operate with $0.1~{\rm K}$ precision.  

X\mbox{-}radiography and tomography with sub-50~nm pixel size were performed on a transmission x\mbox{-}ray microscope at the full\mbox{-}field X\mbox{-}ray imaging (FXI) beamline 18\mbox{-}ID at Brookhaven National Laboratory's NSLS\mbox{-}II  (Brookhaven, New York, USA)~\cite{coburn2019design} (Fig.~\ref{fig06_TXM_method}). in~situ radiography (without rotation) can be done during solidification of immobile samples under a maximum frame rate of $\approx18$~frames per second.  Images with adequate signal\mbox{-}to\mbox{-}noise ratio and good x\mbox{-}ray absorption contrast between the three phases (solid Al, Al\textsubscript{3}Ni, and liquid) were obtained with an 8\mbox{-}12~keV monochromatic beam. After complete solidification, we acquired tomographic data (900 x\mbox{-}ray projection images evenly distributed between 0 and 180$^o$ rotation of the sample about {\bf z}).  This data was then reconstructed using the Tomopy package~\cite{gursoy2014tomopy} to reveal the 3D microstructure of the Al\mbox{-}Al\textsubscript{3}Ni eutectic.  Thanks to a high x-ray flux ($\approx 10^{13}~{\rm ph s^{-1}}$), short exposure time (0.05-0.1~s), and stable sample rotation ($3~{\rm \textdegree s^{-1}}$), tomography data can be collected in less than a minute \cite{ge2018one} (as opposed to the few\mbox{-}hours long tomography scans with a laboratory source). 

In a typical experiment, the micrometer\mbox{-}scale sample is affixed to an alumina rod via boron nitride (BN) paste, and this assembly is placed inside the furnace, at room temperature.  Then, the two blocks of the furnace are heated to the set temperatures of 773~K on the colder size and 1073~K on the hotter side, resulting in a positive thermal gradient along {\bf z} (Fig.~\ref{fig06_TXM_method}).  After a brief 2~min anneal, the sample is directionally melted by raising the temperature of the hot zone at a rate of 0.5\mbox{-}1 Kmin${\rm^{-1}}$.  Subsequently, DS is initiated by lowering the temperature of both blocks at the same, preset rate $\dot{T}$.  Simultaneously, a continuous stream of x\mbox{-}ray images is recorded in transmission. After solidification, the sample is removed from the furnace and imaged via nanotomography at the beamline.

\textcolor{black}{
To bring out contrast of the solid-liquid interfaces from the in~situ videos, the x-ray image frames can be digitally enhanced by dividing each by the initial frame (wherein the FOV is entirely liquid). Additional noise reduction can be achieved by registration of the images prior to division or through  a Total Variation algorithm with an $\ell^1$ data fidelity term \cite{brendt_wohlberg-proc-scipy-2017}. The dynamic range in the images can be further adjusted for clarity.} 

Finally, we used electron backscatter diffraction (EBSD) to characterize the crystallographic orientations of the eutectic solids in the seeded microstructure (created before the synchrotron experiments). EBSD was performed using Thermo Fisher Scientific Helios G5 UX SEM/FIB with an operating voltage and current of 20~keV and 26~nA, respectively. 


\FloatBarrier
\section{Results} \label{sec_results}
\subsection{In situ optical microscopy} \label{subsec_OM_results}

Optical observations during DS were obtained in a thin Al\mbox{-}Al\textsubscript{3}Ni sample with a slightly hypereutectic concentration \textcolor{black}{(see Fig.~\ref{SI_initial})}. The metallic film was solidified over a (relatively long) distance of about 17.5~mm at four different solidification velocities (see Fig.~\ref{panorama}). Two other experiments provided similar observations. This investigation allowed us to (i) exemplify the effect of the orientation of the growing Al\textsubscript{3}Ni crystals on the growth patterns, and (ii) bring experimental evidence for the morphological transformations from decoupled to coupled, and then to regular rod\mbox{-}like patterns upon increasing the solidification velocity. 

\begin{figure}[htbp]
\centering \includegraphics[width=8cm]{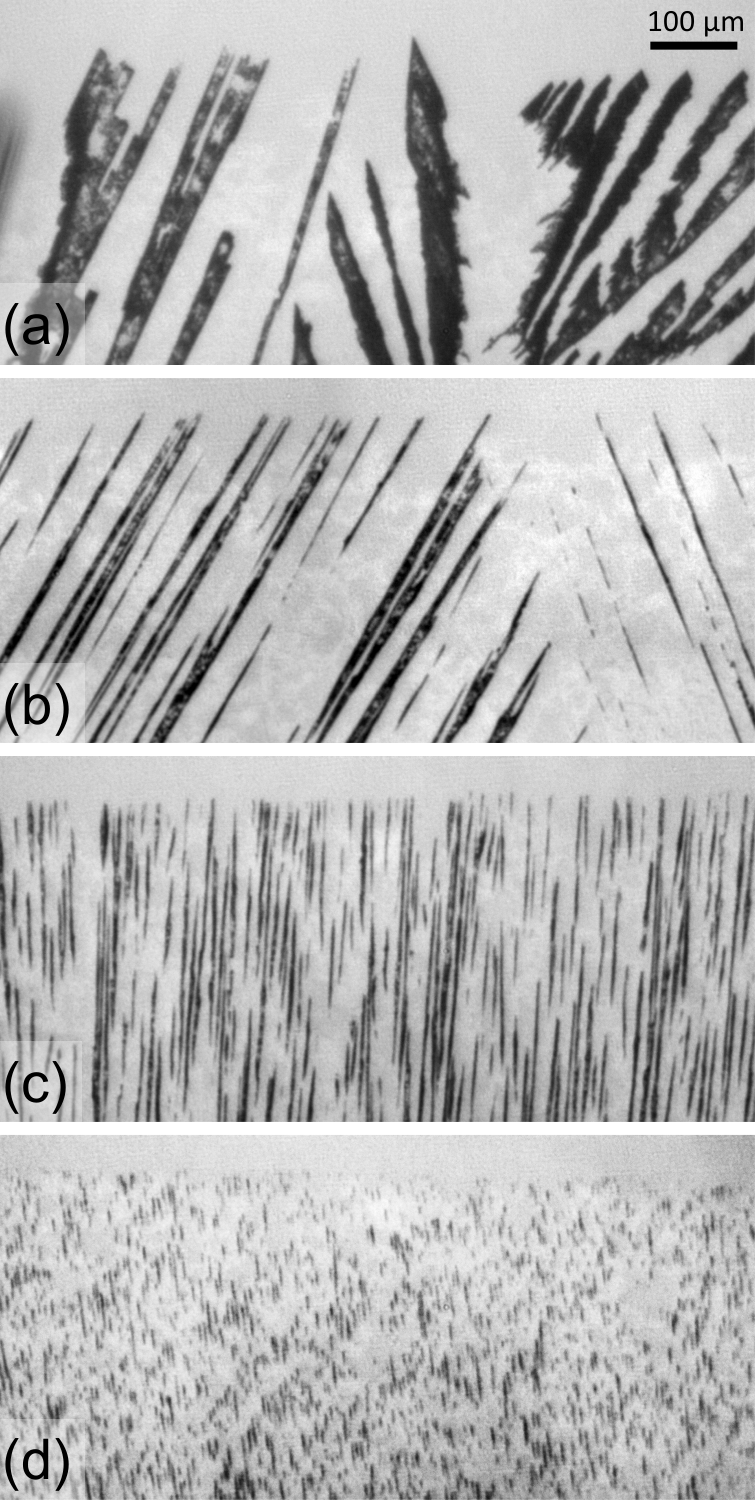}
\caption{In~situ optical images of DS patterns in a thin Al\mbox{-}Al\textsubscript{3}Ni sample of slightly hypereutectic concentration \textcolor{black}{(same sample as in Fig.~\ref{SI_initial})}. The growth direction is upward. The liquid (light grey) is on the top. In the solid (bottom), the Al\textsubscript{3}Ni crystals appear black, and the Al phase has a faint contrast with the liquid. (a) $V=0.5~{\rm \mu ms^{-1}}$; (b) $V=1.0~{\rm \mu ms^{-1}}$; (c) $V=5.0~{\rm \mu ms^{-1}}$; and (d) $V=10.0~{\rm \mu ms^{-1}}$.}
\label{Ech169_InSitu}
\end{figure}

Typical solidification patterns are shown in Fig.~\ref{Ech169_InSitu}. The upper part of the image with a light\mbox{-}grey contrast corresponds to the liquid. In the solid, the  Al\textsubscript{3}Ni crystals appear with a black contrast. The solid Al phase is faintly contrasted with the liquid, and exhibits an imperfect contact with the sapphire plate (a similar dynamic\mbox{-}wetting phenomenon was observed in previous studies using thin\mbox{-}sample DS of metallic eutectic alloys~\cite{bottinrousseau_lockedlamellar_2021,akamatsu2011determination}). At low velocity, the growth morphologies clearly depend on the orientation of the Al\textsubscript{3}Ni crystals. In the image shown in Fig.~\ref{Ech169_InSitu}(a)  ($V=0.5~{\rm \mu ms^{-1}}$), one can distinguish three eutectic grains. On the left part of the image, an array of decoupled\mbox{-}growth patterns involves tilted Al\textsubscript{3}Ni crystals delimited by two parallel low\mbox{-}mobility facets. Their tips protrude and grow slightly ahead of the Al\mbox{-}liquid interface. The Al\textsubscript{3}Ni crystals are prone to splitting, which may signal a transient coupling --- as is confirmed in Sec.~\ref{veloCh}. On the right part of the image, Al\textsubscript{3}Ni crystals with a different orientation also grow tilted, but their sharply pointed tips (within optical resolution) are delimited by two mobile facets in contact with the liquid. It is likely that the rear (low-temperature) edges of the facets coincide with stable trijunctions. This coupled-growth morphology is reminiscent of the pencil-like shape conjectured in Ref.~\cite{jaffrey_lamellar_1969}. The dynamics of the  central eutectic grain is qualitatively similar. In summary, in the low-velocity regime, the irregular eutectic growth dynamics is mostly sensitive to the orientation of the intermetallic crystals, and to the kinetics of the facets exposed to the liquid.

Over long solidification times, a dynamic competition between eutectic grains with tilted patterns leads to the extension of the left\mbox{-}hand grain, and the elimination of the pointed\mbox{-}tip Al\textsubscript{3}Ni crystals before the solidification was changed (see Fig.~\ref{panorama}). When $V$ is increased to $1~{\rm \mu ms^{-1}}$ (Fig.~\ref{Ech169_InSitu}(b)), the typical length\mbox{-}scale of the microstructures decreases, as expected. More interestingly, the distance between the Al\textsubscript{3}Ni crystal tips and the Al-liquid interface also decreases. The evolution toward a quasi isothermal coupled growth is evident at $V = 5~{\rm \mu ms^{-1}}$ (Fig.~\ref{Ech169_InSitu}(c)).  At a higher velocity ($V = 10~{\rm \mu ms^{-1}}$), the faceted character of the intermetallic phase seems to fade out, suggesting a morphological transition from an irregular to a regular coupled-growth dynamics (Fig.~\ref{Ech169_InSitu}(d)). Moreover, as $V$ increases, the Al\textsubscript{3}Ni fibers grow less and less tilted, differences between distinct eutectic grains are smoothed out, and the system operates like a regular, weakly anisotropic eutectic with poor crystal orientation effects. In the high\mbox{-}$V$ regime, the typical size of the Al\textsubscript{3}Ni crystals becomes substantially smaller than the thickness of the metallic film, and the microstructure, which is hardly resolved optically, evolves toward a confined\mbox{-}3D morphology. A combination of finite\mbox{-}size~\cite{akamatsu2007real,cserefouglu2023lamella} and crystal orientation~\cite{medjkoune2023formation} effects may be at the origin of the marked disorder of the microstructure.  For comparison, an evolution from decoupled to coupled growth upon increasing $V$  was also observed previously in a transparent, faceted/nonfaceted eutectic alloy \cite{mohagheghi2020decoupled}. However, in the transparent-alloy case, the thin needles were observed to keep their faceted character, and the growth microstructure remained markedly irregular over the whole range of investigated $V$ values, in contrast to the present system.

\begin{figure}[htbp]
\centering \includegraphics[width=8cm]{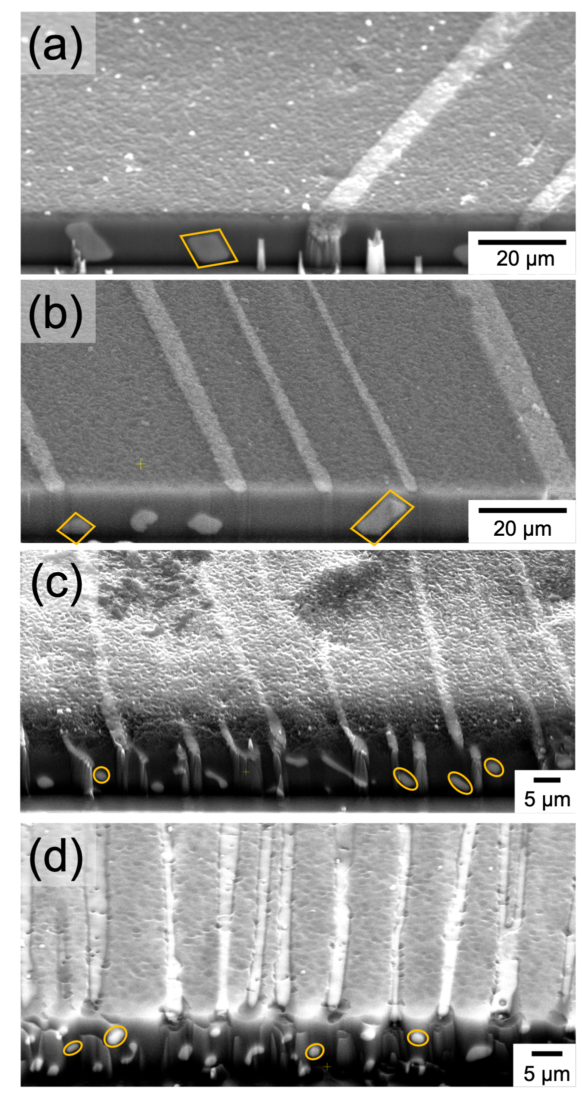}
\caption{Scanning electron microscopy (SEM) images of FIB cross\mbox{-}sections. Same Al\mbox{-}Al\textsubscript{3}Ni film as in Fig.~\ref{Ech169_InSitu}. The growth microstructures in (a), (b), (c) and (d) are representative of the patterns observed at $V=0.5, 1.0, 5.0$  and $10.0~{\rm \mu ms^{-1}}$, respectively. The microstructure becomes more and more three\mbox{-}dimensional, and the crystals more and more rounded as $V$ increases.  The perspective\mbox{-}view effect is due to the sample tilt (\ang{52}) in the SEM. Note that the imaging contrast between the eutectic solids is reversed in comparison to the optical micrographs in Fig.~\ref{Ech169_InSitu}.}
\label{fig05_FIB_Jan24}
\end{figure}

Further information, bringing strong support to the above analysis, has been obtained by ex~situ FIB cross sectioning and SEM observations of the thin metallic film in regions located close to the in~situ patterns of Fig.~\ref{Ech169_InSitu} (Fig.~\ref{fig05_FIB_Jan24}). In the low\mbox{-}velocity microstructures of Figs.~\ref{fig05_FIB_Jan24}(a) and \ref{fig05_FIB_Jan24}(b), the faceted growth of the Al\textsubscript{3}Ni crystals is clearly evidenced, in agreement with the in~situ optical images (note that the imaging contrast between the two eutectic solids is reversed when we compare the SEM images to the optical images given above). The cross-sections also reveal that some of the Al\textsubscript{3}Ni crystals connect the two surfaces of the metallic film --- these ones are visible optically in side view. In contrast, smaller faceted crystals that are fully surrounded by the Al solid do not contact the sapphire plate,  and cannot be seen in optical images --- or solely intermittently (Fig.~\ref{Ech169_InSitu}(b)). More significantly, the transition from an irregular\mbox{-}growth regime to a regular rod\mbox{-}like growth dynamics at higher velocity (Figs.~\ref{fig05_FIB_Jan24}(c\mbox{-}d)) is clearly evidenced. In the $V = 10~{\rm \mu ms^{-1}}$ cross\mbox{-}section (Fig.~\ref{fig05_FIB_Jan24}(d)), the contours of the Al\textsubscript{3}Ni crystals become rounded, and the faceted parts of the interphase boundaries are absent. 

In summary, OM observations during thin\mbox{-}sample DS of the Al\mbox{-}Al\textsubscript{3}Ni eutectic allowed us to identify unambiguously  a transition from irregular to regular growth upon increasing the solidification velocity from $V$ in the $1-{\rm \mu ms^{-1}}$ range to the $10-{\rm \mu ms^{-1}}$ range. In addition, while the decoupled\mbox{-}growth dynamics at low velocity is markedly crystal orientation dependent, the coupled\mbox{-}growth pattern at high velocity exhibits weakly anisotropic features. These findings are insightful, within some limitations. The 2D surface\mbox{-}level observations with intermittent access to the actual shape of the solid\mbox{-}liquid interfaces leaves us with some open questions on the mechanisms at play during the irregular\mbox{-}to\mbox{-}regular growth transition. The FIB sections of the metallic film demonstrate that, even at low velocity, the microstructure in the solid is basically 3D, but information on the detailed shape of the growth patterns during solidification is missing. The small\mbox{-}scale dynamics of trijunction lines also remains inaccessible to the optics. Finally, the above experiments are well designed for evidencing the crystal orientation dependence of the growth patterns at low solidification velocity, but the non\mbox{-}controlled, multiple\mbox{-}grain structure with tilted morphologies is not optimized with regards to the study of steady\mbox{-}state patterns. From this point of view, the seeding method that we employed for the synchrotron experiments (see the next section) represents a major improvement. 

\FloatBarrier

\subsection{Synchrotron transmission x-ray microscopy} \label{subsec_TXM_results}

\subsubsection{Constant-velocity experiments}\label{constvelo}
\begin{figure}[ht!]
\centering \includegraphics[width=\textwidth,height=\textheight,keepaspectratio]{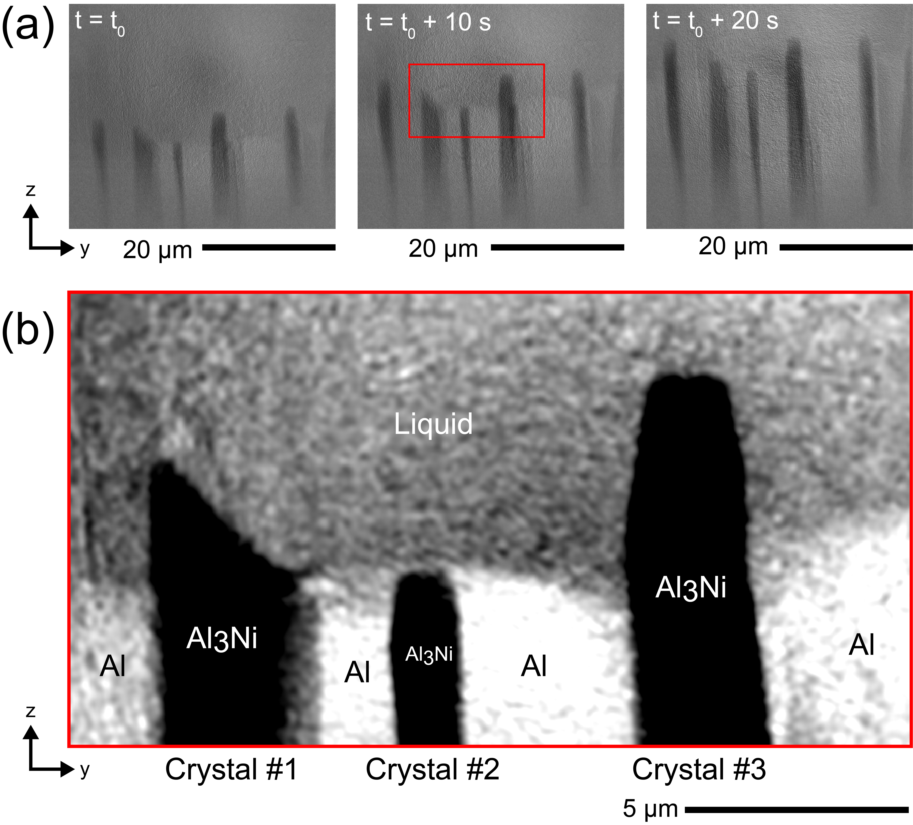}
\caption{(a) Three transmission x\mbox{-}ray microscopy (TXM) projection images, taken 10~s apart, from left to right (see labels), capturing directional eutectic solidification at constant velocity $V= 0.6~{\rm \mu ms^{-1}}$ ($G \approx 5~{\rm Kmm^{-1}}$). The growth direction is vertical. The liquid is dark grey (top part of the images), and the solid is made of solid Al (light grey) and Al\textsubscript{3}Ni crystals (dark). (b) Magnified view of a detail (see red frame in (a)), after image processing. \textcolor{black}{All image frames presented here have been digitally enhanced by dividing each by the initial frame (wherein the FOV was entirely liquid). The contrast in the images has been adjusted for clarity.} Also see Video~\ref{V_S1}.}
\label{fig07_rad}
\end{figure}

Figure \ref{fig07_rad} presents in~situ TXM images taken during near\mbox{-}steady conditions in a rectangular\mbox{-}pillar Al-Al\textsubscript{3}Ni sample (Fig.~\ref{SI_fig03_TXM_prep}). As the temperatures of the heating elements are decreased, the isotherms move upwards and the eutectic growth front progresses across the field\mbox{-}of\mbox{-}view (Fig.~\ref{fig07_rad}(a); also see Video~\ref{V_S1}). The average solidification velocity, measured in~situ, was $V= 0.6~{\rm \mu ms^{-1}}$, thus approximately matching the growth conditions as those of Fig.~\ref{Ech169_InSitu}(a). One can distinguish the liquid (on top), and, in the solid, the Al phase (light grey), and the Al\textsubscript{3}Ni crystals (dark). Five needle\mbox{-}like Al\textsubscript{3}Ni crystals are visible in the FOV --- but two crystals aligned in the path of the x\mbox{-}ray beam can superimpose in the image. The average growth front is slightly tilted from the horizontal, but this instrumental imperfection was of poor influence on the overall dynamics. As observed, the Al\textsubscript{3}Ni crystals protrude in the liquid ahead of the Al-liquid interface, and the lateral Al\textsubscript{3}Ni\mbox{-}liquid facets prolong into the frozen solid.  This is in agreement with the OM observations shown in the previous section. 

The TXM images can be nuanced by closer examination of the magnified region of Fig.~\ref{fig07_rad}(b), which shows three Al\textsubscript{3}Ni crystals (labelled 1 to 3, respectively, from the left to the right) and the Al solid phase surrounding them. Three distinct morphologies can be identified: a decoupled\mbox{-}growth pattern (crystal~\#3), a coupled\mbox{-}growth pattern (crystal~\#2), and an asymmetric pattern (crystal~\#1). In the decoupled\mbox{-}growth pattern, the (leading) Al\textsubscript{3}Ni crystal extends  into the liquid by about $6~{\rm \mu m}$ ahead of the Al\mbox{-}liquid interface. The latter exhibits a substantially curved, meniscus\mbox{-}like shape. It can be reasonably assumed that this configuration is essentially that of the left\mbox{-}side eutectic grain in Fig.~\ref{Ech169_InSitu}(a), the tilt angle apart. In contrast, in the coupled\mbox{-}growth pattern (crystal~\#2), the Al\textsubscript{3}Ni- and Al\mbox{-}liquid interfaces appear at the same level, with trijunctions on both sides. Finally, the asymmetric pattern (crystal~\#1) appears decoupled on the left side, and coupled on the right side, with a  trijunction pinned at the edge of the slanted Al\textsubscript{3}Ni\mbox{-}liquid facet. The time series of Fig. \ref{fig07_rad}(a) shows that the three configurations were stable over the observation time lapse --- except for a possible evolution of crystal~\#2 towards decoupling.  It is worth noting that, as evidenced by EBSD measurements (see below), the three Al\textsubscript{3}Ni crystals in Fig.~\ref{fig07_rad}(b) have the same orientation with respect to the {\bf z} axis, in contrast to the different eutectic grains in  Fig.~\ref{Ech169_InSitu}(a).

\begin{figure}[ht!]
\centering \includegraphics[width=0.75\textwidth,height=0.75\textheight,keepaspectratio]{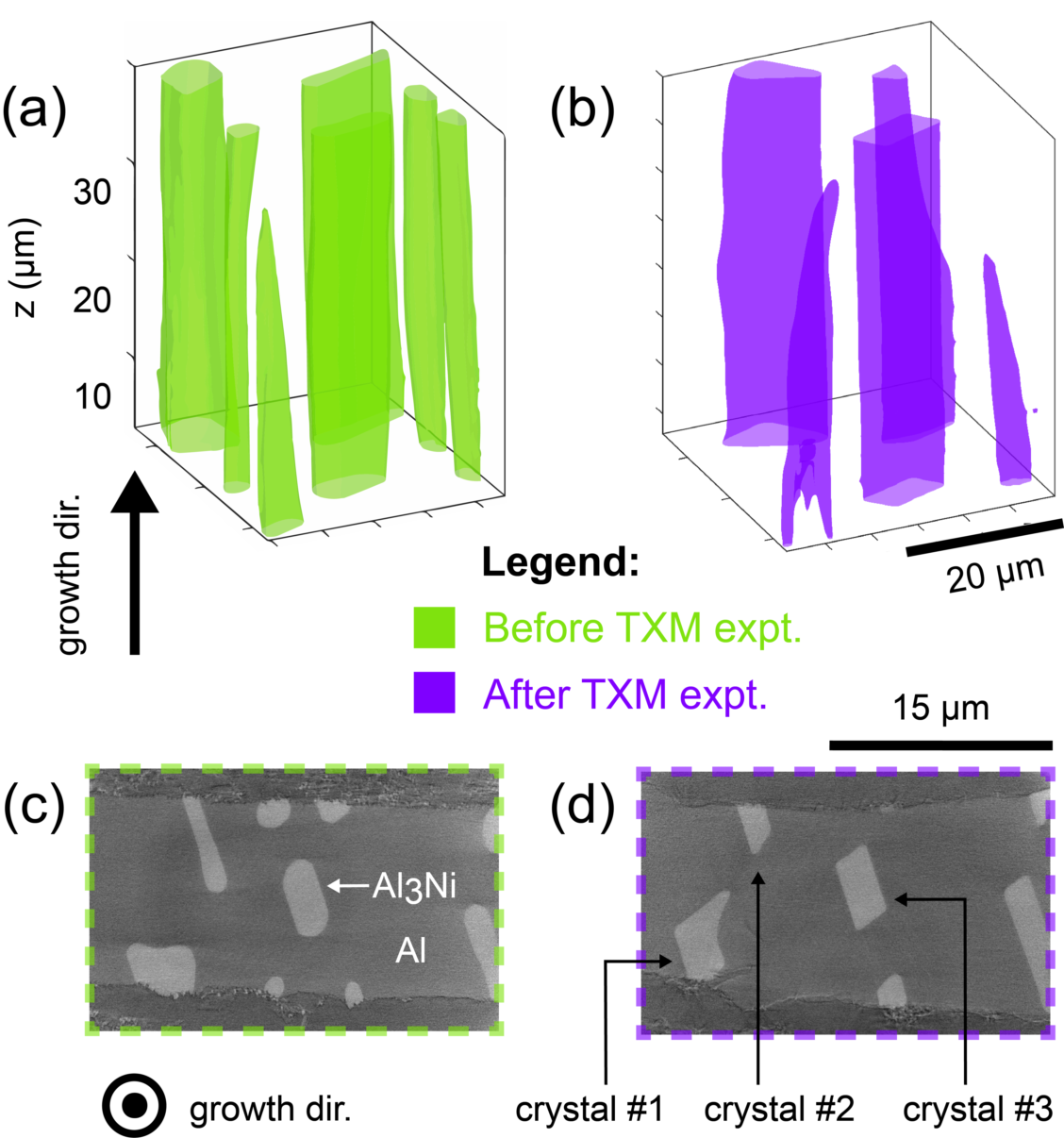}
\caption{Ex~situ, 3D nanotomographic reconstruction of the eutectic solid microstructure, highlighting the shape of the Al\textsubscript{3}Ni crystals: (a) before, and (b) after the same real\mbox{-}time TXM experiment as in Fig.~\ref{fig07_rad}. Selected cross\mbox{-}section images of the reconstruction before (c) and after (d) the solidification experiment. The Al\textsubscript{3}Ni crystals (light-grey contrast) are embedded in the Al matrix (darker contrast). Note that the imaging contrast between the two solids is reversed in comparison to the projection images in Fig.~\ref{fig07_rad}.} 
\label{fig08_recon}
\end{figure}

The Al\textsubscript{3}Ni crystals in Fig.~\ref{fig07_rad} grow parallel to each other, and in the average growth direction. This is obviously not by chance, but a successful result of our seeding method. Figs.~\ref{fig08_recon}(a) and  \ref{fig08_recon}(b) show 3D microstructures of the same region, reconstructed from x\mbox{-}ray nanotomography  data collected before DS in the seed, and after DS (experiment of Fig.~\ref{fig07_rad}), respectively.  Five Al\textsubscript{3}Ni crystals out of the seven ones present in the seed survived the early solidification stages. We checked (by superimposing the two 3D images) that the surviving crystals are in continuity with the seeding ones. In other words, the  solidification process essentially followed the original templated structure, with good alignment and position of the Al\textsubscript{3}Ni crystals before and after the synchrotron experiment. Let us consider the corresponding cross\mbox{-}sections of the tomographic volumes shown in Figs.~\ref{fig08_recon}(c) and \ref{fig08_recon}(d). It appears that the  Al\textsubscript{3}Ni crystals that were eliminated upon resolidification were contacting the free surfaces of the rectangular\mbox{-}pillar sample (note that the imaging contrast between the two eutectic solids is reversed when we compare the tomographic reconstructions to the projection images given above). Their size had been also reduced during fabrication of the micro\mbox{-}pillar.  Interestingly, the remaining  Al\textsubscript{3}Ni crystals adopted roughly similar shapes before and after resolidification, except that the seeding crystals appear much more rounded than the freshly solidified ones. This will be discussed later, in Sec.~\ref{crystSec}. 


By correlating the results from in~situ radiography and ex~situ tomography, we are now positioned to investigate the influence of the solidification front morphology on the microstructures left behind. The decoupled\mbox{-}growth pattern (case \#3) produced a fully faceted morphology (see again Fig.~\ref{fig08_recon}(d)) with a crystal orientation persisting as a remnant from the templated microstructure. The two sets of facets bounding the crystal are not of equal length; the broader one is nearly parallel to the viewing direction in x\mbox{-}radiography. In that case, the shape of the solid microstructure is essentially imposed by the Al\textsubscript{3}Ni\mbox{-}liquid facets during growth. Let us note that, in Fig.~\ref{fig08_recon}(d), one can measure the angle between adjacent facets to be of about 72\textdegree, which coincides with the angle (72.05\textdegree) between the ${\rm (101)}$ and  ${\rm (10\bar 1)}$ lattice planes (see Fig.~\ref{figXTRA_INDandXtal}). It follows that the growth direction is [010], which we  confirm by EBSD~(Fig.~\ref{SI_fig05_EBSD_PF}). In contrast, the asymmetric pattern (case \#1)  delivers a partly faceted microstructure. The two adjacent facets make the same angle as in the previous case, and the two involved Al\textsubscript{3}Ni crystals seem to have the same orientation. However, the shape of crystal \#1 in the solid presents concave, nonfaceted portions, which most likely are a trace of the region at which a trijunction could establish during growth. Surprisingly, the coupled\mbox{-}growth pattern (crystal \#2) yet shows faceted faces in Fig.~\ref{fig08_recon}(d). This would indicate that the corresponding growth pattern was only partly coupled --- the Al\textsubscript{3}Ni crystal was in contact with the sample wall, and the (distorted) 3D solidification dynamics was probably not fully captured in the radiograph. Finally, the above morphologies qualitatively resemble the microstructures revealed in the FIB cross\mbox{-}sections of Figs.~\ref{fig05_FIB_Jan24}(a) and \ref{fig05_FIB_Jan24}(b), which could be expected, considering the similarities in the solidification conditions. 

\begin{figure}[ht!]
\centering \includegraphics[width=\textwidth,height=\textheight,keepaspectratio]{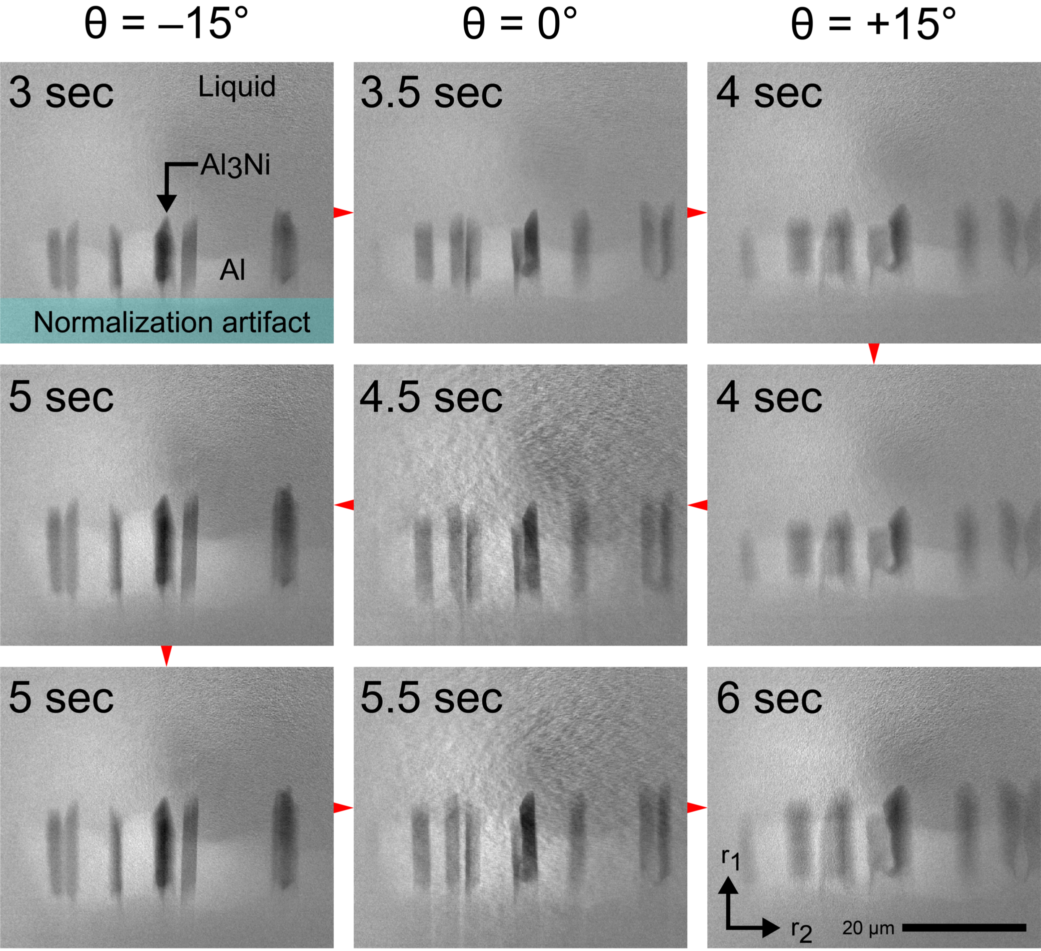}
\caption{Real-time monitoring of eutectic solidification under an imposed perturbation: The series of projection images was recorded during the ``rocking” experiment as the sample was rotated back\mbox{-}and\mbox{-}forth by 30\textdegree~about \textbf{z}. This motion results in a sinusoidal interface trace along one edge of the Al\textsubscript{3}Ni crystal while the other side of the same crystal is fully faceted, see Fig.~\ref{fig11_rock_recon}. The two different interfaces reflect different degrees of coupling between Al\textsubscript{3}Ni and Al phases in DS, refer to crystal-of-interest (black arrow) and the surrounding Al grains. Red arrows indicate the direction of data collection. \textcolor{black}{All image frames presented here have been digitally enhanced by dividing each by the initial frame (wherein the FOV was entirely liquid). The contrast in the images has been adjusted for clarity.}}
\label{fig12_rock_schematic}
\end{figure}

\begin{figure}[ht!]
\centering \includegraphics[width=\textwidth,height=\textheight,keepaspectratio]{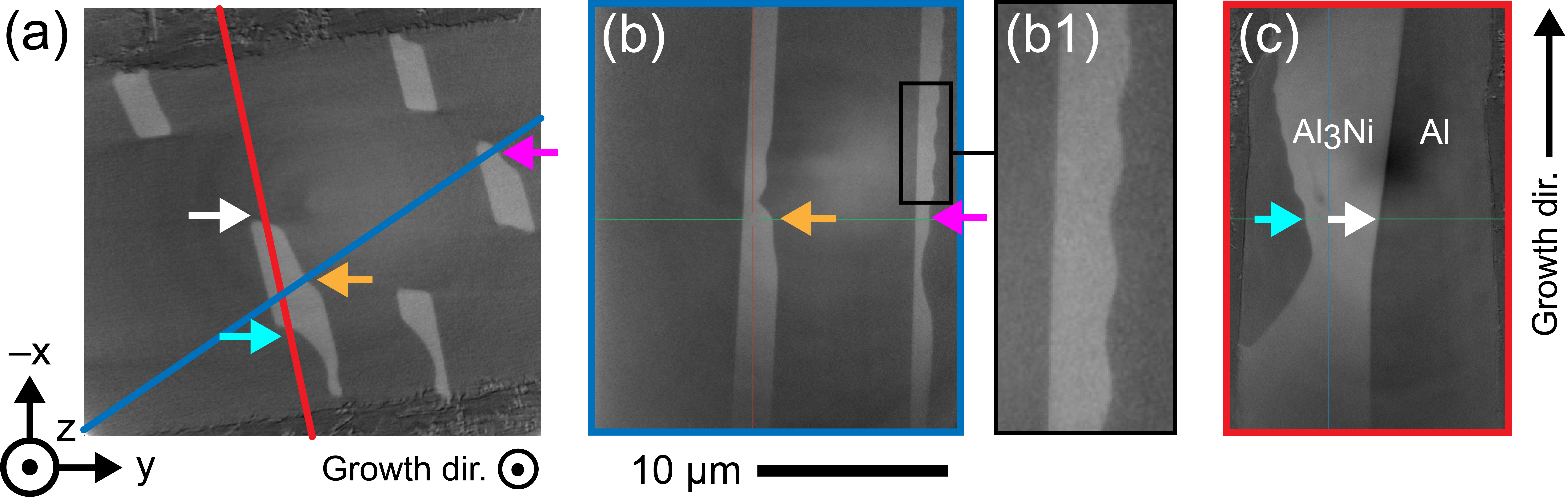}
\caption{Reconstruction of tomographic data from the ``rocking" experiment, shown as representative two\mbox{-}dimensional slices of the sample volume.  $V \approx 0.75~{\rm \mu ms^{-1}}$. The two images (b\mbox{-}c) are cuts of the data along the indicated directions (blue and red), which are orthogonal to the imaged plane in (a). Inset in (b1) shows a large\mbox{-}magnification view of the black\mbox{-}framed details in (b).  Blue and orange arrows in (b\mbox{-}c) point to a sinusoidal interface between Al and Al\textsubscript{3}Ni, which matches the back\mbox{-}and\mbox{-}forth rotation of the sample about {\bf z} during the ``rocking” experiment. The white arrow points to an Al\mbox{-}Al\textsubscript{3}Ni interface that remains flat throughout the experiment. Sample rotation began at a solidification length of around 40~{\textmu}m in the tomographic field\mbox{-}of\mbox{-}view (approximately 10~{\textmu}m from the bottom edge of (b\mbox{-}c)). }
\label{fig11_rock_recon}
\end{figure}

We  used a ``rocking" TXM experiment to test the response of the system to an external perturbation during solidification. During solidification, an oscillating (or rocking) motion is imposed to the sample about the main axis {\bf z} between two angular extremes $\pm 15$\textdegree~about the average position with a frequency $f_{\mathrm{rocking}}=0.5 {\rm Hz}$ (see Video~\ref{V_S3} and Fig.~\ref{fig12_rock_schematic}). \textcolor{black}{In comparison to the recent work of Ref.~\cite{qin2024synchrotron}, the imposed perturbation here is relatively weak.} Technically speaking, the rocking TXM can be viewed as an intermediate between x\mbox{-}radiography and continuous 4D x\mbox{-}ray tomography --- a method for 3D reconstruction from the limited angular views is under prospect. Here, in practice, we exploit a finite\mbox{-}perturbation (secondary) effect due to a modulation of the heat flux. Fig.~\ref{fig11_rock_recon} shows x\mbox{-}ray nanotomography observations (from a full angular range of 0 to 180\textdegree), obtained after complete solidification ($V\approx 0.75~{\rm \mu ms^{-1}}$) of the rectangular-pillar sample.  Let us focus on the largest  Al\textsubscript{3}Ni crystal in the cross\mbox{-}section of Fig.~\ref{fig11_rock_recon}(a), and consider the evolution of its partly faceted shape during the sample oscillation. For this purpose, we show two longitudinal sections along two different planes parallel to {\bf z}, thus revealing the typical motions of the various parts of the interface. In the first one (Fig. \ref{fig11_rock_recon}(b)), the absence of mobility of a facet is evidenced: the intersect line with the image plane is rectilinear. On the opposite side of the crystal, the interface exhibits a wavy shape, which corresponds to an oscillation of the interface with a wavelength of about $1.5~{\rm \mu m}$, which is equal to $Vf_{\mathrm{rocking}}$. Note that, in the same image, a similar behavior is recorded for another Al\textsubscript{3}Ni crystal (also see inset 1). In Fig.~\ref{fig11_rock_recon}(c), the wavy motion of the nonfaceted interface superimposes to a (slower) intrinsic dynamics of the system. It can also be noted that, in this image, the faceted part is not strictly a straight line, which signals a finite mobility of the shorter Al\textsubscript{3}Ni-liquid facets (hence the elongated shape of the crystals). In summary, this finite\mbox{-}perturbation test experiment clearly demonstrates the mobility of curved, concave interfaces of Al\textsubscript{3}Ni crystals, and the low mobility of the facets. The trijunction  is mobile as well, and readily adapts to the local growth conditions in the rocking experiment.

\FloatBarrier

\subsubsection{Velocity changes}\label{veloCh}
\begin{figure}[ht!]
\centering \includegraphics[width=\textwidth,height=\textheight,keepaspectratio]{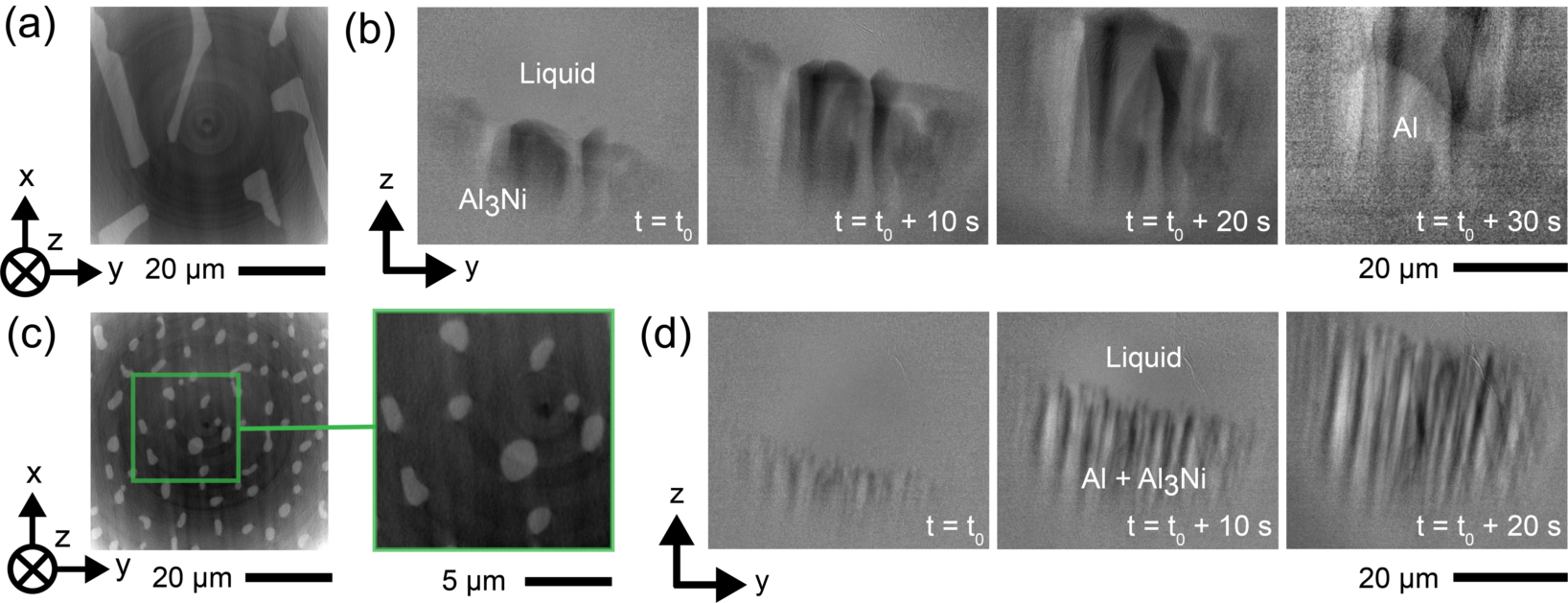}
\caption{Off\mbox{-}eutectic DS in a sample of composition Al\mbox{-}4.9at\%Ni (10wt\%Ni) reveals the relationship between coupled growth at higher velocities and the resulting non\mbox{-}faceted morphology. (a) Cross\mbox{-}sectional view of coarse, faceted microstructure grown under low growth velocity, $V\approx~1~{\rm \mu ms^{-1}}$, with (b) corresponding frames every 10 seconds from the real\mbox{-}time TXM experiment. We observe two fronts that traverse the field\mbox{-}of\mbox{-}view, the first corresponding to the Al\textsubscript{3}Ni\mbox{-}liquid interface, and the second to the Al\mbox{-}liquid interface. \textcolor{black}{The Al phase can be seen 30~s after the start of the experiment; hence, it has been marked only in the fourth image in the sequence.} (c) Finer, non\mbox{-}faceted microstructure (see inset) grown under higher  velocity, $V\approx~10~{\rm \mu ms^{-1}}$, with (d) corresponding frames every 1 second from the real\mbox{-}time TXM experiment. \textcolor{black}{Due to the coupling of the solid phases (at nearly the same isotherm), and the fine scale of the eutectic microstructure, the Al\textsubscript{3}Ni and Al growth fronts are difficult to distinguish.} Both experiments were done with $G= 15~{\rm K mm^{-1}}$ for purpose of comparison. \textcolor{black}{The image frames in (b,d) have been digitally enhanced by dividing each by the initial frame of the image sequence (wherein the FOV was entirely liquid). The contrast in the images has been adjusted for clarity.}}
\label{fig14_hypereut_rad}
\end{figure}

We observed an irregular\mbox{-}to\mbox{-}regular morphological transition upon increasing the solidification velocity in a hypereutectic (Al\mbox{-}10wt\%Ni) alloy --- that is, in conditions that are a priori more favorable for decoupled growth. We combine observations from in~situ x\mbox{-}radiography and ex~situ x\mbox{-}ray nanotomography. In Fig.~\ref{fig14_hypereut_rad}(a), one can see a cross\mbox{-}section of the sample solidified at a low velocity ($V\approx1~{\rm \mu ms^{-1}}$). As expected from the observations reported above, the microstructure is made of large faceted Al\textsubscript{3}Ni crystals in the Al matrix.  The in~situ sequence of TXM projections in Fig.~\ref{fig14_hypereut_rad}(b) (also see Video~\ref{V_S4}) reveals that  the intermetallic crystals grow decoupled, with the Al-liquid interface located relatively far behind the Al\textsubscript{3}Ni crystal tips (the Al front is observed, in practice, after a $\approx$30~s delay time in the FOV). The lead distance $\Delta z$ between the Al\textsubscript{3}Ni tips and the Al\mbox{-}liquid interface is estimated to  be of about $60~{\rm \mu m}$ (it is larger than the vertical dimension of the FOV). This value of $\Delta z$ is much larger than that measured in the near\mbox{-}eutectic alloy of Sec.~\ref{constvelo}, which is consistent with the large steepness of the Al\textsubscript{3}Ni liquidus in the phase diagram.  When $V$ is set to $\approx 10~{\rm \mu ms^{-1}}$, we observe fully non\mbox{-}faceted rods in the microstructure (Fig.~\ref{fig14_hypereut_rad}(c)). This agrees well with the results shown in Section \ref{subsec_OM_results}.  In the in~situ TXM image sequence of  Fig.~\ref{fig14_hypereut_rad}(d) (also see Video~\ref{V_S5}), the Al\mbox{-} and Al\textsubscript{3}Ni\mbox{-}liquid interfaces are indistinguishable, to within experimental resolution. This is a confirmation of the rod\mbox{-}like morphology at large velocity resulting from a coupling between Al and Al\textsubscript{3}Ni at the growth front.


\begin{figure}[ht!]
\centering \includegraphics[width=0.9\textwidth,height=\textheight,keepaspectratio]{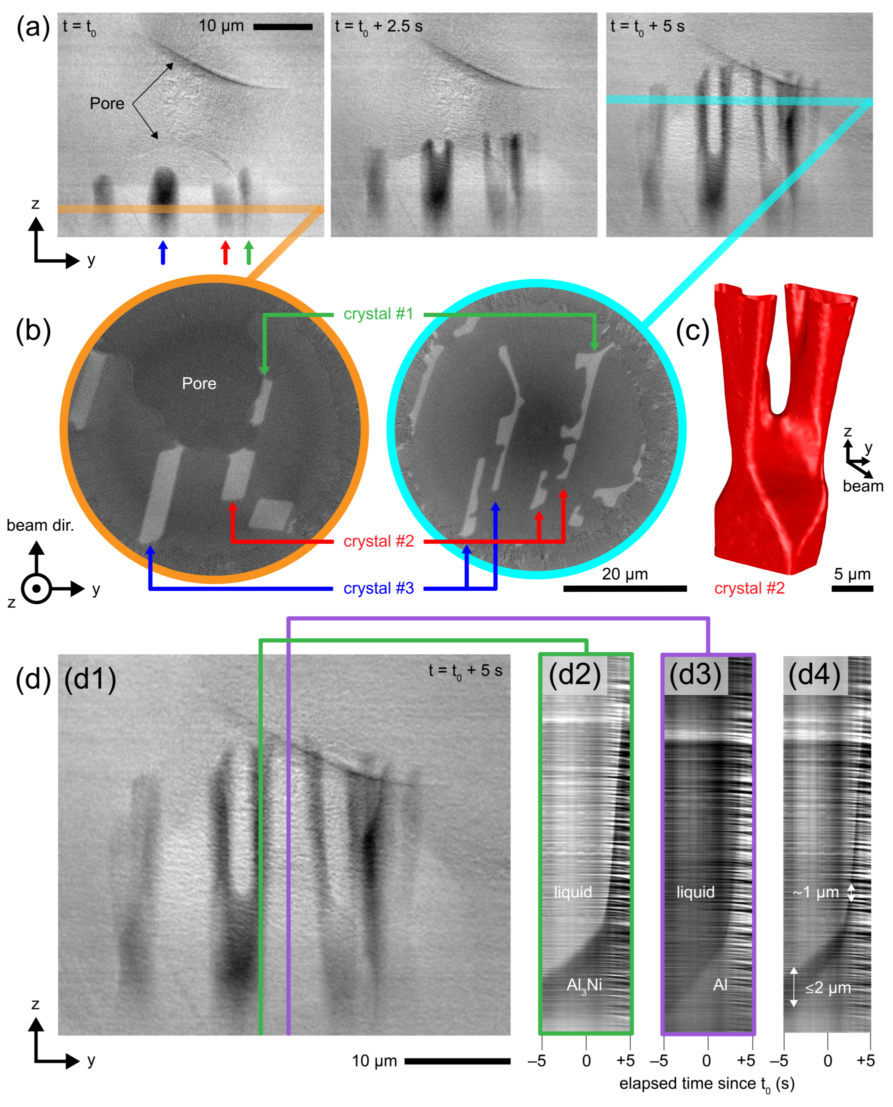}
\caption{Transient solidification dynamics under an imposed velocity jump: (a) Three in~situ TXM projection images. Time $t=t_0$ ($t=t_0+5~{\rm s}$): $V\approx 3~{\rm \mu ms^{-1}}$ ($V\approx 15~{\rm \mu ms^{-1}}$). \textcolor{black}{All image frames presented here have been digitally enhanced by dividing each by the initial frame (wherein the FOV was entirely liquid). The contrast in the images has been adjusted for clarity.} (b) Two cross\mbox{-}sections of the tomographic data obtained after complete solidification, corresponding to the orange- and teal-colored isopleths in (a).  Green, red, and blue arrows point to  Al\textsubscript{3}Ni crystals \#1, \#2, and \#3, respectively (unrelated to the three crystals in Figs.~\ref{fig07_rad}\mbox{-}\ref{fig08_recon}). The x\mbox{-}ray beam is directed into the page in (a) and upwards in (b). \textcolor{black}{(c) 3D reconstruction of crystal \#2, showing  splitting upon increasing $V$ in DS. (d) Spatiotemporal diagrams along reference lines parallel to  {\bf z}, as shown in (d1): the trajectories of tip of Al\textsubscript{3}Ni crystal \#1, and of the Al\mbox{-}liquid interface are visible in panels (d2) and (d3), respectively (weak dark\mbox{-}bright contrast; horizontal lines: image noise). (d4) Image made by superimposing the (d1) and (d2) images.}}
\label{fig09_vChange}
\end{figure}

In Fig.~\ref{fig14_hypereut_rad}, increasing the velocity gives rise to a qualitative change of the microstructure morphology, but also (less surprisingly) to a quantitative decrease of the average interphase spacing. We could examine a mechanism via which this occurs under non\mbox{-}steady conditions, as a consequence of an abrupt increase of the solidification velocity. Following the same method as in Fig.~\ref{fig14_hypereut_rad}, we use both x\mbox{-}radiography and ex~situ x\mbox{-}ray nanotomography. The TXM images of Fig.~\ref{fig09_vChange}(a) (also see Video~\ref{V_S2}) were obtained during solidification of a cylindrical\mbox{-}shaped sample (Fig.~\ref{fig06_TXM_method}). In this experiment, the temperatures of the two\mbox{-}zone furnace were decreased in such a way that the motion of the isotherms was  accelerated, and the actual growth velocity continuously increased from $V\approx 3~{\rm \mu ms^{-1}}$  to $V\approx 15~{\rm \mu ms^{-1}}$  within a short (5~s) interval of time. Initially, the solidification patterns visible in the field of view were of the \textcolor{black}{decoupled}-, or partly coupled\mbox{-}growth type, as expected in the corresponding $V$ range. Upon acceleration, a splitting of the Al\textsubscript{3}Ni crystals is observed. This leads to a neat decrease of the interphase spacing, which is ordinarily expected to accompany an increase of $V$ during eutectic growth. In this experiment, the solidification front adopted a markedly curved shape on the scale of the specimen, due to transient thermal conditions and an accidental perturbation induced by a large gas pore.  In spite of those instrumental imperfections, the observations clearly indicate that the splitting of Al\textsubscript{3}Ni crystals involves the appearance of nonfaceted interfaces characteristic of a partial coupling. Interestingly, a similar splitting dynamics is observed in Fig.~\ref{Ech169_InSitu}(a), that is at constant velocity, but as part of a multigrain, tilted\mbox{-}growth pattern with a markedly nonuniform interphase spacing. 

The 3D details of the splitting events are well captured by the two tomographic cross\mbox{-}sections of Figs.~\ref{fig09_vChange}(b). The first one shows the microstructure delivered during the initial, quasi steady\mbox{-}state dynamics captured in~situ at time $t=t_0$ in Fig.~\ref{fig09_vChange}(a)). The faceted shape of the  Al\textsubscript{3}Ni crystals is in accordance with the observed decoupled\mbox{-}growth dynamics. The presence of the gas pore does not seem to change qualitatively the growth dynamics. In the second tomographic cross\mbox{-}section, it can be seen that most of the crystals (e.g., crystals labelled $\#2$ and $\#3$) underwent a splitting. The details of the transient are quite complex, but a key feature is the appearance of nonfaceted, concave regions of the interfaces. This, again, signals a coupling with the Al solid. In the time of the observation, crystal \#1 did not split, but developed a characteristic ``C\mbox{-}shape" in cross-section. This C\mbox{-}shape has been observed in other systems consisting of a faceted phase \cite{aramanda_exotic_2020, aramanda_exotic_2021, pawlak2010far, kaczkan2011emission}. The transient dynamics shown here is expected to  lead to an irregular-to-regular transition. We could follow it thanks to the outstanding time and space resolution of synchrotron\mbox{-}based TXM. 

\begin{figure}[ht!]
\centering \includegraphics[width=\textwidth,height=\textheight,keepaspectratio]{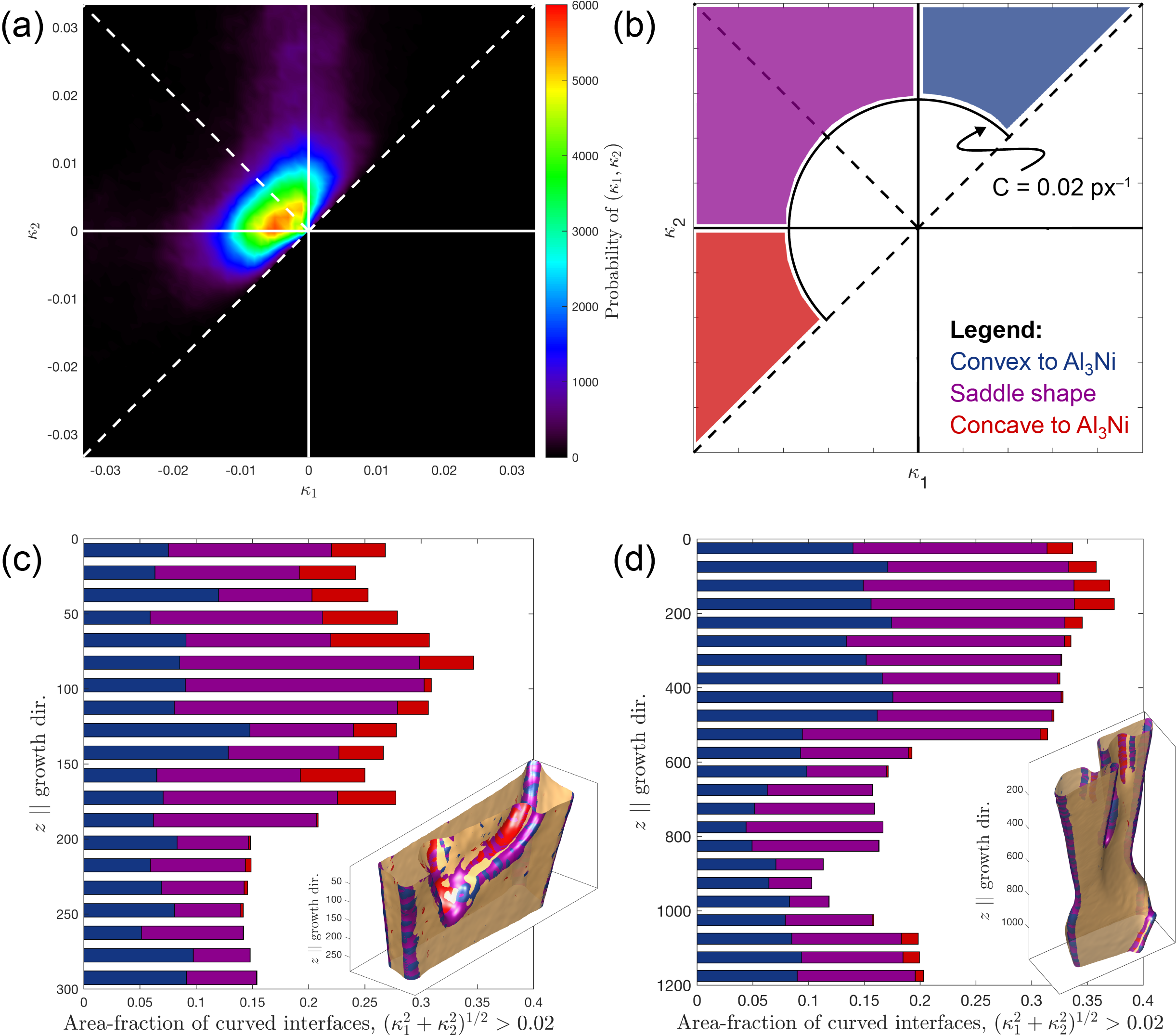}
\caption{Interfacial shape changes during imposed velocity jump: (a) Principal curvatures $\kappa_1$ and $\kappa_2$ measured for all patches of interface along a single Al\textsubscript{3}Ni crystal (pictured at inset in (c)). The distribution in principal curvatures is shown as a bivariate histogram (so\mbox{-}called interfacial shape distribution or ISD). We isolate those patches for which $\mathbb{C} \equiv \left({\kappa_1}^2+{\kappa_2}^2\right)^{1/2} > $~0.02~px\textsuperscript{\mbox{-}1} (or 0.4~nm\textsuperscript{\mbox{-}1}), corresponding to high ``curvedness”, see colored regions in (b). This is further separated into convex shapes (highlighted in blue), saddle shapes (purple), and concave shapes (red) with respect to the Al\textsubscript{3}Ni phase. Evolution of interfacial curvature in (c) crystal \#1 and (d) crystal \#2 from Fig.~\ref{fig09_vChange} along the growth direction {\bf z}. As a result of the velocity jump, the area fraction for which $\mathbb{C} >$~0.02~px\textsuperscript{\mbox{-}1} increases from $\approx$~15 to 30\%. The increase originates from the preponderance of saddle and concave shapes (with respect to the Al\textsubscript{3}Ni crystal). }
\label{fig10_velchange_ISD}
\end{figure}

To delve deeper into the morphological evolution of Al\textsubscript{3}Ni phase in the velocity jump experiment, we quantify the \textit{local} interfacial curvatures. In practice, we calculate the principal curvature components $\kappa_1$ and $\kappa_2$ for all patches of interface along Al\textsubscript{3}Ni and display the data in a so-called Interfacial Shape Distribution (ISD)~\cite{mendoza2003morphological,kammer2006morphological,fife2009morphological,kwon2007coarsening,felberbaum2011curvature,chen2013analysis,chen2013morphological,gibbs2015three,sun2017analytics,guo2017dendritic,wang2020integrated,arai2021coarsening,hu2022synthesis}. By definition, $\kappa_2 \ge \kappa_1$.  Observe in Fig.~\ref{fig10_velchange_ISD}(a) that the probability is concentrated near the origin, corresponding to zero or near\mbox{-}zero curvature associated with the broad facets of Al\textsubscript{3}Ni (as expected). Meanwhile, there is still a low but finite probability of finding highly curved interfaces along Al\textsubscript{3}Ni.  We suppose that interfaces have relatively high curvature when $\mathbb{C} \ge 0.4$~nm$^{-1}$, where the interfacial \textit{curvedness} $\mathbb{C}$ is given as $\mathbb{C} \equiv \left({\kappa_1}^2+{\kappa_2}^2\right)^{1/2}$~\cite{koenderink1992surface}.  We can further categorize such high-curvedness regions of the ISD as concave, convex or saddle\mbox{-}shaped~\cite{spanos2010three} with respect to the Al\textsubscript{3}Ni phase, if $\kappa_1$ and $\kappa_2$ are both negative, positive, or of mixed parity, respectively (each type is assigned a unique color in Fig.~\ref{fig10_velchange_ISD}(b)). Based on these definitions, we quantify the area fractions of concave, convex, and saddle shapes as a function of distance (or time) along the solidification direction, for the same Al\textsubscript{3}Ni crystals $\#1$ (Fig.~\ref{fig10_velchange_ISD}(c)) and $\#2$ (Fig.~\ref{fig10_velchange_ISD}(d)) from Fig.~\ref{fig09_vChange}.  For both crystals, the data reveal that the area fraction of highly curved interfaces increases from $\approx$15 to 30\% as the eutectic adapts to the velocity jump. Ultimately, acceleration leads to not only a microstructural refinement but also a concomitant proliferation of concave curvatures on Al\textsubscript{3}Ni, especially for the C\mbox{-}shaped crystals. 

\textcolor{black}{In a similar vein, ISDs in Fig.~\ref{SI_fig_9_hypereut_ISD} corroborate the qualitative observations in Figs.~\ref{fig14_hypereut_rad}(a,c). For the hyper\mbox{-}eutectic sample solidified at the higher growth rate, the ISD shows a high probability tail ($\kappa_1 \approx 0$, $\kappa_2 > 0$) corresponding to cylinder\mbox{-}like interfaces with respect to the Al\textsubscript{3}Ni. In contrast, at the lower growth rate, the ISD shows mostly planar interfaces. Thus, the interfacial morphology is strongly influenced by the solidification conditions.}

\FloatBarrier

\section{Discussion}\label{sec_disc}

\subsection{Growth front patterns and solidified microstructures}

\begin{figure}[ht!]
\centering \includegraphics[width=\textwidth,height=\textheight,keepaspectratio]{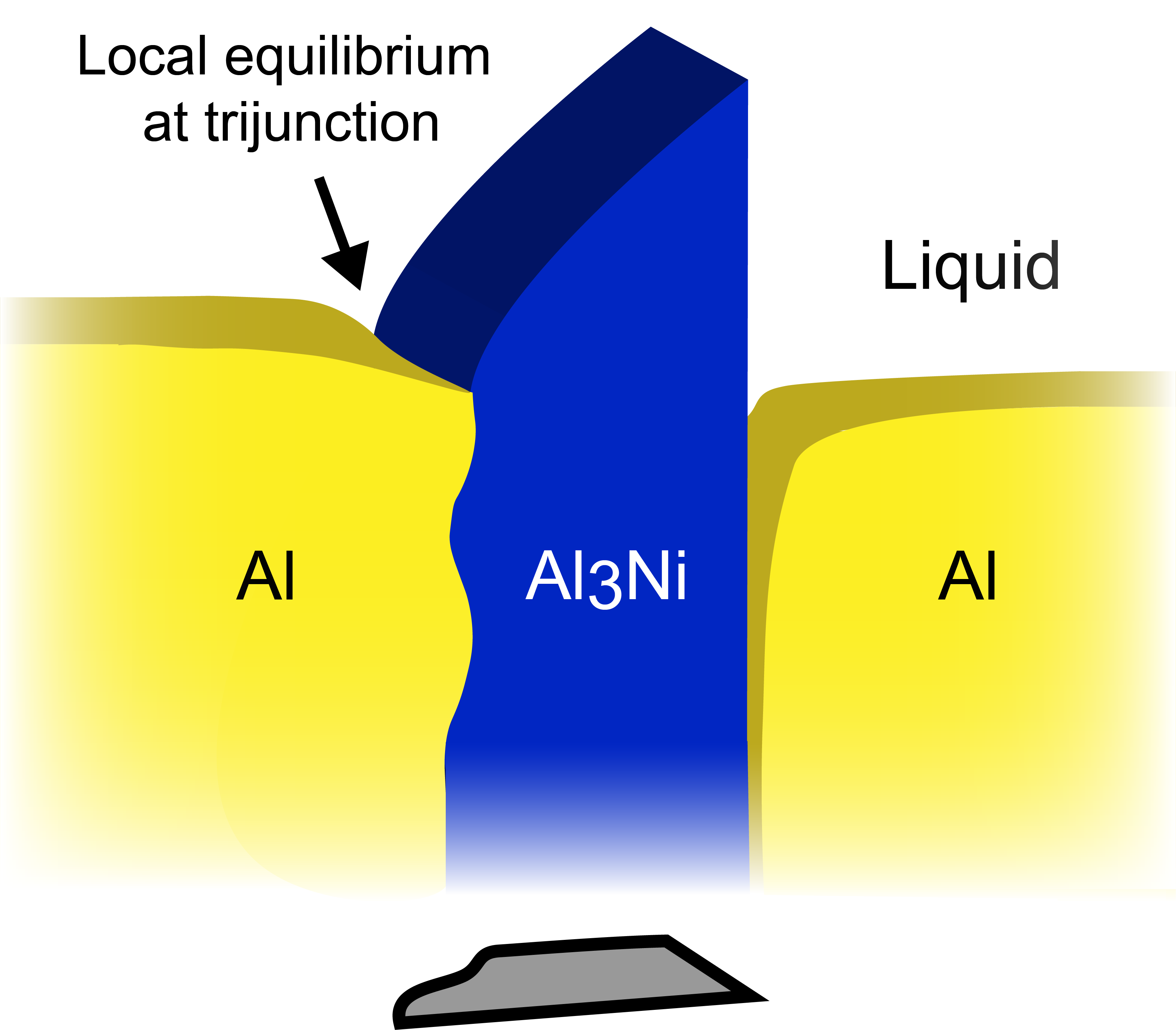}
\caption{Graphical illustration of the same Al\textsubscript{3}Ni crystal from Fig.~\ref{fig12_rock_schematic} represented in 3D. During growth, decoupling between Al and Al\textsubscript{3}Ni at the right edge of the intermetallic will result in a facet, whereas coupling between the two solids at the left edge will give rise to a mobile solid\mbox{-}solid\mbox{-}liquid trijunction. The grey shape at the bottom shows a 2D cut of the Al\textsubscript{3}Ni crystal perpendicular to its growth direction, cf.~Fig.~\ref{fig11_rock_recon}(a).}
\label{fig13_sketch}
\end{figure}

Through optical microscopy, FIB cross sectioning, and synchrotron\mbox{-}based TXM, we can correlate the real\mbox{-}time growth dynamics with the 3D structures left behind in the solid after solidification.  As a central result, our observations of the triangular tipped Al\textsubscript{3}Ni crystals in Figs.~\ref{fig07_rad} and \ref{fig08_recon} (also see Fig.~\ref{fig12_rock_schematic}) are the first to reconcile the previous reports that specify \textit{either} a faceted or non\mbox{-}faceted Al\textsubscript{3}Ni morphology.  In fact, smooth and rough interfaces can coexist (on the same crystal) at relatively low velocities in near\mbox{-}eutectic alloys as a direct consequence of the transient solid\mbox{-}liquid interfacial morphology. The existence of coupling on one side of a triangular tipped Al\textsubscript{3}Ni crystal gives rise to a mobile trijunction, as shown in the rocking experiment (Fig.~\ref{fig11_rock_recon}). The converse is true for the uncoupled side, which is resistant to any such perturbation and presents a blocked facet to the liquid. This facet is most likely wetted by a microscopically thin liquid layer, as suggested by Fig.~\ref{fig07_rad}(b).  It has not (yet) been possible to reconstruct the in~situ projection data from the rocking experiment by conventional approaches because of the limited angular range (the missing wedge problem in computed tomography~\cite{frank2008electron}). However, this type of data may be amenable to advanced post\mbox{-}processing and algorithms such as Discrete Algebraic Reconstruction Tomography (DART) where the known attenuation coefficients can be used a priori for the reconstruction~\cite{batenburg2007dart}. Nevertheless, we provide a sketch of the envisioned structure in Fig.~\ref{fig13_sketch}. This figure conveys our current understanding of how a mobile trijunction at a coupled growth front facilitates the development of non\mbox{-}faceted morphologies. 

At higher velocities, we find that the two solids are generally coupled at the growth front and hence the shape of Al\textsubscript{3}Ni ceases to be governed by faceted growth in the liquid (see Figs.~\ref{fig05_FIB_Jan24}(c-d) or \ref{fig14_hypereut_rad}(c-d)). The following scenario can be proposed. It is more than likely that the kinetic undercooling $\Delta T_k$ of Al\textsubscript{3}Ni facets increases with \textcolor{black}{the imposed} $V$ at a faster rate than the growth undercooling of the Al phase.  
\textcolor{black}{To test this idea semi quantitatively, we examine spatiotemporal diagrams of the Al\textsubscript{3}Ni-liquid interface and the Al-liquid interface (Fig.~\ref{fig09_vChange}(d)). The dark\mbox{-}bright boundary represents the growth front of the solid phases. The sharp change in slope of both trajectories coincides with the change in the growth velocity imposed at time $t \approx t_0 + 2.5~{\rm s}$. The combination (union) of the spatiotemporal plots in Fig.~\ref{fig09_vChange}(d3) provides evidence that the lead distance of the intermetallic nearly vanishes after the $V$ jump, at the resolution of TXM, thus signaling a coupled\mbox{-}growth dynamics.  A similar behavior has been reported in Ref.~\cite{mohagheghi2020decoupled}. That is, eventually}  the \textcolor{black}{side} Al\textsubscript{3}Ni\mbox{-}liquid facets would be completely covered by Al, the growth would become quasi isothermal, and the pinning of a closed-looped trijunction  between Al, Al\textsubscript{3}Ni, and liquid would be made possible.  In the high-$V$ regime, the eutectic growth would be indistinguishable from a regular, diffusion controlled rod\mbox{-}like pattern in solidification, and the Al\textsubscript{3}Ni crystal rods would lose their faceted cross\mbox{-}section. The role of the triple junctions in the roughening of the Al\textsubscript{3}Ni tips exposed to the liquid still remains to be assessed. Even so, this mechanism can help to explain the observations from the velocity jump experiment (Sec.~\ref{veloCh}).  As the velocity is progressively increased in DS, it is plausible that the lead distance between the Al\textsubscript{3}Ni tips and the Al growth front is gradually reduced until the two solids form a diffusively\mbox{-}coupled front with equilibrated trijunctions.  The trijunctions are free to move laterally in response to the velocity change, in turn splitting the Al\textsubscript{3}Ni crystal apart at the higher $V$ and simultaneously transforming its shape from faceted to concave (Fig.~\ref{fig10_velchange_ISD}(c\mbox{-}d)).  That is, diffusive coupling is likely a prerequisite for the ensuing morphological transformation. 

We may use a similar logic to understand the influence of composition (hypereutecticity) on the solidified eutectic microstructures.  In hypereutectic (or Ni-enriched) alloys, the tips of the Al\textsubscript{3}Ni crystals will grow behind (but near\mbox{-}to) the Al\textsubscript{3}Ni liquidus isotherm, i.e., well ahead of the Al front due to the steep liquidus of Al\textsubscript{3}Ni ($\approx 20 {\rm K~at\%^{-1}}$).  That is, the lead distance of Al\textsubscript{3}Ni will increase with a higher degree of hypereutecticity.  Concomittantly, the degree of faceting will also increase, since now more of the Al\textsubscript{3}Ni crystal is exposed to the liquid (see Figs.~\ref{fig14_hypereut_rad}(a-b)). In this scenario, the two phases would be weakly coupled (if at all). In the solid, the interphase boundaries would be contiguous with the blocked facets. 

\subsection{Impact of crystalline anisotropy on crystal shape}\label{crystSec}

The experimental data also gives us important clues about the  crystal orientation effects  on irregular Al\mbox{-}Al\textsubscript{3}Ni eutectic solidification. In general, one must consider two distinct anisotropies in interfacial energy, the first associated with the solid Al\textsubscript{3}Ni-liquid interfaces and the second with the solid Al\textsubscript{3}Ni-solid Al interfaces. We confirm previously reported observations \cite{jaffrey_lamellar_1969,cantor_crystallography_1975, wang_alignment_2008, li_effect_2010, jardine_effect_1986,farag_effect_1976} that Al\textsubscript{3}Ni fibers are typically oriented along the [010] direction, nearly parallel to the thermal gradient via EBSD, see Fig.~\ref{SI_fig05_EBSD_PF}. However, the details from the constant\mbox{-}velocity experiment (Sec.~\ref{constvelo}) reveal a nuanced distinction between the relative importance of the two anisotropies on the eutectic pattern that solidifies. 

Originally, the Al\mbox{-}Al\textsubscript{3}Ni alloy was a candidate system for high temperature environments \cite{farag_structure_1975,smartt_elevated_1971,pandey_development_2017, pandey_origin_2019, uan_study_1997,uan_extrusion_2001,uan_superplasticity-like_1996,uan_microstructural_2001,suwanpreecha_new_2018, suwanpreecha_strengthening_2019, koutsoukis_alternatives_2016} (up to $\approx 673$~K~\cite{nakagawa_thermal_1972} but below the eutectic temperature of 913~K) because the Al\textsubscript{3}Ni phase was thought to resist coarsening and maintain a fine microstructure suitable for structural applications \cite{houghton_thermal_1979,smartt_kinetics_1976,mclean_microstructural_1977,davies_microstructural_1980,suwanpreecha_effects_2019,kim_outstanding_2020, belov_improving_2004,czerwinski_strengthening_2022}. However, spheroidization of the Al\textsubscript{3}Ni crystals is actually inevitable, and occurs rapidly within 18 hours when annealed at 873~K or 0.93~$T_E$ \cite{smartt_elevated_1971,smartt_kinetics_1976}. In this context, the elliptical cross\mbox{-}sectional morphology of the Al\textsubscript{3}Ni rods in the original, templated microstructure (Fig.~\ref{fig08_recon}(b)), likely developed during a comparatively prolonged anneal (on the order of a few hours, while the sample was slowly drawn through the large heating zones of the furnace in the seeding experiment). This would suggest that the anisotropy in the solid-solid interfacial energy is relatively weak (i.e., it does not stabilize the facets).  Contrastingly, the as-solidified pattern, following the \textit{same} crystallographic template, does not exhibit a rounded morphology. Instead, it displays distinct facets that bound the the Al\textsubscript{3}Ni crystals, as shown postmortem in Fig.~\ref{fig08_recon}(d), only a few minutes after the real\mbox{-}time solidification experiment. Provided that the Al\textsubscript{3}Ni crystal surfaces are exposed to the liquid (as in Figs.~\ref{Ech169_InSitu}(a) or \ref{fig07_rad}(b)), the as\mbox{-}solidified patterns ``inherit" the anisotropy in the intermetallic-liquid interfacial energy. According to our calculations of the interfacial orientations (Fig.~\ref{figXTRA_INDandXtal}), the broadest facets are \{$101$\}. This is in agreement with the theoretical work of Tassoni and coworkers \cite{tassoni_morphologie}, who show that \{$101$\} is the lowest energy plane of Al\textsubscript{3}Ni.  Even though their focus was on examining the solid\mbox{-}vacuum interface rather than the solid\mbox{-}liquid interface, the relative trends are expected to persist \cite{jackson2010kinetic, sekerka2005equilibrium}. 


\section*{Conclusions}\label{sec_conc}
In this study, we conduct a series of in~situ experiments to visualize in real time the spatio-temporal evolution of coupled- and decoupled-growth patterns during DS of the Al\mbox{-}Al\textsubscript{3}Ni eutectic. By combining synchrotron\mbox{-}based transmission x\mbox{-}ray microscopy and regular optical microscopy, the following conclusions can be drawn from our work:
\begin{enumerate}
  \item Real\mbox{-}time optical microscopy (OM) brings to light a morphological transition from irregular eutectic growth at low solidification velocity $V$ to regular growth at higher $V$ in a near\mbox{-}eutectic Al\mbox{-}Al\textsubscript{3}Ni alloy --- with a cross-over value of $V$ of about $10~{\rm \mu ms^{-1}}$. \textcolor{black}{In general, this velocity is depends on several factors, including the alloy composition, thermal gradient, and the crystallographic orientation of the Al\textsubscript{3}Ni-liquid interfaces.} Postmortem FIB cross\mbox{-}sectioning reveals a concomitant transition in the shapes of the intermetallic crystals from faceted to rounded, respectively.
  \item Transmission x-ray microscopy (TXM) enables us to visualize at unprecedented, nanoscale resolution the detailed shape of the eutectic growth front, including the dynamics of the trijunction lines.  The experiments prove that diffusive coupling between the two eutectic solids leads to non-faceted interfaces while decoupling leads to the emergence of Al\textsubscript{3}Ni facets that persist into the solid. If the two solids are coupled at the growth front, the resulting triple junction is free to move laterally (as in the velocity jump and ``rocking" experiments). In this scenario, the eutectic behaves as a regular rod\mbox{-}like pattern in solidification.  
  \item The results from OM and TXM are self\mbox{-}consistent. Real-time observations of the growth patterns at the moving solid-liquid interface are the key to a good, unambiguous understanding of the formation of frozen microstructures in the bulk. This is particularly relevant for irregular eutectic growth involving non-isothermal growth shapes during DS.

\end{enumerate}  
 
It is indeed worth insisting --- and our work leaves little doubt on that --- that the selection of microstructure in an irregular eutectic is largely determined by the solid\mbox{-}liquid growth front morphology. The dynamic mechanisms, as revealed here, are far more complicated than that implied by the Jackson criterion (evaluated for the faceted phase alone) that we mentioned in the introduction (Sec.~\ref{sec_introduction}). For instance, the solidification dynamics is determined by the kinetics of faceted growth in a diffusion field --- which itself depends on the presence of crystallographic defects and the geometry and orientation of the intermetallic crystals --- as well as the diffusive interaction between the two eutectic solids (or lack thereof).  

The experimental results herein point to general principles of irregular eutectic solidification, that transcend material system.  For example, our real\mbox{-}time imaging data shows coupling and decoupling in the same way as that recently reported for an organic eutectic~\cite{mohagheghi2020decoupled}. The insights we gleaned may also be important for semi\mbox{-}metallic eutectics, e.g., that between Al and Si  \cite{hellawell_growth_1970}. Like Al\textsubscript{3}Ni, the Si phase is faceted with respect to the liquid~\cite{beatty2000monte}. Its mechanical properties such as tensile strength and ductility can be further improved by avoiding the formation of the broad Si facets \cite{tiwary_based_2015, lien_plastic_2022,ferrarini2004microstructure,suryawanshi2016simultaneous}. In light of our results, it is plausible that Al and Si phases could be fully coupled at the growth front under conditions of relatively high $V$ and high $G$, such that the shape of the Si phase would no longer be governed by faceted growth~\cite{major1989effect, pierantoni_coupled_1992, zhou2023three}. This morphological transformation from irregular to regular growth may well occur at undercoolings that are \textit{below} that required for the kinetic roughening of the Si crystal (on the order of 100~K~\cite{herlach2018crystal}). 

Finally, these complexities call for a systematic, theoretical (or numerical) investigation on pattern formation in eutectic systems with partly or fully faceted solid\mbox{-}liquid interfaces.  \textcolor{black}{Importantly, the simulations must reconcile the role of interfacial kinetics on the growth of the intermetallic phase, which itself is a function of the crystallographic orientations and the solidification conditions.} Such an investigation can answer some of the still\mbox{-}open questions, including the existence of steady\mbox{-}state decoupled patterns and their stability.


\section*{Acknowledgements}

We gratefully acknowledge financial support from the National Science Foundation (NSF) CAREER program under Award No.~1847855 and the Air Force Office of Scientific Research (AFOSR) under Award No.~FA9550-21-1-0260. We appreciate George R. Lindemann, Soumyadeep Dasgupta, Jonathan Goettsch, and Jaime Perez-Coronado for assisting in the synchrotron\mbox{-}based experiment. We thank Kyle Farmer, Dr.~Elizabeth Holm, Dr.~Jeffrey A. Fessler, and Dr.~Jonathan Schwartz for the insightful discussions on data processing and reconstruction. We extend our thanks to Kent Pruss from the University of Michigan Machine Shop for assisting with sample preparation. We also thank Jerry Kieffer, a micromaching expert for his guidance and advice during the sample preparation process (TXM) experiments, as well as Maria Achache, Lo\"ic Becerra and Erwan Dandeu for their help on the plasma sputtering. We acknowledge beamline scientists: Dr.~Mingyuan Ge, Dr.~Wah\mbox{-}Keat Lee, and beamline engineers: Stephen Antonelli, and Steve Bennett at the National Synchrotron Light Source II (NSLS\mbox{-}II). We also thank Dr.~Fernando Camino at the Center for Functional Nanomaterials (CFN). This research used resources 18\mbox{-}ID of the National Synchrotron Light Source II (NSLS\mbox{-}II) and the nanofabrication facilities at CFN, a U.S. Department of Energy (DOE) Office of Science User Facility operated for the DOE Office of Science by Brookhaven National Laboratory under Contract No.~DE\mbox{-}SC0012704. Additional support comes from the US Department of Energy, Office of Science, Office of Workforce Development for Teachers and Scientists, Office of Science Graduate Student Research (SCGSR) program, 2021 Solicitation 1 Award. Finally we thank the Michigan Center for Materials Characterization for use of the instruments and Bobby Kerns, Dr.~Allen H. Hunter, Dr.~Nancy Senabulya Muyanja, Dr.~Tao Ma, and Dr.~Haiping Sun for their assistance in that regard. One of us (AJS) benefited from a stay at INSP as an invited professor in the Physics department of Sorbonne University.

\newpage
\bibliography{more_refs}

\begin{thebibliography}{125}
\expandafter\ifx\csname natexlab\endcsname\relax\def\natexlab#1{#1}\fi
\providecommand{\url}[1]{\texttt{#1}}
\providecommand{\href}[2]{#2}
\providecommand{\path}[1]{#1}
\providecommand{\DOIprefix}{doi:}
\providecommand{\ArXivprefix}{arXiv:}
\providecommand{\URLprefix}{URL: }
\providecommand{\Pubmedprefix}{pmid:}
\providecommand{\doi}[1]{\href{http://dx.doi.org/#1}{\path{#1}}}
\providecommand{\Pubmed}[1]{\href{pmid:#1}{\path{#1}}}
\providecommand{\bibinfo}[2]{#2}
\ifx\xfnm\relax \def\xfnm[#1]{\unskip,\space#1}\fi
\bibitem[{Kurz et~al.(2022)Kurz, Fisher, and Rappaz}]{kurz_fundamentals_1992}
\bibinfo{author}{W.~Kurz}, \bibinfo{author}{D.~J. Fisher}, \bibinfo{author}{M.~Rappaz}, \bibinfo{title}{Fundamentals of solidification}, \bibinfo{edition}{5. rev. ed.} ed., \bibinfo{publisher}{Trans Tech Publications}, \bibinfo{year}{2022}.
\bibitem[{Dantzig and Rappaz(2016)}]{dantzig2016solidification}
\bibinfo{author}{J.~A. Dantzig}, \bibinfo{author}{M.~Rappaz}, \bibinfo{title}{{Solidification}}, \bibinfo{publisher}{EPFL press}, \bibinfo{year}{2016}.
\bibitem[{Jackson and Hunt(1966)}]{Jackson_Hunt_1966}
\bibinfo{author}{K.~Jackson}, \bibinfo{author}{J.~Hunt},
\newblock \bibinfo{title}{Lamellar and rod eutectic growth},
\newblock \bibinfo{journal}{Transactions of the Metallurgical Society of AIME} \bibinfo{volume}{236} (\bibinfo{year}{1966}) \bibinfo{pages}{1129--1142}.
\bibitem[{Akamatsu and Plapp(2016)}]{akamatsu_eutectic_2016}
\bibinfo{author}{S.~Akamatsu}, \bibinfo{author}{M.~Plapp},
\newblock \bibinfo{title}{Eutectic and peritectic solidification patterns},
\newblock \bibinfo{journal}{Current Opinion in Solid State and Materials Science} \bibinfo{volume}{20} (\bibinfo{year}{2016}) \bibinfo{pages}{46--54}.
\bibitem[{Decker(1973)}]{decker1973alloy}
\bibinfo{author}{R.~F. Decker},
\newblock \bibinfo{title}{Alloy design, using second phases},
\newblock \bibinfo{journal}{Metallurgical Transactions} \bibinfo{volume}{4} (\bibinfo{year}{1973}) \bibinfo{pages}{2495--2518}.
\bibitem[{Varin(2002)}]{varin2002intermetallic}
\bibinfo{author}{R.~Varin},
\newblock \bibinfo{title}{Intermetallic-reinforced light-metal matrix in-situ composites},
\newblock \bibinfo{journal}{Metallurgical and Materials Transactions A} \bibinfo{volume}{33} (\bibinfo{year}{2002}) \bibinfo{pages}{193--201}.
\bibitem[{Aikin(1997)}]{aikin_mechanical_1997}
\bibinfo{author}{R.~M. Aikin},
\newblock \bibinfo{title}{The mechanical properties of in-situ composites},
\newblock \bibinfo{journal}{{JOM}} \bibinfo{volume}{49} (\bibinfo{year}{1997}) \bibinfo{pages}{35}.
\bibitem[{Kaufman and Rooy(2004)}]{kaufman2004aluminum}
\bibinfo{author}{J.~G. Kaufman}, \bibinfo{author}{E.~L. Rooy}, \bibinfo{title}{Aluminum alloy castings: properties, processes, and applications}, \bibinfo{publisher}{Asm International}, \bibinfo{year}{2004}.
\bibitem[{Chadwick(1963)}]{chadwick_eutectic_1963}
\bibinfo{author}{G.~Chadwick},
\newblock \bibinfo{title}{Eutectic alloy solidification},
\newblock \bibinfo{journal}{Progress in Materials Science} \bibinfo{volume}{12} (\bibinfo{year}{1963}) \bibinfo{pages}{99--182}.
\bibitem[{Jaffrey and Chadwick(1970)}]{jaffrey_nucleation_1970}
\bibinfo{author}{D.~Jaffrey}, \bibinfo{author}{G.~A. Chadwick},
\newblock \bibinfo{title}{The nucleation, growth morphology, and thermal stability of {Sn-Zn} and {Al-Al\textsubscript{3}Ni} eutectic alloys},
\newblock \bibinfo{journal}{Metallurgical Transactions}  (\bibinfo{year}{1970}) \bibinfo{pages}{8}.
\bibitem[{Hertzberg et~al.(1965)Hertzberg, Ford, and Lemkey}]{hertzberg_microstructure_1965}
\bibinfo{author}{R.~W. Hertzberg}, \bibinfo{author}{J.~A. Ford}, \bibinfo{author}{F.~D. Lemkey},
\newblock \bibinfo{title}{The microstructure, crystallography and mechanical behavior of unidirectionally solidified {Al-Al\textsubscript{3}Ni} eutectic},
\newblock \bibinfo{journal}{Transactions of the Metallurgical Society of AIME}  (\bibinfo{year}{1965}).
\bibitem[{Ratke and Alkemper(2005)}]{ratke_ordering_2000}
\bibinfo{author}{L.~Ratke}, \bibinfo{author}{J.~Alkemper},
\newblock \bibinfo{title}{Ordering of the fibrous eutectic microstructure of {Al–Al\textsubscript{3}Ni} due to accelerated solidification conditions},
\newblock \bibinfo{journal}{Acta Materialia} \bibinfo{volume}{48} (\bibinfo{year}{2005}) \bibinfo{pages}{1939--1948}.
\bibitem[{Kaya et~al.(2010)Kaya, Boyuk, Cadirli, and Marasli}]{kaya_unidirectional_2010}
\bibinfo{author}{H.~Kaya}, \bibinfo{author}{U.~Boyuk}, \bibinfo{author}{E.~Cadirli}, \bibinfo{author}{N.~Marasli},
\newblock \bibinfo{title}{Unidirectional solidification of aluminium-nickel eutectic alloy},
\newblock \bibinfo{journal}{Metallic Materials} \bibinfo{volume}{48} (\bibinfo{year}{2010}) \bibinfo{pages}{291--300}.
\bibitem[{Cantor and Chadwick(1975)}]{cantor_crystallography_1975}
\bibinfo{author}{B.~Cantor}, \bibinfo{author}{G.~Chadwick},
\newblock \bibinfo{title}{Crystallography of {Al-Al\textsubscript{3}Ni, Al-Al\textsubscript{2}Cu and Al-$\zeta$(AlAg)} eutectics during nucleation and the early stages of growth},
\newblock \bibinfo{journal}{Journal of Crystal Growth} \bibinfo{volume}{30} (\bibinfo{year}{1975}) \bibinfo{pages}{101--108}. \bibinfo{note}{Publisher: North-Holland}.
\bibitem[{Lemkey et~al.(1965)Lemkey, Hertzberg, and Ford}]{lemkey_microstructure_1965}
\bibinfo{author}{F.~D. Lemkey}, \bibinfo{author}{R.~W. Hertzberg}, \bibinfo{author}{J.~A. Ford},
\newblock \bibinfo{title}{The microstructure, crystallography and mechanical behavior of unidirectionally solidified {Al-Al\textsubscript{3}Ni} eutectic},
\newblock \bibinfo{journal}{Transactions of the Metallurgical Society of AIME}  (\bibinfo{year}{1965}).
\bibitem[{Knowles and Goodhew(1983{\natexlab{a}})}]{knowles_structure_1983}
\bibinfo{author}{K.~M. Knowles}, \bibinfo{author}{P.~J. Goodhew},
\newblock \bibinfo{title}{The structure of interphase boundaries in an {Al-Al\textsubscript{3}Ni} directionally solidified eutectic alloy {I.} experimental observations},
\newblock \bibinfo{journal}{Philosophical Magazine A} \bibinfo{volume}{48} (\bibinfo{year}{1983}{\natexlab{a}}) \bibinfo{pages}{527--553}.
\bibitem[{Knowles and Goodhew(1983{\natexlab{b}})}]{knowles_structure_1983-1}
\bibinfo{author}{K.~M. Knowles}, \bibinfo{author}{P.~J. Goodhew},
\newblock \bibinfo{title}{The structure of interphase boundaries in an {Al-Al\textsubscript{3}Ni} directionally solidified eutectic alloy {II}. geometrical modelling},
\newblock \bibinfo{journal}{Philosophical Magazine A} \bibinfo{volume}{48} (\bibinfo{year}{1983}{\natexlab{b}}) \bibinfo{pages}{555--570}.
\bibitem[{Zhuang et~al.(2001)Zhuang, Zhang, Zhu, and Hu}]{zhuang_eutectic_2001}
\bibinfo{author}{Y.~Zhuang}, \bibinfo{author}{X.~Zhang}, \bibinfo{author}{L.~Zhu}, \bibinfo{author}{Z.~Hu},
\newblock \bibinfo{title}{Eutectic spacing and faults of directionally solidified {Al–Al\textsubscript{3}Ni} eutectic},
\newblock \bibinfo{journal}{Science and Technology of Advanced Materials} \bibinfo{volume}{2} (\bibinfo{year}{2001}) \bibinfo{pages}{37--39}.
\bibitem[{Smartt and Courtney(1972)}]{smartt_rod_1972}
\bibinfo{author}{H.~B. Smartt}, \bibinfo{author}{T.~H. Courtney},
\newblock \bibinfo{title}{On the rod to blade transition in the {Al-Al\textsubscript{3}Ni} eutectic},
\newblock \bibinfo{journal}{Metallurgical and Materials Transactions B} \bibinfo{volume}{3} (\bibinfo{year}{1972}) \bibinfo{pages}{2000--2002}.
\bibitem[{Jaffrey and Chadwick(1969)}]{jaffrey_lamellar_1969}
\bibinfo{author}{D.~Jaffrey}, \bibinfo{author}{G.~A. Chadwick},
\newblock \bibinfo{title}{The ``lamellar to fibrous transition" and orientation relationships in the {Sn-Zn} and {Al-Al\textsubscript{3}Ni} eutectic systems},
\newblock \bibinfo{journal}{Metallurgical Transactions}  (\bibinfo{year}{1969}).
\bibitem[{Hosch et~al.(2009)Hosch, England, and Napolitano}]{hosch_analysis_2009}
\bibinfo{author}{T.~Hosch}, \bibinfo{author}{L.~G. England}, \bibinfo{author}{R.~E. Napolitano},
\newblock \bibinfo{title}{Analysis of the high growth-rate transition in {Al}–{Si} eutectic solidification},
\newblock \bibinfo{journal}{Journal of Materials Science} \bibinfo{volume}{44} (\bibinfo{year}{2009}) \bibinfo{pages}{4892--4899}.
\bibitem[{Woodruff(1973)}]{woodruff1973solid}
\bibinfo{author}{D.~P. Woodruff}, \bibinfo{title}{The solid-liquid interface}, \bibinfo{publisher}{Cambridge University Press}, \bibinfo{year}{1973}.
\bibitem[{Beatty and Jackson(2000)}]{beatty2000monte}
\bibinfo{author}{K.~M. Beatty}, \bibinfo{author}{K.~A. Jackson},
\newblock \bibinfo{title}{Monte carlo modeling of silicon crystal growth},
\newblock \bibinfo{journal}{Journal of crystal growth} \bibinfo{volume}{211} (\bibinfo{year}{2000}) \bibinfo{pages}{13--17}.
\bibitem[{Jackson(2002)}]{jackson2002interface}
\bibinfo{author}{K.~A. Jackson},
\newblock \bibinfo{title}{The interface kinetics of crystal growth processes},
\newblock \bibinfo{journal}{Interface Science} \bibinfo{volume}{10} (\bibinfo{year}{2002}) \bibinfo{pages}{159--169}.
\bibitem[{Glicksman(2011)}]{glicksman_principles_2011}
\bibinfo{author}{M.~E. Glicksman}, \bibinfo{title}{Principles of Solidification}, \bibinfo{publisher}{Springer New York}, \bibinfo{year}{2011}. \DOIprefix\doi{10.1007/978-1-4419-7344-3}.
\bibitem[{Mara{\c{s}}li and Hunt(1996)}]{maracsli1996solid}
\bibinfo{author}{N.~Mara{\c{s}}li}, \bibinfo{author}{J.~Hunt},
\newblock \bibinfo{title}{Solid-liquid surface energies in the {Al-CuAl\textsubscript{2}, Al-NiAl\textsubscript{3} and Al-Ti} systems},
\newblock \bibinfo{journal}{Acta Materialia} \bibinfo{volume}{44} (\bibinfo{year}{1996}) \bibinfo{pages}{1085--1096}.
\bibitem[{Fan and Makhlouf(2015)}]{fan_al3ni_2015}
\bibinfo{author}{Y.~Fan}, \bibinfo{author}{M.~M. Makhlouf},
\newblock \bibinfo{title}{The {Al-Al\textsubscript{3}Ni} eutectic reaction: Crystallography and mechanism of formation},
\newblock \bibinfo{journal}{Metallurgical and Materials Transactions A} \bibinfo{volume}{46} (\bibinfo{year}{2015}) \bibinfo{pages}{3808--3812}.
\bibitem[{Peng et~al.(2020)Peng, Zhang, and Yue}]{peng_competitive_2020}
\bibinfo{author}{P.~Peng}, \bibinfo{author}{A.~Zhang}, \bibinfo{author}{J.~Yue},
\newblock \bibinfo{title}{Competitive growth of leading phase and tensile properties of directionally solidified eutectic {Al–Ni} alloy},
\newblock \bibinfo{journal}{Materials Science and Engineering: A} \bibinfo{volume}{773} (\bibinfo{year}{2020}) \bibinfo{pages}{138887}.
\bibitem[{Akamatsu and Nguyen-Thi(2016)}]{akamatsu_situ_2016}
\bibinfo{author}{S.~Akamatsu}, \bibinfo{author}{H.~Nguyen-Thi},
\newblock \bibinfo{title}{In situ observation of solidification patterns in diffusive conditions},
\newblock \bibinfo{journal}{Acta Materialia} \bibinfo{volume}{108} (\bibinfo{year}{2016}) \bibinfo{pages}{325--346}.
\bibitem[{Shahani et~al.(2020)Shahani, Xiao, Lauridsen, and Voorhees}]{shahani_characterization_2020}
\bibinfo{author}{A.~J. Shahani}, \bibinfo{author}{X.~Xiao}, \bibinfo{author}{E.~M. Lauridsen}, \bibinfo{author}{P.~W. Voorhees},
\newblock \bibinfo{title}{Characterization of metals in four dimensions},
\newblock \bibinfo{journal}{Materials Research Letters} \bibinfo{volume}{8} (\bibinfo{year}{2020}) \bibinfo{pages}{462--476}.
\bibitem[{Mohagheghi et~al.(2020)Mohagheghi, Bottin-Rousseau, Akamatsu, and {\c{S}}erefo{\u{g}}lu}]{mohagheghi2020decoupled}
\bibinfo{author}{S.~Mohagheghi}, \bibinfo{author}{S.~Bottin-Rousseau}, \bibinfo{author}{S.~Akamatsu}, \bibinfo{author}{M.~{\c{S}}erefo{\u{g}}lu},
\newblock \bibinfo{title}{Decoupled versus coupled growth dynamics of an irregular eutectic alloy},
\newblock \bibinfo{journal}{Scripta Materialia} \bibinfo{volume}{189} (\bibinfo{year}{2020}) \bibinfo{pages}{11--15}.
\bibitem[{Hou et~al.(2018)Hou, Belyakov, Pay, Sugiyama, Yasuda, and Gourlay}]{hou2018competition}
\bibinfo{author}{N.~Hou}, \bibinfo{author}{S.~Belyakov}, \bibinfo{author}{L.~Pay}, \bibinfo{author}{A.~Sugiyama}, \bibinfo{author}{H.~Yasuda}, \bibinfo{author}{C.~Gourlay},
\newblock \bibinfo{title}{Competition between stable and metastable eutectic growth in {Sn-Ni} alloys},
\newblock \bibinfo{journal}{Acta Materialia} \bibinfo{volume}{149} (\bibinfo{year}{2018}) \bibinfo{pages}{119--131}.
\bibitem[{Shahani et~al.(2016)Shahani, Xiao, and Voorhees}]{shahani_mechanism_2016}
\bibinfo{author}{A.~J. Shahani}, \bibinfo{author}{X.~Xiao}, \bibinfo{author}{P.~W. Voorhees},
\newblock \bibinfo{title}{The mechanism of eutectic growth in highly anisotropic materials},
\newblock \bibinfo{journal}{Nature Communications} \bibinfo{volume}{7} (\bibinfo{year}{2016}) \bibinfo{pages}{1--7}.
\bibitem[{Barclay et~al.(1971)Barclay, Kerr, and Niessen}]{barclay_off-eutectic_1971}
\bibinfo{author}{R.~S. Barclay}, \bibinfo{author}{H.~W. Kerr}, \bibinfo{author}{P.~Niessen},
\newblock \bibinfo{title}{Off-eutectic composite solidification and properties in {Al-Ni} and {Al-Co} alloys},
\newblock \bibinfo{journal}{Journal of Materials Science} \bibinfo{volume}{6} (\bibinfo{year}{1971}) \bibinfo{pages}{1168--1173}.
\bibitem[{Gwyer(1908)}]{gwyer1908alloys}
\bibinfo{author}{A.~Gwyer},
\newblock \bibinfo{title}{On the alloys of {Al} with {Cu}, {Fe}, {Ni}, {Co}, {Pb}, and {Cd}},
\newblock \bibinfo{journal}{Zeitschrift f\"{u}r anorganische und allgemeine Chemie} \bibinfo{volume}{57} (\bibinfo{year}{1908}) \bibinfo{pages}{113--153}.
\bibitem[{Fink and Freche(1934)}]{fink1934correlation}
\bibinfo{author}{W.~L. Fink}, \bibinfo{author}{H.~Freche},
\newblock \bibinfo{title}{Correlation of equilibrium relations in binary aluminum alloys of high purity},
\newblock \bibinfo{journal}{Trans. AIME} \bibinfo{volume}{111} (\bibinfo{year}{1934}) \bibinfo{pages}{14}.
\bibitem[{Fink and Willey(1934)}]{fink1934equilibrium}
\bibinfo{author}{W.~L. Fink}, \bibinfo{author}{L.~Willey},
\newblock \bibinfo{title}{Equilibrium relations in aluminium--nickel alloys of high purity},
\newblock \bibinfo{journal}{Trans. AIME} \bibinfo{volume}{111} (\bibinfo{year}{1934}) \bibinfo{pages}{293--303}.
\bibitem[{Okamoto and Okamoto(2000)}]{okamoto2000phase}
\bibinfo{author}{H.~Okamoto}, \bibinfo{author}{H.~Okamoto}, \bibinfo{title}{Phase diagrams for binary alloys}, volume~\bibinfo{volume}{44}, \bibinfo{publisher}{ASM international Materials Park, OH}, \bibinfo{year}{2000}.
\bibitem[{Du and Clavaguera(1996)}]{du1996thermodynamic}
\bibinfo{author}{Y.~Du}, \bibinfo{author}{N.~Clavaguera},
\newblock \bibinfo{title}{Thermodynamic assessment of the {Al-Ni} system},
\newblock \bibinfo{journal}{Journal of alloys and compounds} \bibinfo{volume}{237} (\bibinfo{year}{1996}) \bibinfo{pages}{20--32}.
\bibitem[{Ansara et~al.(1997)Ansara, Dupin, Lukas, and Sundman}]{ansara1997thermodynamic}
\bibinfo{author}{I.~Ansara}, \bibinfo{author}{N.~Dupin}, \bibinfo{author}{H.~L. Lukas}, \bibinfo{author}{B.~Sundman},
\newblock \bibinfo{title}{Thermodynamic assessment of the {Al-Ni} system},
\newblock \bibinfo{journal}{Journal of Alloys and Compounds} \bibinfo{volume}{247} (\bibinfo{year}{1997}) \bibinfo{pages}{20--30}.
\bibitem[{Huang and Chang(1998)}]{huang1998thermodynamic}
\bibinfo{author}{W.~Huang}, \bibinfo{author}{Y.~Chang},
\newblock \bibinfo{title}{A thermodynamic analysis of the {Ni-Al} system},
\newblock \bibinfo{journal}{Intermetallics} \bibinfo{volume}{6} (\bibinfo{year}{1998}) \bibinfo{pages}{487--498}.
\bibitem[{Chen et~al.(2011)Chen, Doernberg, Svoboda, and Schmid-Fetzer}]{chen2011thermodynamics}
\bibinfo{author}{H.-L. Chen}, \bibinfo{author}{E.~Doernberg}, \bibinfo{author}{P.~Svoboda}, \bibinfo{author}{R.~Schmid-Fetzer},
\newblock \bibinfo{title}{Thermodynamics of the {Al\textsubscript{3}Ni} phase and revision of the {Al--Ni} system},
\newblock \bibinfo{journal}{Thermochimica Acta} \bibinfo{volume}{512} (\bibinfo{year}{2011}) \bibinfo{pages}{189--195}.
\bibitem[{Yang et~al.(2022)Yang, Wang, Huang, and Zhang}]{yang2022thermodynamic}
\bibinfo{author}{W.~Yang}, \bibinfo{author}{P.~Wang}, \bibinfo{author}{X.~Huang}, \bibinfo{author}{S.~Zhang},
\newblock \bibinfo{title}{Thermodynamic analysis of the {Al--Ni} system},
\newblock \bibinfo{journal}{Intermetallics} \bibinfo{volume}{149} (\bibinfo{year}{2022}) \bibinfo{pages}{107647}.
\bibitem[{Bradley and Taylor(1937)}]{bradley1937crystal}
\bibinfo{author}{A.~Bradley}, \bibinfo{author}{A.~Taylor},
\newblock \bibinfo{title}{The crystal structures of {Ni\textsubscript{2}Al\textsubscript{3}} and {NiAl\textsubscript{3}}},
\newblock \bibinfo{journal}{Philosophical Magazine} \bibinfo{volume}{23} (\bibinfo{year}{1937}) \bibinfo{pages}{1049--67}.
\bibitem[{Pearson(2013)}]{pearson2013handbook}
\bibinfo{author}{W.~B. Pearson}, \bibinfo{title}{A handbook of lattice spacings and structures of metals and alloys: International series of monographs on metal physics and physical metallurgy}, volume~\bibinfo{volume}{4}, \bibinfo{publisher}{Elsevier}, \bibinfo{year}{2013}.
\bibitem[{Wang et~al.(2008)Wang, Wang, Liu, Wang, Zhang, and He}]{wang_alignment_2008}
\bibinfo{author}{Q.~Wang}, \bibinfo{author}{Z.~Wang}, \bibinfo{author}{T.~Liu}, \bibinfo{author}{C.~Wang}, \bibinfo{author}{C.~Zhang}, \bibinfo{author}{J.~He},
\newblock \bibinfo{title}{Alignment of primary {Al\textsubscript{3}Ni} phases in hypereutectic {Al-Ni} alloys with various compositions under high magnetic fields},
\newblock \bibinfo{journal}{Science in China Series E: Technological Sciences} \bibinfo{volume}{52} (\bibinfo{year}{2008}) \bibinfo{pages}{857}.
\bibitem[{Li et~al.(2010)Li, Fautrelle, Ren, Zhang, and Esling}]{li_effect_2010}
\bibinfo{author}{X.~Li}, \bibinfo{author}{Y.~Fautrelle}, \bibinfo{author}{Z.~Ren}, \bibinfo{author}{Y.~Zhang}, \bibinfo{author}{C.~Esling},
\newblock \bibinfo{title}{Effect of a high magnetic field on the {Al–Al\textsubscript{3}Ni} fiber eutectic during directional solidification},
\newblock \bibinfo{journal}{Acta Materialia} \bibinfo{volume}{58} (\bibinfo{year}{2010}) \bibinfo{pages}{2430--2441}.
\bibitem[{Jardine and Cantor(1986)}]{jardine_effect_1986}
\bibinfo{author}{F.~S. Jardine}, \bibinfo{author}{B.~Cantor},
\newblock \bibinfo{title}{The effect of hot-rolling on chill-cast {Al-Al\textsubscript{3}Ni}, chill-cast {Al-Al\textsubscript{2}Cu}, and unidirectionally solidified {Al-Al\textsubscript{3}Ni} eutectic alloys},
\newblock \bibinfo{journal}{Metall. Trans. A} \bibinfo{volume}{17A} (\bibinfo{year}{1986}).
\bibitem[{Farag and Flemings(1976)}]{farag_effect_1976}
\bibinfo{author}{M.~M. Farag}, \bibinfo{author}{M.~C. Flemings},
\newblock \bibinfo{title}{Effect of shape variations on the structure of directionally solidified {Al-Al\textsubscript{3}Ni} composites},
\newblock \bibinfo{journal}{Metallurgical Transactions A} \bibinfo{volume}{7} (\bibinfo{year}{1976}) \bibinfo{pages}{215--220}.
\bibitem[{Hunt and Jackson(1966)}]{hunt1966binary}
\bibinfo{author}{J.~Hunt}, \bibinfo{author}{K.~Jackson},
\newblock \bibinfo{title}{Binary eutectic solidification},
\newblock \bibinfo{journal}{Transactions of the Metallurgical Society of AIME} \bibinfo{volume}{236} (\bibinfo{year}{1966}) \bibinfo{pages}{843}.
\bibitem[{Witusiewicz et~al.(2013)Witusiewicz, Hecht, and Rex}]{witusiewicz2013situ}
\bibinfo{author}{V.~Witusiewicz}, \bibinfo{author}{U.~Hecht}, \bibinfo{author}{S.~Rex},
\newblock \bibinfo{title}{In-situ observation of eutectic growth in {Al}-based alloys by light microscopy},
\newblock \bibinfo{journal}{Journal of crystal growth} \bibinfo{volume}{372} (\bibinfo{year}{2013}) \bibinfo{pages}{57--64}.
\bibitem[{Akamatsu et~al.(2011)Akamatsu, Bottin-Rousseau, and Faivre}]{akamatsu2011determination}
\bibinfo{author}{S.~Akamatsu}, \bibinfo{author}{S.~Bottin-Rousseau}, \bibinfo{author}{G.~Faivre},
\newblock \bibinfo{title}{Determination of the {J}ackson--{H}unt constants of the {In}--{In\textsubscript{2}Bi} eutectic alloy based on in situ observation of its solidification dynamics},
\newblock \bibinfo{journal}{Acta materialia} \bibinfo{volume}{59} (\bibinfo{year}{2011}) \bibinfo{pages}{7586--7591}.
\bibitem[{Bottin-Rousseau et~al.(2021)Bottin-Rousseau, Medjkoune, Senninger, Carroz, Soucek, Hecht, and Akamatsu}]{bottinrousseau_lockedlamellar_2021}
\bibinfo{author}{S.~Bottin-Rousseau}, \bibinfo{author}{M.~Medjkoune}, \bibinfo{author}{O.~Senninger}, \bibinfo{author}{L.~Carroz}, \bibinfo{author}{R.~Soucek}, \bibinfo{author}{U.~Hecht}, \bibinfo{author}{S.~Akamatsu},
\newblock \bibinfo{title}{Locked-lamellar eutectic growth in thin {Al–Al\textsubscript{2}Cu} samples: In situ directional solidification and crystal orientation analysis},
\newblock \bibinfo{journal}{Journal of Crystal Growth} \bibinfo{volume}{570} (\bibinfo{year}{2021}) \bibinfo{pages}{126203}.
\bibitem[{Akamatsu et~al.(2007)Akamatsu, Bottin-Rousseau, Perrut, Faivre, Witusiewicz, and Sturz}]{akamatsu2007real}
\bibinfo{author}{S.~Akamatsu}, \bibinfo{author}{S.~Bottin-Rousseau}, \bibinfo{author}{M.~Perrut}, \bibinfo{author}{G.~Faivre}, \bibinfo{author}{V.~Witusiewicz}, \bibinfo{author}{L.~Sturz},
\newblock \bibinfo{title}{Real-time study of thin and bulk eutectic growth in succinonitrile--(d) camphor alloys},
\newblock \bibinfo{journal}{Journal of crystal growth} \bibinfo{volume}{299} (\bibinfo{year}{2007}) \bibinfo{pages}{418--428}.
\bibitem[{Akamatsu et~al.(2001)Akamatsu, Faivre, and Moulinet}]{akamatsu2001formation}
\bibinfo{author}{S.~Akamatsu}, \bibinfo{author}{G.~Faivre}, \bibinfo{author}{S.~Moulinet},
\newblock \bibinfo{title}{The formation of lamellar-eutectic grains in thin samples},
\newblock \bibinfo{journal}{Metallurgical and Materials Transactions A} \bibinfo{volume}{32} (\bibinfo{year}{2001}) \bibinfo{pages}{2039--2048}.
\bibitem[{Mohagheghi et~al.(2024)Mohagheghi, {\c{S}}erefo{\u{g}}lu, Akamatsu, and Bottin-Rousseau}]{mohagheghi2024pinning}
\bibinfo{author}{S.~Mohagheghi}, \bibinfo{author}{M.~{\c{S}}erefo{\u{g}}lu}, \bibinfo{author}{S.~Akamatsu}, \bibinfo{author}{S.~Bottin-Rousseau},
\newblock \bibinfo{title}{Pinning/depinning dynamics of trijunction lines during faceted/nonfaceted eutectic growth},
\newblock \bibinfo{journal}{Journal of Crystal Growth} \bibinfo{volume}{636} (\bibinfo{year}{2024}) \bibinfo{pages}{127705}.
\bibitem[{Chao et~al.(2022)Chao, Lindemann, Hunter, and Shahani}]{chao2022pseudo}
\bibinfo{author}{P.~Chao}, \bibinfo{author}{G.~R. Lindemann}, \bibinfo{author}{A.~H. Hunter}, \bibinfo{author}{A.~J. Shahani},
\newblock \bibinfo{title}{{Pseudo-4D} view of the growth and form of locked eutectic colonies},
\newblock \bibinfo{journal}{Acta Materialia} \bibinfo{volume}{240} (\bibinfo{year}{2022}) \bibinfo{pages}{118335}.
\bibitem[{Scheil and Zimmermann(1957)}]{scheil1957untersuchungen}
\bibinfo{author}{E.~Scheil}, \bibinfo{author}{R.~Zimmermann},
\newblock \bibinfo{title}{Untersuchungen {\"u}ber die eutektische kristallisation},
\newblock \bibinfo{journal}{International Journal of Materials Research} \bibinfo{volume}{48} (\bibinfo{year}{1957}) \bibinfo{pages}{509--516}.
\bibitem[{Kurz and Fisher(1979)}]{kurz1979dendrite}
\bibinfo{author}{W.~Kurz}, \bibinfo{author}{D.~Fisher},
\newblock \bibinfo{title}{Dendrite growth in eutectic alloys: the coupled zone},
\newblock \bibinfo{journal}{International Metals Reviews} \bibinfo{volume}{24} (\bibinfo{year}{1979}) \bibinfo{pages}{177--204}.
\bibitem[{Juarez-Hernandez and Jones(1998)}]{juarez-hernandez_growth_1998}
\bibinfo{author}{A.~Juarez-Hernandez}, \bibinfo{author}{H.~Jones},
\newblock \bibinfo{title}{Growth temperature measurements and solidification microstructure selection of primary {Al\textsubscript{3}Ni} and eutectic in the $\alpha${Al-Al\textsubscript{3}Ni} system},
\newblock \bibinfo{journal}{Scripta Materialia} \bibinfo{volume}{38} (\bibinfo{year}{1998}) \bibinfo{pages}{729--734}.
\bibitem[{Sun et~al.(2018)Sun, Hu, Lu, Ding, Xu, Xia, and Li}]{sun_situ_2018}
\bibinfo{author}{S.-Y. Sun}, \bibinfo{author}{Q.-D. Hu}, \bibinfo{author}{W.-Q. Lu}, \bibinfo{author}{Z.-Y. Ding}, \bibinfo{author}{M.-Q. Xu}, \bibinfo{author}{M.-X. Xia}, \bibinfo{author}{J.-G. Li},
\newblock \bibinfo{title}{In situ study on the growth behavior of primary {Al\textsubscript{3}Ni} phase in solidifying {Al–Ni} alloy by synchrotron radiography},
\newblock \bibinfo{journal}{Acta Metallurgica Sinica (English Letters)} \bibinfo{volume}{31} (\bibinfo{year}{2018}) \bibinfo{pages}{668--672}.
\bibitem[{Ding et~al.(2020{\natexlab{a}})Ding, Hu, Lu, Yang, Zhou, Zhang, Cao, Yu, and Li}]{ding_intergrowth_2020}
\bibinfo{author}{Z.~Ding}, \bibinfo{author}{Q.~Hu}, \bibinfo{author}{W.~Lu}, \bibinfo{author}{F.~Yang}, \bibinfo{author}{Y.~Zhou}, \bibinfo{author}{N.~Zhang}, \bibinfo{author}{S.~Cao}, \bibinfo{author}{L.~Yu}, \bibinfo{author}{J.~Li},
\newblock \bibinfo{title}{Intergrowth mechanism and morphology prediction of faceted {Al\textsubscript{3}Ni} formed during solidification by a spatial geometric model},
\newblock \bibinfo{journal}{Journal of Materials Science \& Technology} \bibinfo{volume}{54} (\bibinfo{year}{2020}{\natexlab{a}}) \bibinfo{pages}{40--47}.
\bibitem[{Ding et~al.(2020{\natexlab{b}})Ding, Hu, Yang, Lu, Yang, Cao, and Li}]{ding_new_2020}
\bibinfo{author}{Z.~Ding}, \bibinfo{author}{Q.~Hu}, \bibinfo{author}{F.~Yang}, \bibinfo{author}{W.~Lu}, \bibinfo{author}{T.~Yang}, \bibinfo{author}{S.~Cao}, \bibinfo{author}{J.~Li},
\newblock \bibinfo{title}{A new sight of the growth characteristics of solidified {Al\textsubscript{3}Ni} at the liquid–solid interface by synchrotron radiography and {3D} tomography},
\newblock \bibinfo{journal}{Metallurgical and Materials Transactions A} \bibinfo{volume}{51} (\bibinfo{year}{2020}{\natexlab{b}}) \bibinfo{pages}{2689--2696}.
\bibitem[{Yu et~al.(2021)Yu, Hu, Ding, Yang, Lu, Zhang, Cao, and Li}]{yu_effect_2021}
\bibinfo{author}{L.~Yu}, \bibinfo{author}{Q.~Hu}, \bibinfo{author}{Z.~Ding}, \bibinfo{author}{F.~Yang}, \bibinfo{author}{W.~Lu}, \bibinfo{author}{N.~Zhang}, \bibinfo{author}{S.~Cao}, \bibinfo{author}{J.~Li},
\newblock \bibinfo{title}{Effect of cooling rate on the {3D} morphology of the proeutectic {Al\textsubscript{3}Ni} intermetallic compound formed at the {Al/Ni} interface after solidification},
\newblock \bibinfo{journal}{Journal of Materials Science \& Technology} \bibinfo{volume}{69} (\bibinfo{year}{2021}) \bibinfo{pages}{60--68}.
\bibitem[{Antonelli et~al.(2020)Antonelli, Ronne, Han, Ge, Layne, Shahani, Iwamatsu, Wishart, Hulbert, Lee et~al.}]{antonelli2020versatile}
\bibinfo{author}{S.~Antonelli}, \bibinfo{author}{A.~Ronne}, \bibinfo{author}{I.~Han}, \bibinfo{author}{M.~Ge}, \bibinfo{author}{B.~Layne}, \bibinfo{author}{A.~J. Shahani}, \bibinfo{author}{K.~Iwamatsu}, \bibinfo{author}{J.~F. Wishart}, \bibinfo{author}{S.~L. Hulbert}, \bibinfo{author}{W.-K. Lee}, et~al.,
\newblock \bibinfo{title}{Versatile compact heater design for in situ nano-tomography by transmission {X-ray} microscopy},
\newblock \bibinfo{journal}{Journal of Synchrotron Radiation} \bibinfo{volume}{27} (\bibinfo{year}{2020}) \bibinfo{pages}{746--752}.
\bibitem[{Coburn et~al.(2019)Coburn, Nazaretski, Xu, Ge, Longo, Xu, Gofron, Yin, Chen, Hwu et~al.}]{coburn2019design}
\bibinfo{author}{D.~S. Coburn}, \bibinfo{author}{E.~Nazaretski}, \bibinfo{author}{W.~Xu}, \bibinfo{author}{M.~Ge}, \bibinfo{author}{C.~Longo}, \bibinfo{author}{H.~Xu}, \bibinfo{author}{K.~Gofron}, \bibinfo{author}{Z.~Yin}, \bibinfo{author}{H.~H. Chen}, \bibinfo{author}{Y.~Hwu}, et~al.,
\newblock \bibinfo{title}{Design, characterization, and performance of a hard {X-ray} transmission microscope at the {National Synchrotron Light Source II 18-ID} beamline},
\newblock \bibinfo{journal}{Review of Scientific Instruments} \bibinfo{volume}{90} (\bibinfo{year}{2019}) \bibinfo{pages}{053701}.
\bibitem[{G{\"u}rsoy et~al.(2014)G{\"u}rsoy, De~Carlo, Xiao, and Jacobsen}]{gursoy2014tomopy}
\bibinfo{author}{D.~G{\"u}rsoy}, \bibinfo{author}{F.~De~Carlo}, \bibinfo{author}{X.~Xiao}, \bibinfo{author}{C.~Jacobsen},
\newblock \bibinfo{title}{Tomopy: a framework for the analysis of synchrotron tomographic data},
\newblock \bibinfo{journal}{Journal of Synchrotron Radiation} \bibinfo{volume}{21} (\bibinfo{year}{2014}) \bibinfo{pages}{1188--1193}.
\bibitem[{Ge et~al.(2018)Ge, Coburn, Nazaretski, Xu, Gofron, Xu, Yin, and Lee}]{ge2018one}
\bibinfo{author}{M.~Ge}, \bibinfo{author}{D.~S. Coburn}, \bibinfo{author}{E.~Nazaretski}, \bibinfo{author}{W.~Xu}, \bibinfo{author}{K.~Gofron}, \bibinfo{author}{H.~Xu}, \bibinfo{author}{Z.~Yin}, \bibinfo{author}{W.-K. Lee},
\newblock \bibinfo{title}{One-minute nano-tomography using hard {X-ray} full-field transmission microscope},
\newblock \bibinfo{journal}{Applied Physics Letters} \bibinfo{volume}{113} (\bibinfo{year}{2018}) \bibinfo{pages}{083109}.
\bibitem[{{B}rendt {W}ohlberg(2017)}]{brendt_wohlberg-proc-scipy-2017}
\bibinfo{author}{{B}rendt {W}ohlberg},
\newblock \bibinfo{title}{{S}{P}{O}{R}{C}{O}: {A} {P}ython package for standard and convolutional sparse representations},
\newblock in: \bibinfo{editor}{{K}aty {H}uff}, \bibinfo{editor}{{D}avid {L}ippa}, \bibinfo{editor}{{D}illon {N}iederhut}, \bibinfo{editor}{M.~{P}acer} (Eds.), \bibinfo{booktitle}{{P}roceedings of the 16th {P}ython in {S}cience {C}onference}, \bibinfo{year}{2017}, pp. \bibinfo{pages}{1 -- 8}. \DOIprefix\doi{10.25080/shinma-7f4c6e7-001}.
\bibitem[{{\c{S}}erefo{\u{g}}lu et~al.(2023){\c{S}}erefo{\u{g}}lu, Bottin-Rousseau, and Akamatsu}]{cserefouglu2023lamella}
\bibinfo{author}{M.~{\c{S}}erefo{\u{g}}lu}, \bibinfo{author}{S.~Bottin-Rousseau}, \bibinfo{author}{S.~Akamatsu},
\newblock \bibinfo{title}{Lamella-rod pattern transition and confinement effects during eutectic growth},
\newblock \bibinfo{journal}{Acta Materialia} \bibinfo{volume}{242} (\bibinfo{year}{2023}) \bibinfo{pages}{118425}.
\bibitem[{Medjkoune et~al.(2023)Medjkoune, Bottin-Rousseau, Carroz, Pr{\'e}vot, Croset, Micha, and Akamatsu}]{medjkoune2023formation}
\bibinfo{author}{M.~Medjkoune}, \bibinfo{author}{S.~Bottin-Rousseau}, \bibinfo{author}{L.~Carroz}, \bibinfo{author}{G.~Pr{\'e}vot}, \bibinfo{author}{B.~Croset}, \bibinfo{author}{J.-S. Micha}, \bibinfo{author}{S.~Akamatsu},
\newblock \bibinfo{title}{Formation of locked-lamellar grains in a slightly hypoeutectic al-al2cu alloy during thin-sample directional solidification},
\newblock in: \bibinfo{booktitle}{IOP Conference Series: Materials Science and Engineering}, volume \bibinfo{volume}{1274}, \bibinfo{organization}{IOP Publishing}, \bibinfo{year}{2023}, p. \bibinfo{pages}{012037}.
\bibitem[{Qin et~al.(2024)Qin, Du, Cipiccia, Bodey, Rau, and Mi}]{qin2024synchrotron}
\bibinfo{author}{L.~Qin}, \bibinfo{author}{W.~Du}, \bibinfo{author}{S.~Cipiccia}, \bibinfo{author}{A.~J. Bodey}, \bibinfo{author}{C.~Rau}, \bibinfo{author}{J.~Mi},
\newblock \bibinfo{title}{Synchrotron x-ray operando study and multiphysics modelling of the solidification dynamics of intermetallic phases under electromagnetic pulses},
\newblock \bibinfo{journal}{Acta Materialia} \bibinfo{volume}{265} (\bibinfo{year}{2024}) \bibinfo{pages}{119593}.
\bibitem[{Aramanda et~al.(2020)Aramanda, Salapaka, Khanna, Chattopadhyay, and Choudhury}]{aramanda_exotic_2020}
\bibinfo{author}{S.~K. Aramanda}, \bibinfo{author}{S.~K. Salapaka}, \bibinfo{author}{S.~Khanna}, \bibinfo{author}{K.~Chattopadhyay}, \bibinfo{author}{A.~Choudhury},
\newblock \bibinfo{title}{Exotic colony formation in {Sn-Te} eutectic system},
\newblock \bibinfo{journal}{Acta Materialia} \bibinfo{volume}{197} (\bibinfo{year}{2020}) \bibinfo{pages}{108--121}.
\bibitem[{Aramanda et~al.(2021)Aramanda, Chattopadhyay, and Choudhury}]{aramanda_exotic_2021}
\bibinfo{author}{S.~K. Aramanda}, \bibinfo{author}{K.~Chattopadhyay}, \bibinfo{author}{A.~Choudhury},
\newblock \bibinfo{title}{Exotic three-phase microstructures in the ternary {Ag-Cu-Sb} eutectic system},
\newblock \bibinfo{journal}{Acta Materialia} \bibinfo{volume}{221} (\bibinfo{year}{2021}) \bibinfo{pages}{117400}.
\bibitem[{Pawlak et~al.(2010)Pawlak, Turczynski, Gajc, Kolodziejak, Diduszko, Rozniatowski, Smalc, and Vendik}]{pawlak2010far}
\bibinfo{author}{D.~A. Pawlak}, \bibinfo{author}{S.~Turczynski}, \bibinfo{author}{M.~Gajc}, \bibinfo{author}{K.~Kolodziejak}, \bibinfo{author}{R.~Diduszko}, \bibinfo{author}{K.~Rozniatowski}, \bibinfo{author}{J.~Smalc}, \bibinfo{author}{I.~Vendik},
\newblock \bibinfo{title}{How far are we from making metamaterials by self-organization? the microstructure of highly anisotropic particles with an {SRR-like} geometry},
\newblock \bibinfo{journal}{Advanced Functional Materials} \bibinfo{volume}{20} (\bibinfo{year}{2010}) \bibinfo{pages}{1116--1124}.
\bibitem[{Kaczkan et~al.(2011)Kaczkan, Pawlak, Turczy{\'n}ski, and Malinowski}]{kaczkan2011emission}
\bibinfo{author}{M.~Kaczkan}, \bibinfo{author}{D.~Pawlak}, \bibinfo{author}{S.~Turczy{\'n}ski}, \bibinfo{author}{M.~Malinowski},
\newblock \bibinfo{title}{Emission properties of {(SrTiO\textsubscript{3}--TiO\textsubscript{2})}: Pr\textsuperscript{3+} eutectic with self-organized fractal microstructure},
\newblock \bibinfo{journal}{Optical Materials} \bibinfo{volume}{33} (\bibinfo{year}{2011}) \bibinfo{pages}{1519--1524}.
\bibitem[{Mendoza et~al.(2003)Mendoza, Alkemper, and Voorhees}]{mendoza2003morphological}
\bibinfo{author}{R.~Mendoza}, \bibinfo{author}{J.~Alkemper}, \bibinfo{author}{P.~Voorhees},
\newblock \bibinfo{title}{The morphological evolution of dendritic microstructures during coarsening},
\newblock \bibinfo{journal}{Metallurgical and Materials Transactions} \bibinfo{volume}{34} (\bibinfo{year}{2003}) \bibinfo{pages}{481}.
\bibitem[{Kammer and Voorhees(2006)}]{kammer2006morphological}
\bibinfo{author}{D.~Kammer}, \bibinfo{author}{P.~Voorhees},
\newblock \bibinfo{title}{The morphological evolution of dendritic microstructures during coarsening},
\newblock \bibinfo{journal}{Acta materialia} \bibinfo{volume}{54} (\bibinfo{year}{2006}) \bibinfo{pages}{1549--1558}.
\bibitem[{Fife and Voorhees(2009)}]{fife2009morphological}
\bibinfo{author}{J.~Fife}, \bibinfo{author}{P.~Voorhees},
\newblock \bibinfo{title}{The morphological evolution of equiaxed dendritic microstructures during coarsening},
\newblock \bibinfo{journal}{Acta Materialia} \bibinfo{volume}{57} (\bibinfo{year}{2009}) \bibinfo{pages}{2418--2428}.
\bibitem[{Kwon et~al.(2007)Kwon, Thornton, and Voorhees}]{kwon2007coarsening}
\bibinfo{author}{Y.~Kwon}, \bibinfo{author}{K.~Thornton}, \bibinfo{author}{P.~W. Voorhees},
\newblock \bibinfo{title}{Coarsening of bicontinuous structures via nonconserved and conserved dynamics},
\newblock \bibinfo{journal}{Physical Review E} \bibinfo{volume}{75} (\bibinfo{year}{2007}) \bibinfo{pages}{021120}.
\bibitem[{Felberbaum and Rappaz(2011)}]{felberbaum2011curvature}
\bibinfo{author}{M.~Felberbaum}, \bibinfo{author}{M.~Rappaz},
\newblock \bibinfo{title}{Curvature of micropores in {Al--Cu} alloys: An {X-ray} tomography study},
\newblock \bibinfo{journal}{Acta Materialia} \bibinfo{volume}{59} (\bibinfo{year}{2011}) \bibinfo{pages}{6849--6860}.
\bibitem[{Chen-Wiegart et~al.(2013{\natexlab{a}})Chen-Wiegart, Liu, Faber, Barnett, and Wang}]{chen2013analysis}
\bibinfo{author}{Y.-c.~K. Chen-Wiegart}, \bibinfo{author}{Z.~Liu}, \bibinfo{author}{K.~T. Faber}, \bibinfo{author}{S.~A. Barnett}, \bibinfo{author}{J.~Wang},
\newblock \bibinfo{title}{{3D} analysis of a {LiCoO\textsubscript{2}--Li (Ni\textsubscript{1/3}Mn\textsubscript{1/3}Co\textsubscript{1/3}) O\textsubscript{2}} {Li-ion} battery positive electrode using {X-ray} nano-tomography},
\newblock \bibinfo{journal}{Electrochemistry Communications} \bibinfo{volume}{28} (\bibinfo{year}{2013}{\natexlab{a}}) \bibinfo{pages}{127--130}.
\bibitem[{Chen-Wiegart et~al.(2013{\natexlab{b}})Chen-Wiegart, Wada, Butakov, Xiao, De~Carlo, Kato, Wang, Dunand, and Maire}]{chen2013morphological}
\bibinfo{author}{Y.-c.~K. Chen-Wiegart}, \bibinfo{author}{T.~Wada}, \bibinfo{author}{N.~Butakov}, \bibinfo{author}{X.~Xiao}, \bibinfo{author}{F.~De~Carlo}, \bibinfo{author}{H.~Kato}, \bibinfo{author}{J.~Wang}, \bibinfo{author}{D.~C. Dunand}, \bibinfo{author}{E.~Maire},
\newblock \bibinfo{title}{{3D} morphological evolution of porous titanium by {X-ray} micro- and nano-tomography},
\newblock \bibinfo{journal}{Journal of Materials Research} \bibinfo{volume}{28} (\bibinfo{year}{2013}{\natexlab{b}}) \bibinfo{pages}{2444--2452}.
\bibitem[{Gibbs et~al.(2015)Gibbs, Mohan, Gulsoy, Shahani, Xiao, Bouman, De~Graef, and Voorhees}]{gibbs2015three}
\bibinfo{author}{J.~W. Gibbs}, \bibinfo{author}{K.~A. Mohan}, \bibinfo{author}{E.~Gulsoy}, \bibinfo{author}{A.~Shahani}, \bibinfo{author}{X.~Xiao}, \bibinfo{author}{C.~Bouman}, \bibinfo{author}{M.~De~Graef}, \bibinfo{author}{P.~Voorhees},
\newblock \bibinfo{title}{The three-dimensional morphology of growing dendrites},
\newblock \bibinfo{journal}{Scientific reports} \bibinfo{volume}{5} (\bibinfo{year}{2015}) \bibinfo{pages}{11824}.
\bibitem[{Sun et~al.(2017)Sun, Cecen, Gibbs, Kalidindi, and Voorhees}]{sun2017analytics}
\bibinfo{author}{Y.~Sun}, \bibinfo{author}{A.~Cecen}, \bibinfo{author}{J.~W. Gibbs}, \bibinfo{author}{S.~R. Kalidindi}, \bibinfo{author}{P.~W. Voorhees},
\newblock \bibinfo{title}{Analytics on large microstructure datasets using two-point spatial correlations: Coarsening of dendritic structures},
\newblock \bibinfo{journal}{Acta Materialia} \bibinfo{volume}{132} (\bibinfo{year}{2017}) \bibinfo{pages}{374--388}.
\bibitem[{Guo et~al.(2017)Guo, Phillion, Cai, Shuai, Kazantsev, Jing, and Lee}]{guo2017dendritic}
\bibinfo{author}{E.~Guo}, \bibinfo{author}{A.~Phillion}, \bibinfo{author}{B.~Cai}, \bibinfo{author}{S.~Shuai}, \bibinfo{author}{D.~Kazantsev}, \bibinfo{author}{T.~Jing}, \bibinfo{author}{P.~D. Lee},
\newblock \bibinfo{title}{Dendritic evolution during coarsening of {Mg-Zn} alloys via {4D} synchrotron tomography},
\newblock \bibinfo{journal}{Acta Materialia} \bibinfo{volume}{123} (\bibinfo{year}{2017}) \bibinfo{pages}{373--382}.
\bibitem[{Wang et~al.(2020)Wang, Chao, Moniri, Gao, Volkenandt, De~Andrade, and Shahani}]{wang2020integrated}
\bibinfo{author}{Y.~Wang}, \bibinfo{author}{P.~Chao}, \bibinfo{author}{S.~Moniri}, \bibinfo{author}{J.~Gao}, \bibinfo{author}{T.~Volkenandt}, \bibinfo{author}{V.~De~Andrade}, \bibinfo{author}{A.~J. Shahani},
\newblock \bibinfo{title}{Integrated three-dimensional characterization of reactive phase formation and coarsening during isothermal annealing of metastable {Zn--3Mg--4Al} eutectic},
\newblock \bibinfo{journal}{Materials Characterization} \bibinfo{volume}{170} (\bibinfo{year}{2020}) \bibinfo{pages}{110685}.
\bibitem[{Arai et~al.(2021)Arai, Stan, Macfarland, Voorhees, Muyanja, Shahani, and Faber}]{arai2021coarsening}
\bibinfo{author}{N.~Arai}, \bibinfo{author}{T.~Stan}, \bibinfo{author}{S.~Macfarland}, \bibinfo{author}{P.~W. Voorhees}, \bibinfo{author}{N.~S. Muyanja}, \bibinfo{author}{A.~J. Shahani}, \bibinfo{author}{K.~T. Faber},
\newblock \bibinfo{title}{Coarsening of dendrites in solution-based freeze-cast ceramic systems},
\newblock \bibinfo{journal}{Acta Materialia} \bibinfo{volume}{215} (\bibinfo{year}{2021}) \bibinfo{pages}{117039}.
\bibitem[{Hu et~al.(2022)Hu, Liu, Zou, Shao, Wang, and Jin}]{hu2022synthesis}
\bibinfo{author}{W.-K. Hu}, \bibinfo{author}{L.-Z. Liu}, \bibinfo{author}{L.~Zou}, \bibinfo{author}{J.-C. Shao}, \bibinfo{author}{S.-G. Wang}, \bibinfo{author}{H.-J. Jin},
\newblock \bibinfo{title}{Synthesis and mechanical properties of porous metals with inverted dealloying structure},
\newblock \bibinfo{journal}{Scripta Materialia} \bibinfo{volume}{210} (\bibinfo{year}{2022}) \bibinfo{pages}{114483}.
\bibitem[{Koenderink and Van~Doorn(1992)}]{koenderink1992surface}
\bibinfo{author}{J.~J. Koenderink}, \bibinfo{author}{A.~J. Van~Doorn},
\newblock \bibinfo{title}{Surface shape and curvature scales},
\newblock \bibinfo{journal}{Image and vision computing} \bibinfo{volume}{10} (\bibinfo{year}{1992}) \bibinfo{pages}{557--564}.
\bibitem[{Spanos et~al.(2010)Spanos, Rowenhorst, Kral, Voorhees, and Kammer}]{spanos2010three}
\bibinfo{author}{G.~Spanos}, \bibinfo{author}{D.~Rowenhorst}, \bibinfo{author}{M.~Kral}, \bibinfo{author}{P.~Voorhees}, \bibinfo{author}{D.~Kammer},
\newblock \bibinfo{title}{{Three-Dimensional Microstructure Representation}},
\newblock in: \bibinfo{booktitle}{{Metals Process Simulation}}, \bibinfo{publisher}{{ASM} International}, \bibinfo{year}{2010}, pp. \bibinfo{pages}{100--114}.
\bibitem[{Frank(2008)}]{frank2008electron}
\bibinfo{author}{J.~Frank}, \bibinfo{title}{Electron tomography: methods for three-dimensional visualization of structures in the cell}, \bibinfo{publisher}{Springer}, \bibinfo{year}{2008}.
\bibitem[{Batenburg and Sijbers(2007)}]{batenburg2007dart}
\bibinfo{author}{K.~J. Batenburg}, \bibinfo{author}{J.~Sijbers},
\newblock \bibinfo{title}{{DART}: a fast heuristic algebraic reconstruction algorithm for discrete tomography},
\newblock in: \bibinfo{booktitle}{2007 IEEE International Conference on Image Processing}, volume~\bibinfo{volume}{4}, \bibinfo{organization}{IEEE}, \bibinfo{year}{2007}, pp. \bibinfo{pages}{IV--133}.
\bibitem[{Farag and Flemings(1975)}]{farag_structure_1975}
\bibinfo{author}{M.~M. Farag}, \bibinfo{author}{M.~C. Flemings},
\newblock \bibinfo{title}{Structure and strength of {Al, Cu--Al\textsubscript{3}Ni} directionally solidified composites},
\newblock \bibinfo{journal}{Metallurgical Transactions A} \bibinfo{volume}{6} (\bibinfo{year}{1975}) \bibinfo{pages}{1009}.
\bibitem[{Smartt et~al.(1971)Smartt, Tu, and Courtney}]{smartt_elevated_1971}
\bibinfo{author}{H.~B. Smartt}, \bibinfo{author}{L.~K. Tu}, \bibinfo{author}{T.~H. Courtney},
\newblock \bibinfo{title}{Elevated temperature stability of the {Al-Al\textsubscript{3}Ni} eutectic composite},
\newblock \bibinfo{journal}{Metallurgical Transactions} \bibinfo{volume}{2} (\bibinfo{year}{1971}) \bibinfo{pages}{2717--2727}.
\bibitem[{Pandey et~al.(2017)Pandey, Kashyap, Tiwary, and Chattopadhyay}]{pandey_development_2017}
\bibinfo{author}{P.~Pandey}, \bibinfo{author}{S.~Kashyap}, \bibinfo{author}{C.~S. Tiwary}, \bibinfo{author}{K.~Chattopadhyay},
\newblock \bibinfo{title}{Development of high-strength high-temperature cast {Al-Ni-Cr} alloys through evolution of a novel composite eutectic structure},
\newblock \bibinfo{journal}{Metallurgical and Materials Transactions A} \bibinfo{volume}{48} (\bibinfo{year}{2017}) \bibinfo{pages}{5940--5950}.
\bibitem[{Pandey et~al.(2019)Pandey, Makineni, Gault, and Chattopadhyay}]{pandey_origin_2019}
\bibinfo{author}{P.~Pandey}, \bibinfo{author}{S.~K. Makineni}, \bibinfo{author}{B.~Gault}, \bibinfo{author}{K.~Chattopadhyay},
\newblock \bibinfo{title}{On the origin of a remarkable increase in the strength and stability of an {Al rich Al-Ni} eutectic alloy by {Zr} addition},
\newblock \bibinfo{journal}{Acta Materialia} \bibinfo{volume}{170} (\bibinfo{year}{2019}) \bibinfo{pages}{205--217}.
\bibitem[{Uan et~al.(1997)Uan, Chen, and Lui}]{uan_study_1997}
\bibinfo{author}{J.~Y. Uan}, \bibinfo{author}{L.~H. Chen}, \bibinfo{author}{T.~S. Lui},
\newblock \bibinfo{title}{A study on the subgrain superplasticity of extruded {Al-Al\textsubscript{3}Ni} eutectic alloy},
\newblock \bibinfo{journal}{Metallurgical and Materials Transactions A} \bibinfo{volume}{28} (\bibinfo{year}{1997}) \bibinfo{pages}{401--409}.
\bibitem[{Uan et~al.(2001)Uan, Chen, and Lui}]{uan_extrusion_2001}
\bibinfo{author}{J.~Y. Uan}, \bibinfo{author}{L.~H. Chen}, \bibinfo{author}{T.~S. Lui},
\newblock \bibinfo{title}{On the extrusion microstructural evolution of {Al-Al\textsubscript{3}Ni} in situ composite},
\newblock \bibinfo{journal}{Acta Materialia} \bibinfo{volume}{49} (\bibinfo{year}{2001}) \bibinfo{pages}{313--320}.
\bibitem[{Uan et~al.(1996)Uan, Lui, and Chen}]{uan_superplasticity-like_1996}
\bibinfo{author}{J.~Y. Uan}, \bibinfo{author}{T.~S. Lui}, \bibinfo{author}{L.~H. Chen},
\newblock \bibinfo{title}{Superplasticity-like behavior of {Al-Al\textsubscript{3}Ni} eutectic alloy},
\newblock \bibinfo{journal}{Materials Chemistry and Physics} \bibinfo{volume}{43} (\bibinfo{year}{1996}) \bibinfo{pages}{278--282}.
\bibitem[{Uan et~al.(2001)Uan, Lui, and Chen}]{uan_microstructural_2001}
\bibinfo{author}{J.~Y. Uan}, \bibinfo{author}{T.~S. Lui}, \bibinfo{author}{L.~H. Chen},
\newblock \bibinfo{title}{Microstructural evolution in the superplastic-like deformation of {Al-Al\textsubscript{3}Ni} eutectic alloy with [111] fiber texture},
\newblock \bibinfo{journal}{Metallurgical and Materials Transactions A} \bibinfo{volume}{32} (\bibinfo{year}{2001}) \bibinfo{pages}{547--555}.
\bibitem[{Suwanpreecha et~al.(2018)Suwanpreecha, Pandee, Patakham, and Limmaneevichitr}]{suwanpreecha_new_2018}
\bibinfo{author}{C.~Suwanpreecha}, \bibinfo{author}{P.~Pandee}, \bibinfo{author}{U.~Patakham}, \bibinfo{author}{C.~Limmaneevichitr},
\newblock \bibinfo{title}{New generation of eutectic {Al-Ni} casting alloys for elevated temperature services},
\newblock \bibinfo{journal}{Materials Science and Engineering: A} \bibinfo{volume}{709} (\bibinfo{year}{2018}) \bibinfo{pages}{46--54}.
\bibitem[{Suwanpreecha et~al.(2019)Suwanpreecha, Toinin, Michi, Pandee, Dunand, and Limmaneevichitr}]{suwanpreecha_strengthening_2019}
\bibinfo{author}{C.~Suwanpreecha}, \bibinfo{author}{J.~P. Toinin}, \bibinfo{author}{R.~A. Michi}, \bibinfo{author}{P.~Pandee}, \bibinfo{author}{D.~C. Dunand}, \bibinfo{author}{C.~Limmaneevichitr},
\newblock \bibinfo{title}{Strengthening mechanisms in {Al--Ni--Sc} alloys containing {Al\textsubscript{3}Ni} microfibers and {Al\textsubscript{3}Sc} nanoprecipitates},
\newblock \bibinfo{journal}{Acta Materialia} \bibinfo{volume}{164} (\bibinfo{year}{2019}) \bibinfo{pages}{334--346}.
\bibitem[{Koutsoukis and Makhlouf(2016)}]{koutsoukis_alternatives_2016}
\bibinfo{author}{T.~Koutsoukis}, \bibinfo{author}{M.~M. Makhlouf},
\newblock \bibinfo{title}{Alternatives to the {Al–Si} eutectic system in aluminum casting alloys},
\newblock \bibinfo{journal}{International Journal of Metalcasting} \bibinfo{volume}{10} (\bibinfo{year}{2016}) \bibinfo{pages}{342--347}.
\bibitem[{Nakagawa and Weatherly(1972)}]{nakagawa_thermal_1972}
\bibinfo{author}{Y.~G. Nakagawa}, \bibinfo{author}{G.~C. Weatherly},
\newblock \bibinfo{title}{The thermal stability of the rod {Al\textsubscript{3}Ni-Al} eutectic},
\newblock \bibinfo{journal}{Acta Metallurgica} \bibinfo{volume}{20} (\bibinfo{year}{1972}) \bibinfo{pages}{345--350}.
\bibitem[{Houghton and Jones(1979)}]{houghton_thermal_1979}
\bibinfo{author}{D.~C. Houghton}, \bibinfo{author}{D.~R.~H. Jones},
\newblock \bibinfo{title}{The thermal stabilities of in situ composites in a temperature gradient—{II}. the {Al-Al\textsubscript{3}Ni} and {Cu-Cr} eutectics},
\newblock \bibinfo{journal}{Acta Metallurgica} \bibinfo{volume}{27} (\bibinfo{year}{1979}) \bibinfo{pages}{1031--1039}.
\bibitem[{Smartt and Courtney(1976)}]{smartt_kinetics_1976}
\bibinfo{author}{H.~B. Smartt}, \bibinfo{author}{T.~H. Courtney},
\newblock \bibinfo{title}{The kinetics of coarsening in the {Al-Al\textsubscript{3}Ni} system},
\newblock \bibinfo{journal}{Metallurgical Transactions A} \bibinfo{volume}{7} (\bibinfo{year}{1976}) \bibinfo{pages}{123--126}.
\bibitem[{{McLean}(1977)}]{mclean_microstructural_1977}
\bibinfo{author}{M.~{McLean}},
\newblock \bibinfo{title}{Microstructural degradation of the {Al–Al\textsubscript{3}Ni} directionally solidified eutectic in the presence of a temperature gradient},
\newblock \bibinfo{journal}{Acta Metallurgica} \bibinfo{volume}{25} (\bibinfo{year}{1977}) \bibinfo{pages}{1209--1215}.
\bibitem[{Davies et~al.(1980)Davies, Courtney, and Przystupa}]{davies_microstructural_1980}
\bibinfo{author}{J.~R. Davies}, \bibinfo{author}{T.~H. Courtney}, \bibinfo{author}{M.~A. Przystupa},
\newblock \bibinfo{title}{Microstructural changes as a result of temperature gradients in the {Al-Al\textsubscript{3}Ni} eutectic},
\newblock \bibinfo{journal}{Metallurgical Transactions A} \bibinfo{volume}{11} (\bibinfo{year}{1980}) \bibinfo{pages}{323--332}.
\bibitem[{Suwanpreecha et~al.(2019)Suwanpreecha, Pandee, Patakham, Dunand, and Limmaneevichitr}]{suwanpreecha_effects_2019}
\bibinfo{author}{C.~Suwanpreecha}, \bibinfo{author}{P.~Pandee}, \bibinfo{author}{U.~Patakham}, \bibinfo{author}{D.~C. Dunand}, \bibinfo{author}{C.~Limmaneevichitr},
\newblock \bibinfo{title}{Effects of {Zr} additions on structure and microhardness evolution of eutectic {Al-6Ni} alloy},
\newblock in: \bibinfo{editor}{C.~Chesonis} (Ed.), \bibinfo{booktitle}{Light Metals 2019}, The Minerals, Metals \& Materials Series, \bibinfo{publisher}{Springer International Publishing}, \bibinfo{year}{2019}, pp. \bibinfo{pages}{373--377}. \DOIprefix\doi{10.1007/978-3-030-05864-7_47}.
\bibitem[{Kim et~al.(2020)Kim, Soprunyuk, Chawake, Zheng, Spieckermann, Hong, Kim, and Eckert}]{kim_outstanding_2020}
\bibinfo{author}{J.~T. Kim}, \bibinfo{author}{V.~Soprunyuk}, \bibinfo{author}{N.~Chawake}, \bibinfo{author}{Y.~H. Zheng}, \bibinfo{author}{F.~Spieckermann}, \bibinfo{author}{S.~H. Hong}, \bibinfo{author}{K.~B. Kim}, \bibinfo{author}{J.~Eckert},
\newblock \bibinfo{title}{Outstanding strengthening behavior and dynamic mechanical properties of in-situ {Al–Al\textsubscript{3}Ni} composites by {Cu} addition},
\newblock \bibinfo{journal}{Composites Part B: Engineering} \bibinfo{volume}{189} (\bibinfo{year}{2020}) \bibinfo{pages}{107891}.
\bibitem[{Belov et~al.(2004)Belov, Alabin, and Eskin}]{belov_improving_2004}
\bibinfo{author}{N.~A. Belov}, \bibinfo{author}{A.~N. Alabin}, \bibinfo{author}{D.~G. Eskin},
\newblock \bibinfo{title}{Improving the properties of cold-rolled {Al–6\%Ni} sheets by alloying and heat treatment},
\newblock \bibinfo{journal}{Scripta Materialia} \bibinfo{volume}{50} (\bibinfo{year}{2004}) \bibinfo{pages}{89--94}.
\bibitem[{Czerwinski et~al.(2022)Czerwinski, Aniolek, and Li}]{czerwinski_strengthening_2022}
\bibinfo{author}{F.~Czerwinski}, \bibinfo{author}{M.~Aniolek}, \bibinfo{author}{J.~Li},
\newblock \bibinfo{title}{Strengthening retention and structural stability of the {Al-Al\textsubscript{3}Ni} eutectic at high temperatures},
\newblock \bibinfo{journal}{Scripta Materialia} \bibinfo{volume}{214} (\bibinfo{year}{2022}) \bibinfo{pages}{114679}.
\bibitem[{Tassoni et~al.(1980)Tassoni, Riquet, and Durand}]{tassoni_morphologie}
\bibinfo{author}{P.~D. Tassoni}, \bibinfo{author}{J.~P. Riquet}, \bibinfo{author}{F.~Durand},
\newblock \bibinfo{title}{Morphologie theorique du compose {Al\textsubscript{3}Ni} et comparaison avec les formes observ6es},
\newblock \bibinfo{journal}{Acta Crystallographica} \bibinfo{volume}{36} (\bibinfo{year}{1980}) \bibinfo{pages}{420--428}.
\bibitem[{Jackson(2010)}]{jackson2010kinetic}
\bibinfo{author}{K.~A. Jackson}, \bibinfo{title}{Kinetic processes: crystal growth, diffusion, and phase transitions in materials}, \bibinfo{publisher}{John Wiley \& Sons}, \bibinfo{year}{2010}.
\bibitem[{Sekerka(2005)}]{sekerka2005equilibrium}
\bibinfo{author}{R.~F. Sekerka},
\newblock \bibinfo{title}{Equilibrium and growth shapes of crystals: how do they differ and why should we care?},
\newblock \bibinfo{journal}{Crystal Research and Technology: Journal of Experimental and Industrial Crystallography} \bibinfo{volume}{40} (\bibinfo{year}{2005}) \bibinfo{pages}{291--306}.
\bibitem[{Hellawell(1970)}]{hellawell_growth_1970}
\bibinfo{author}{A.~Hellawell},
\newblock \bibinfo{title}{The growth and structure of eutectics with silicon and germanium},
\newblock \bibinfo{journal}{Progress in Materials Science} \bibinfo{volume}{15} (\bibinfo{year}{1970}) \bibinfo{pages}{3--78}.
\bibitem[{Tiwary et~al.(2015)Tiwary, Kashyap, Kim, and Chattopadhyay}]{tiwary_based_2015}
\bibinfo{author}{C.~S. Tiwary}, \bibinfo{author}{S.~Kashyap}, \bibinfo{author}{D.~H. Kim}, \bibinfo{author}{K.~Chattopadhyay},
\newblock \bibinfo{title}{Al based ultra-fine eutectic with high room temperature plasticity and elevated temperature strength},
\newblock \bibinfo{journal}{Materials Science and Engineering: A} \bibinfo{volume}{639} (\bibinfo{year}{2015}) \bibinfo{pages}{359--369}.
\bibitem[{Lien et~al.(2022)Lien, Wang, and Misra}]{lien_plastic_2022}
\bibinfo{author}{H.-H. Lien}, \bibinfo{author}{J.~Wang}, \bibinfo{author}{A.~Misra},
\newblock \bibinfo{title}{Plastic deformation induced microstructure transition in nano-fibrous {Al}-{Si} eutectics},
\newblock \bibinfo{journal}{Materials \& Design}  (\bibinfo{year}{2022}) \bibinfo{pages}{110701}.
\bibitem[{Ferrarini et~al.(2004)Ferrarini, Bolfarini, Kiminami et~al.}]{ferrarini2004microstructure}
\bibinfo{author}{C.~Ferrarini}, \bibinfo{author}{C.~Bolfarini}, \bibinfo{author}{C.~Kiminami}, et~al.,
\newblock \bibinfo{title}{Microstructure and mechanical properties of spray deposited hypoeutectic {Al--Si} alloy},
\newblock \bibinfo{journal}{Materials Science and Engineering: A} \bibinfo{volume}{375} (\bibinfo{year}{2004}) \bibinfo{pages}{577--580}.
\bibitem[{Suryawanshi et~al.(2016)Suryawanshi, Prashanth, Scudino, Eckert, Prakash, and Ramamurty}]{suryawanshi2016simultaneous}
\bibinfo{author}{J.~Suryawanshi}, \bibinfo{author}{K.~Prashanth}, \bibinfo{author}{S.~Scudino}, \bibinfo{author}{J.~Eckert}, \bibinfo{author}{O.~Prakash}, \bibinfo{author}{U.~Ramamurty},
\newblock \bibinfo{title}{Simultaneous enhancements of strength and toughness in an {Al-12Si} alloy synthesized using selective laser melting},
\newblock \bibinfo{journal}{Acta Materialia} \bibinfo{volume}{115} (\bibinfo{year}{2016}) \bibinfo{pages}{285--294}.
\bibitem[{Major and Rutter(1989)}]{major1989effect}
\bibinfo{author}{J.~Major}, \bibinfo{author}{J.~Rutter},
\newblock \bibinfo{title}{Effect of strontium and phosphorus on solid/liquid interface of {Al--Si} eutectic},
\newblock \bibinfo{journal}{Materials science and technology} \bibinfo{volume}{5} (\bibinfo{year}{1989}) \bibinfo{pages}{645--656}.
\bibitem[{Pierantoni et~al.(1992)Pierantoni, Gremaud, Magnin, Stoll, and Kurz}]{pierantoni_coupled_1992}
\bibinfo{author}{M.~Pierantoni}, \bibinfo{author}{M.~Gremaud}, \bibinfo{author}{P.~Magnin}, \bibinfo{author}{D.~Stoll}, \bibinfo{author}{W.~Kurz},
\newblock \bibinfo{title}{The coupled zone of rapidly solidified {A}l\mbox{-}{S}i alloys in laser treatment},
\newblock \bibinfo{journal}{Acta metallurgica et materialia} \bibinfo{volume}{40} (\bibinfo{year}{1992}) \bibinfo{pages}{1637--1644}.
\bibitem[{Zhou et~al.(2023)Zhou, Chao, Sloan, Lien, Hunter, Misra, and Shahani}]{zhou2023three}
\bibinfo{author}{X.~Zhou}, \bibinfo{author}{P.~Chao}, \bibinfo{author}{L.~Sloan}, \bibinfo{author}{H.-H. Lien}, \bibinfo{author}{A.~H. Hunter}, \bibinfo{author}{A.~Misra}, \bibinfo{author}{A.~J. Shahani},
\newblock \bibinfo{title}{Three-dimensional morphology of an ultrafine {Al-Si} eutectic produced via laser rapid solidification},
\newblock \bibinfo{journal}{Scripta Materialia} \bibinfo{volume}{232} (\bibinfo{year}{2023}) \bibinfo{pages}{115471}.
\bibitem[{Herlach et~al.(2018)Herlach, Simons, and Pichon}]{herlach2018crystal}
\bibinfo{author}{D.~M. Herlach}, \bibinfo{author}{D.~Simons}, \bibinfo{author}{P.-Y. Pichon},
\newblock \bibinfo{title}{Crystal growth kinetics in undercooled melts of pure {Ge, Si and Ge--Si} alloys},
\newblock \bibinfo{journal}{Philosophical Transactions of the Royal Society A: Mathematical, Physical and Engineering Sciences} \bibinfo{volume}{376} (\bibinfo{year}{2018}) \bibinfo{pages}{20170205}.

\end{thebibliography}
\processdelayedfloats
\clearpage
\newpage

\section*{Supplemental Information}

\renewcommand{\thefigure}{S\arabic{figure}}
\renewcommand{\thepostfigure}{S\arabic{postfigure}}
\setcounter{figure}{0}
\setcounter{postfigure}{0}


\renewcommand{\theVideo}{V\arabic{Video}}
\setcounter{Video}{0}

\begin{figure}[ht!]
\centering \includegraphics[width=\textwidth,height=\textheight,keepaspectratio]{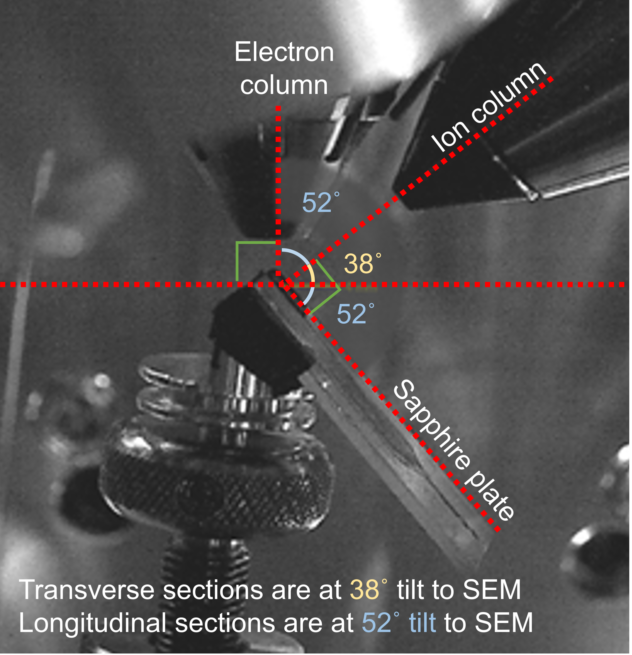}
\caption{Configuration employed within the Thermo Fisher Helios G4 PFIB UXe dual-beam microscope used for cross-sectional milling with Xe ions and imaging with SEM. The annotated angles specified sample orientation for the excavation and characterization. Please note that, due to the camera position relative to the electron and ion columns, annotated angles are not to scale. The process involves positioning the sample within a 45\textdegree~pre\mbox{-}tilt holder and subsequently rotating it to 52\textdegree. }
\label{SI_fig06_fib}
\end{figure}

\begin{figure}[ht!]
\centering \includegraphics[width=\textwidth,height=\textheight,keepaspectratio]{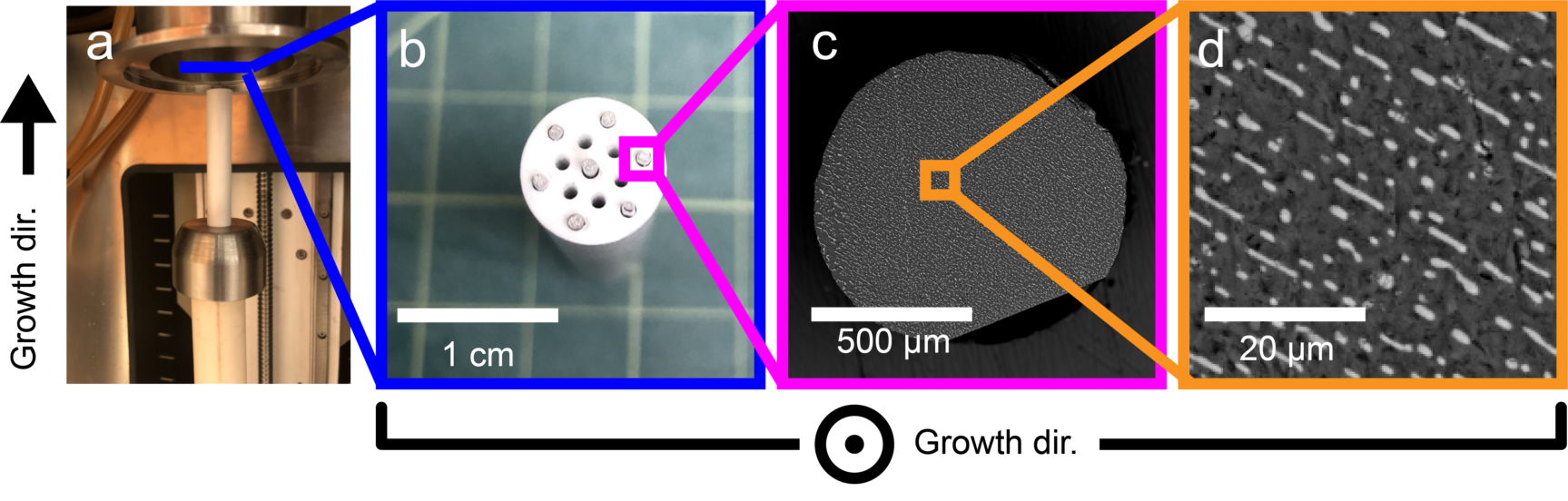}
\caption{Fabrication of templated specimens: (a) Side view of the boron nitride (BN) crucible (white) containing seven 4~cm length rods of the eutectic alloy prepared by electrical discharge machining (EDM), prior to insertion into the DS furnace. (b) Bird’s\mbox{-}eye view of the same BN crucible. (c) SEM\mbox{-}BSE image of a cross\mbox{-}sectional slice of the rod sample after DS showing a uniform eutectic microstructure. This sample was sectioned at roughly 50\% of its height. (Right) magnified view showing Al\textsubscript{3}Ni phase (bright contrast) in the form of elongated rods embedded within an Al matrix (dark contrast). }
\label{SI_fig01_template}
\end{figure}

\begin{figure}[ht!]
\centering \includegraphics[width=\textwidth,height=\textheight,keepaspectratio]{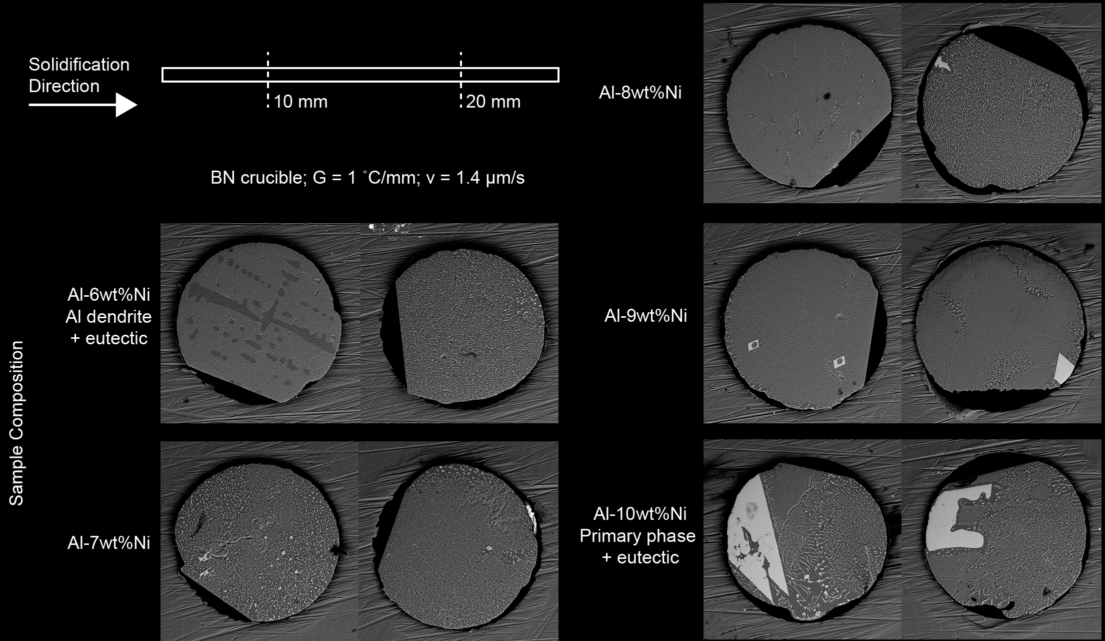}
\caption{Metallographs for various alloy compositions: The eutectic microstructure exhibits elongated rods. Primary Al\textsubscript{3}Ni crystals (large light-grey regions) are present in hypereutectic compositions. In each row, the left image was obtained at 10~mm along the solidification direction and the right at 20~mm. }
\label{SI_fig02_comp_series}
\end{figure}

\begin{figure}[ht!]
\centering \includegraphics[width=\textwidth,height=\textheight,keepaspectratio]{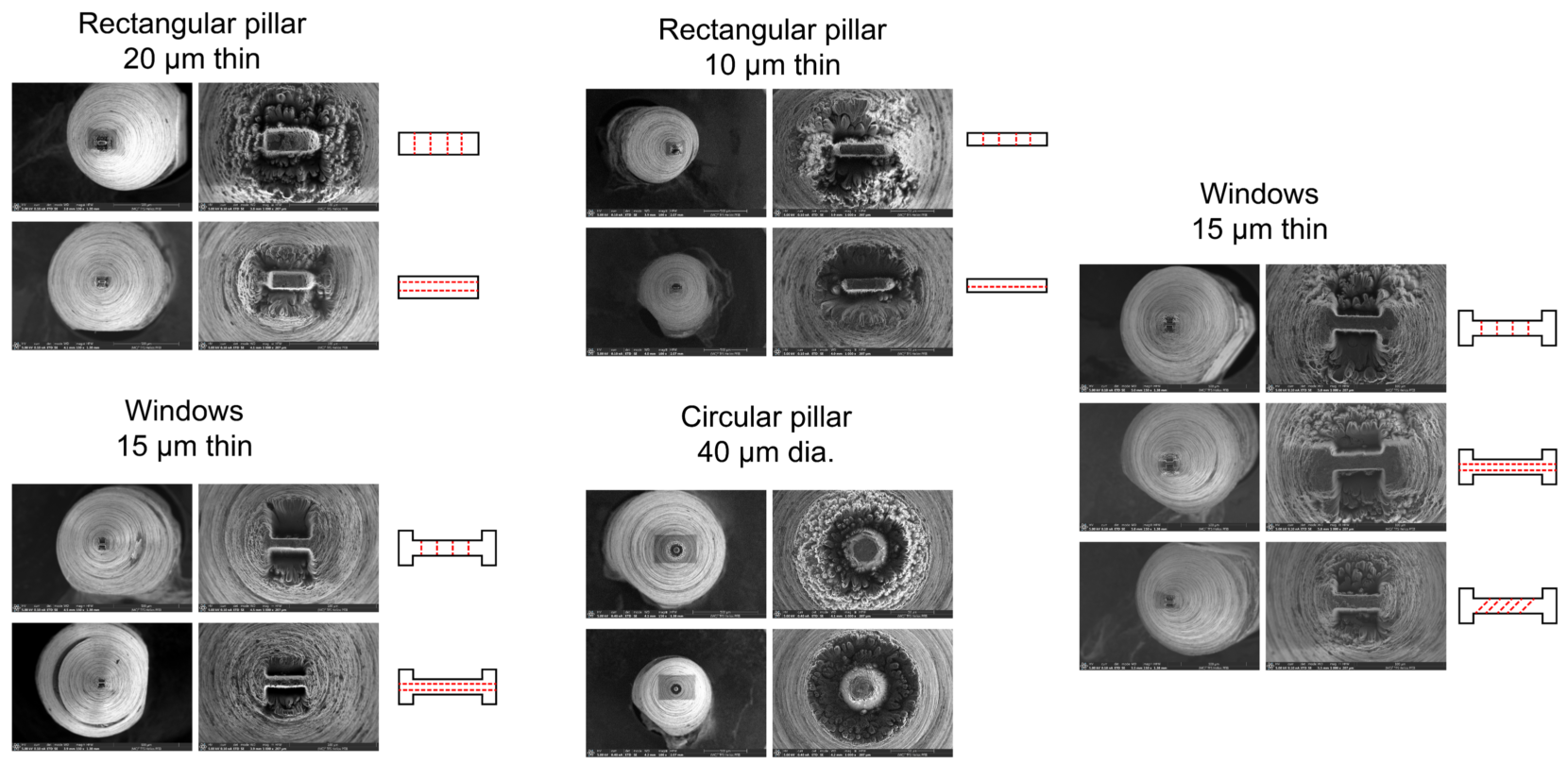}
\caption{Geometries of the investigated samples for TXM: 11 different samples are shown with both low (left) and high (right) magnification SEM images from a ``bird's\mbox{-}eye view." They include rectangular pillars of thickness 20~{\textmu}m and 10~{\textmu}m; ``windows" of thickness 15~{\textmu}m (resembling dog-bones in cross-section); and (d) near-cylindrical pillars 40~{\textmu}m in diameter. The orientation of the Al\textsubscript{3}Ni crystals (red) are included within the schematics. These crystals grow out of the plane of the page during the real\mbox{-}time TXM experiments.}
\label{SI_fig03_TXM_prep}
\end{figure}

\begin{figure}[ht!]
\centering \includegraphics[width=\textwidth,height=\textheight,keepaspectratio]{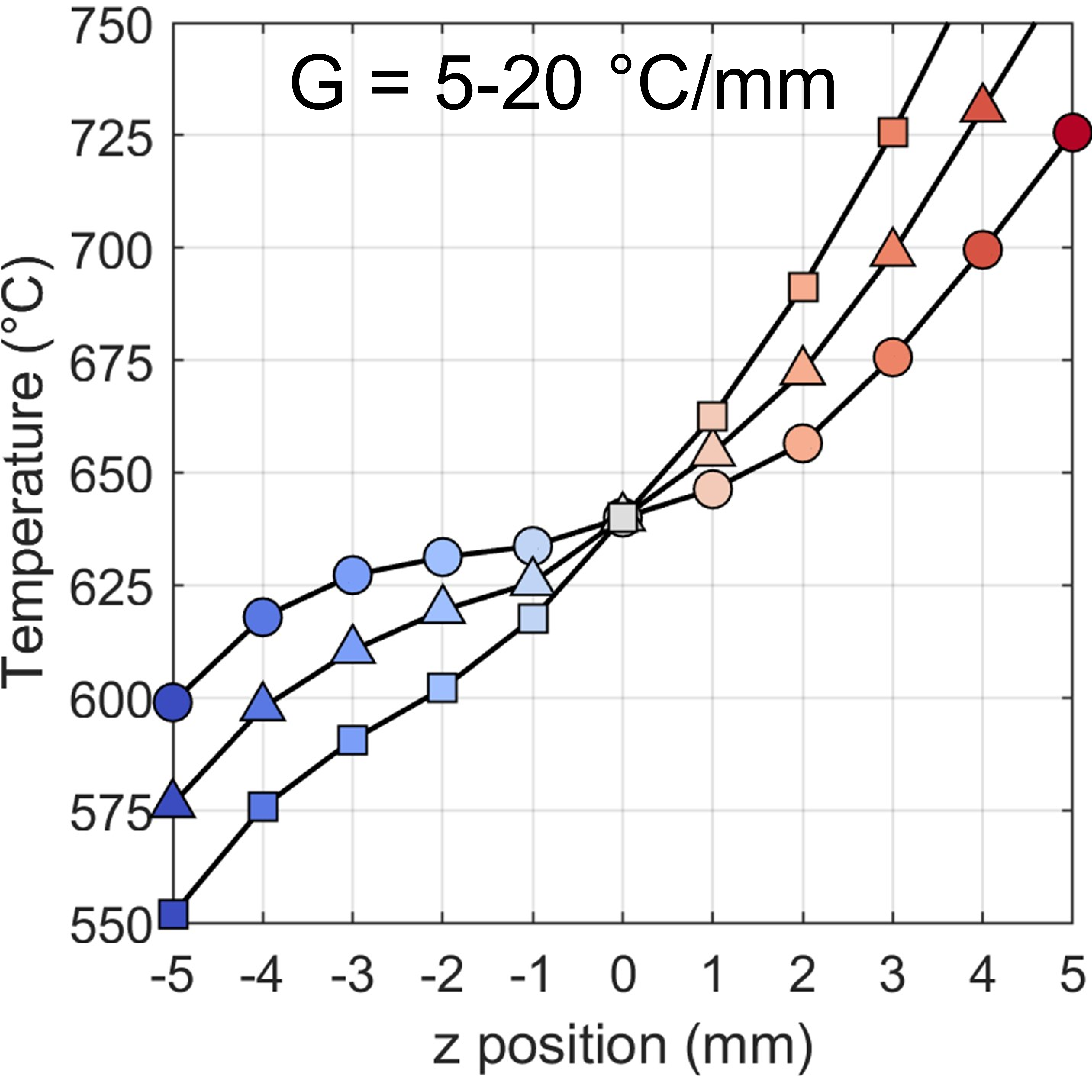}
\caption{\textcolor{black}{Axial temperature distributions (along {\bf z} axis), near the center of the two zones, where x-rays illuminate the sample. Data were obtained using a thermocouple at different set temperatures for each zone. Thermal gradients (evaluated at the position  $z=0~{\rm mm}$) range from 5 to 20 ${\rm Kmm^{-1}}$.}}
\label{SI_fig02_temp}
\end{figure}

\begin{figure}[ht!]
\centering \includegraphics[width=\textwidth,height=\textheight,keepaspectratio]{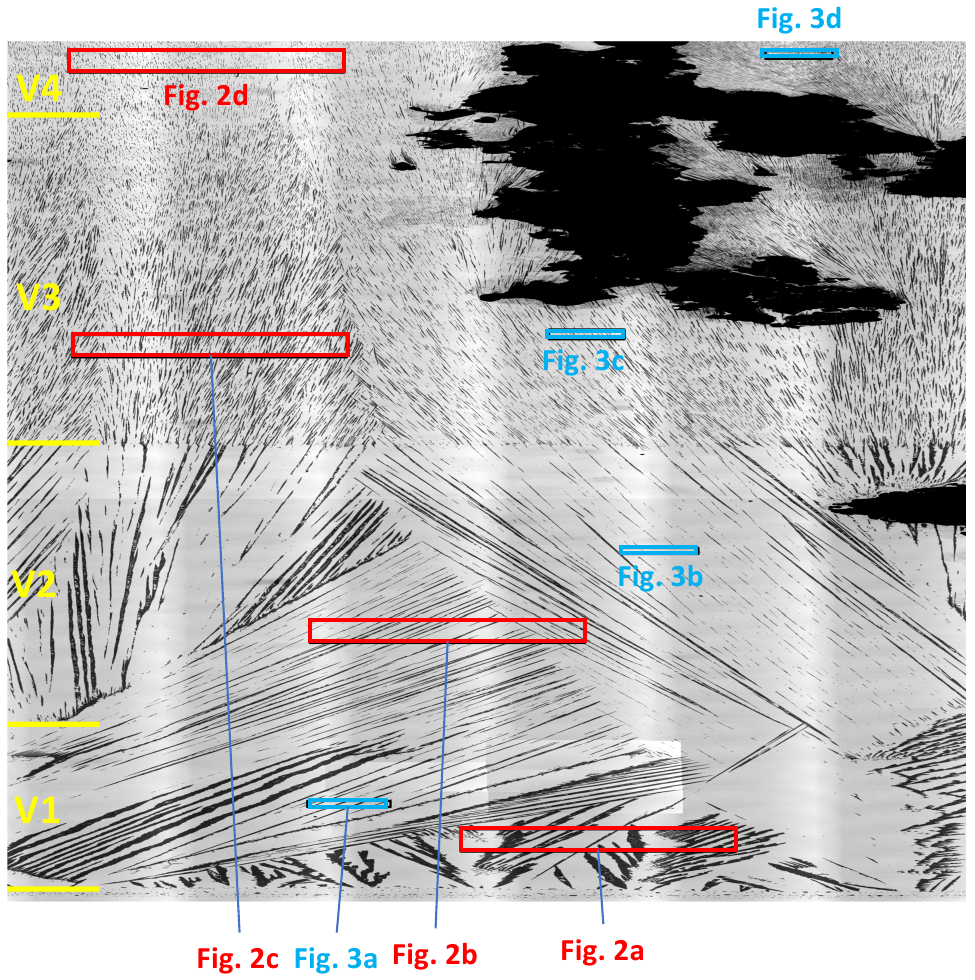}
\caption{a) Large-scale (panorama) optical view of the same thin Al\mbox{-}Al\textsubscript{3}Ni sample as in Fig.~\ref{Ech169_InSitu}. The image is contracted by a factor of 4 vertically.  Yellow ticks: changes of solidification velocity ($V_1=0.5~{\rm \mu ms^{-1}}$;  $V_2=1.0~{\rm \mu ms^{-1}}$; $V_3=5.0~{\rm \mu ms^{-1}}$; $V_4=10.0~{\rm \mu ms^{-1}}$). Red (blue) frames correspond to the optical (SEM) images shown in Fig.~\ref{Ech169_InSitu} (Fig.~\ref{fig05_FIB_Jan24}). Black regions of large extension correspond to dewetting.}
\label{panorama}
\end{figure}



\begin{figure}[ht!]
\centering \includegraphics[width=\textwidth,height=\textheight,keepaspectratio]{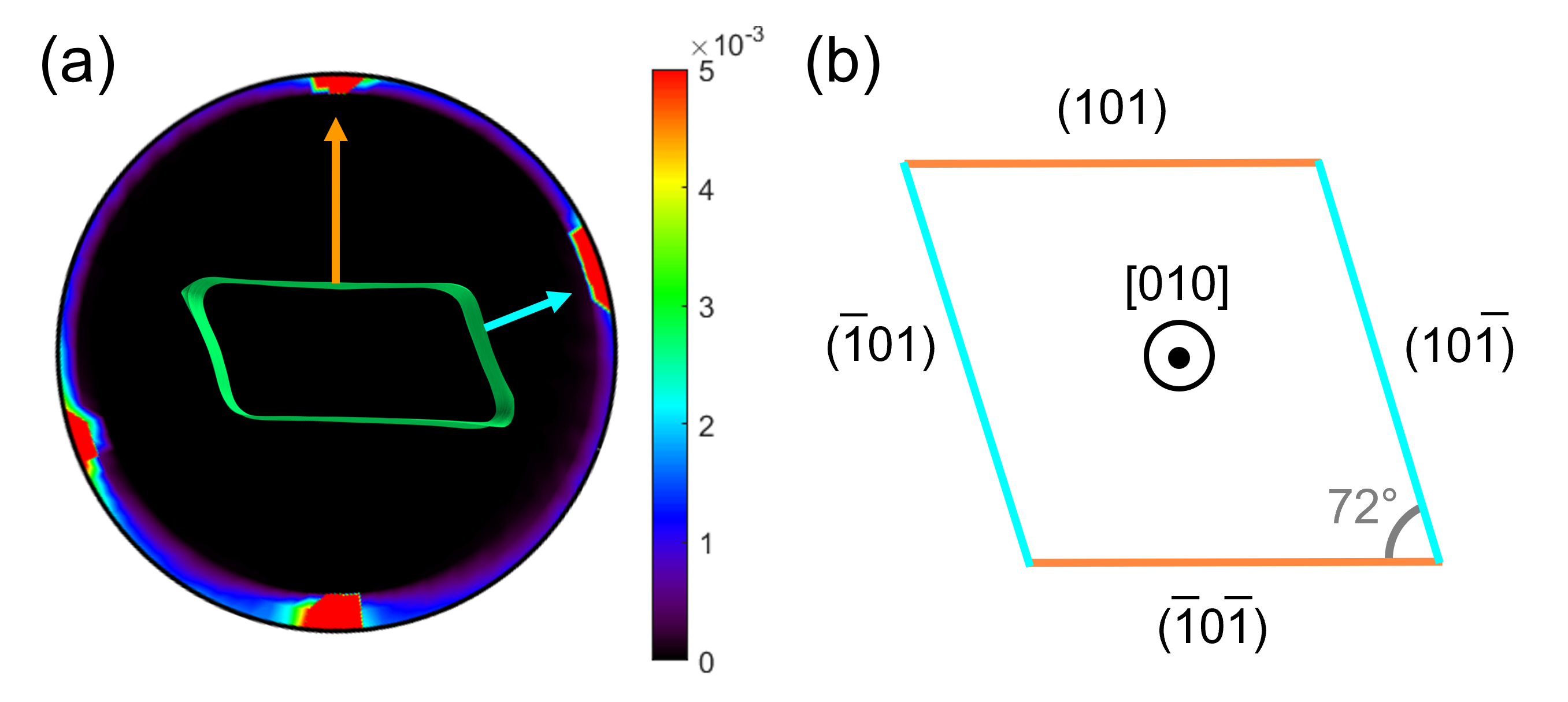}
\caption{Interfacial orientations of Al\textsubscript{3}Ni. We calculate normal vectors of patches of interface bounding the intermetallic crystal and display them stereographically in (a).  The zone axis of the projection corresponds to the growth direction of the intermetallic crystal.  The four peaks in the probability distribution are due to two sets of normal vectors (indicated by cyan and orange arrows). The angle between these two facets is approximately 72\textdegree. For reference, in (b) we show the Wulff shape for the Al\textsubscript{3}Ni along the [010] direction~\cite{tassoni_morphologie}.  It is bounded by \{101\} facets with an interplanar angle of 72.0547\textdegree.}
\label{figXTRA_INDandXtal}
\end{figure}

\begin{figure}[ht!]
\centering \includegraphics[width=\textwidth,height=\textheight,keepaspectratio]{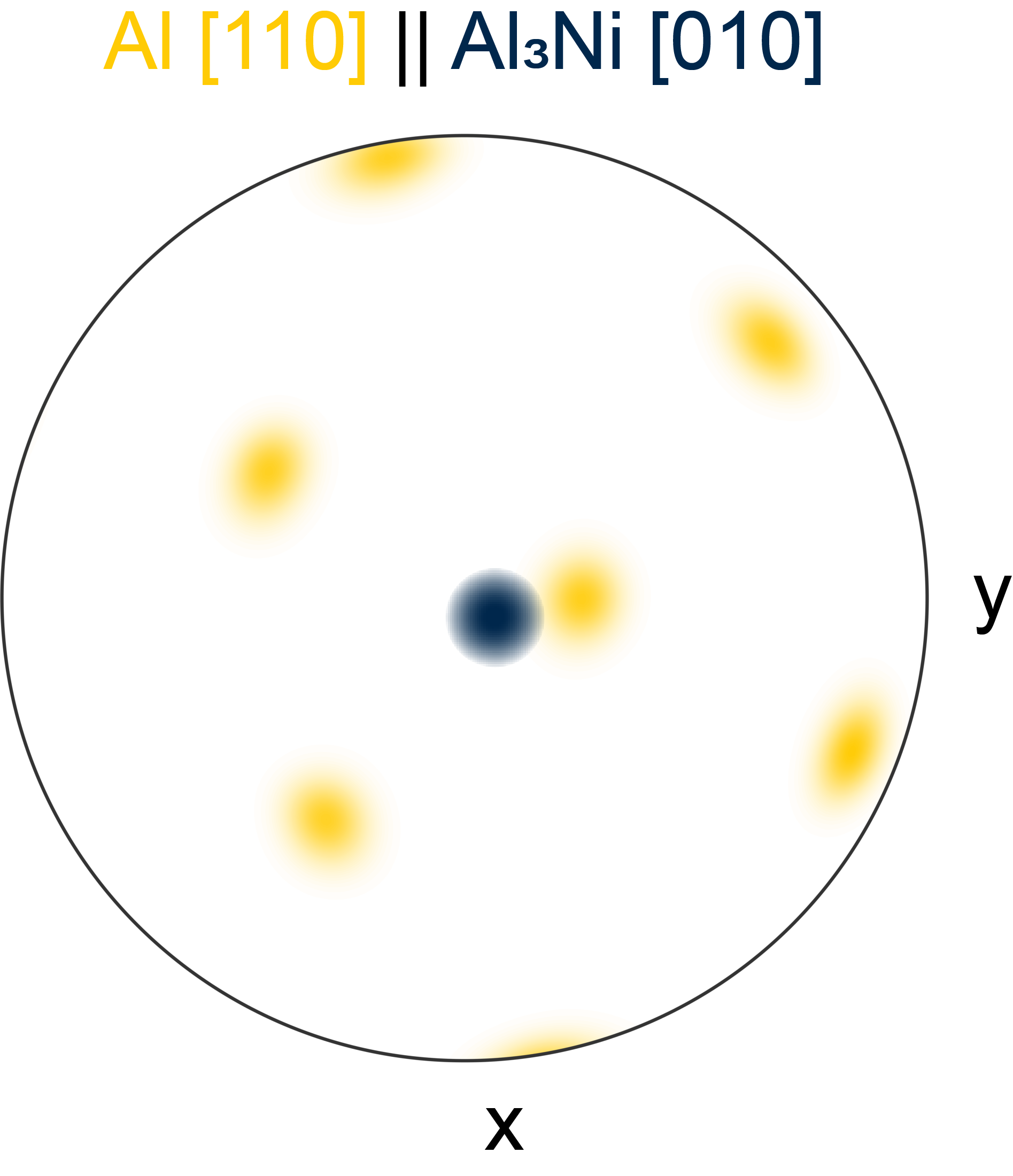}
\caption{Crystallographic growth direction identified by EBSD. The pole figures are viewing down the growth direction. The Al\textsubscript{3}Ni (010) and Al (110) are nearby, with a misorientation of 16\textdegree.}
\label{SI_fig05_EBSD_PF}
\end{figure}

\begin{figure}[ht!]
\centering \includegraphics[width=\textwidth,height=\textheight,keepaspectratio]{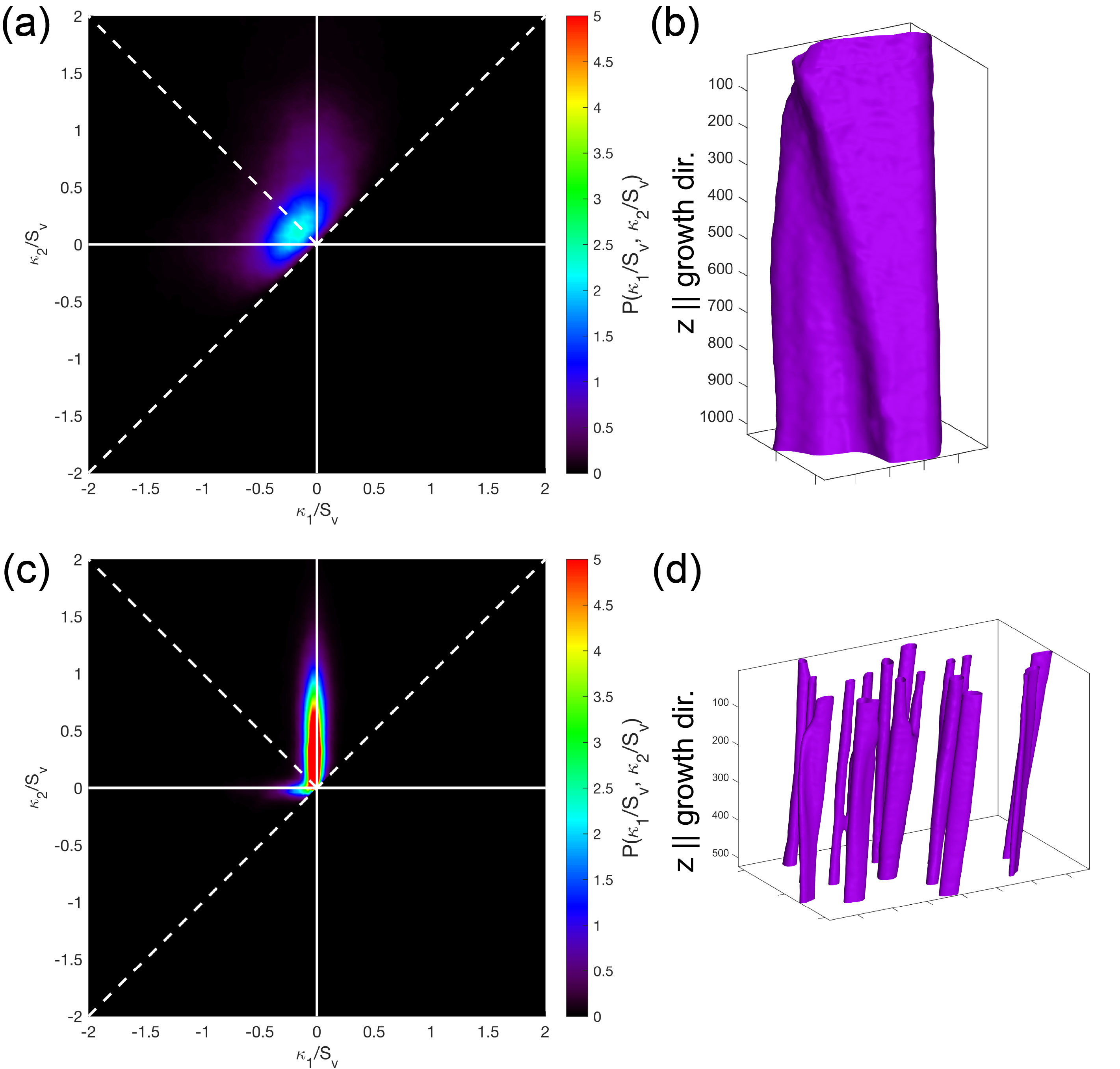}
\caption{\textcolor{black}{Quantification of interfacial morphologies in hypereutectic alloy. Interfacial shape distributions (ISDs) of the Al\textsubscript{3}Ni structure formed at (a\mbox{-}b) low velocity and (c\mbox{-}d) high velocity (cf.~Fig.~\ref{fig14_hypereut_rad}). The latter distribution shows a high\mbox{-}probability tail corresponding to curved, cylinder\mbox{-}like interfaces ($\kappa_1 \sim 0$, $\kappa_2 > 0$), which is absent in the former. In both cases, principal curvatures were scaled by the surface area per unit volume of Al\textsubscript{3}Ni in order to probe morphology independent of length\mbox{-}scale.}}
\label{SI_fig_9_hypereut_ISD}
\end{figure}



\FloatBarrier


\begin{Video}
\captionsetup{font=normalsize,justification=raggedright,format=hang}
\caption{Near\mbox{-}steady growth, TXM images, combined as a video with contrast levels adjusted for clarity. $V =~0.6~{\rm \mu ms^{-1}}$ and $G =~5~{\rm Kmm^{-1}}$.  Images have been processed by dividing each frame by an image of the molten liquid alloy. Playback speed is 30~fps frames per second (fps), nearly twice the acquisition speed (16.67~fps). [gif video file of 300 frames is 34.7MB].}
\label{V_S1}  
\end{Video} 

\begin{Video}
\captionsetup{font=normalsize,justification=raggedright,format=hang}
\caption{Growth under a velocity jump, TXM images, combined as a video with contrast levels adjusted for clarity. $V =~3.3~\rightarrow~14.7~{\rm \mu ms^{-1}}$ and $G =~5~{\rm Kmm^{-1}}$.  Images have been processed by dividing each frame by an image of the molten liquid alloy. Playback speed is 30 fps frames per second (fps) nearly twice the acquisition speed (16.67 fps).  [gif video file of 300 frames is 39.2MB]. }
\label{V_S2}  
\end{Video} 

\begin{Video}
\captionsetup{font=normalsize,justification=raggedright,format=hang}
\caption{Growth under external perturbation, TXM images with different angular views, combined as a video with contrast levels adjusted for clarity. Images have been processed by dividing each frame by an image of the molten liquid alloy at the corresponding view angle. Note, image artifacts arise from the BN that coats the sample exterior and the minor misalignment between frames. Playback speed is 16.67~fps frames per second (fps), equal to the the acquisition speed (16.67 fps). [gif video file of 544 frames is 71.47MB].}
\label{V_S3}  
\end{Video} 

\begin{Video}
\captionsetup{font=normalsize,justification=raggedright,format=hang}
\caption{Hypereutectic alloy, low velocity, TXM images, combined as a video with contrast levels adjusted for clarity. $V =~1~{\rm \mu ms^{-1}}$ and $G =~15~\rm{Kmm^{-1}}$. Images have been processed by dividing each frame by an image of the molten liquid alloy. A second, Al\mbox{-}liquid front  is visible after $t_0+30~{\rm s}$.  Playback speed is 10~frames~per~second (fps). [gif video file of 181 frames is 29.1MB].}
\label{V_S4}  
\end{Video} 

\begin{Video}
\captionsetup{font=normalsize,justification=raggedright,format=hang}
\caption{Hypereutectic alloy, high velocity, TXM images, combined as a video with contrast levels adjusted for clarity. $V =~10~{\rm \mu ms^{-1}}$ and $G =~15~{\rm K mm^{-1}}$. Images have been processed by dividing each frame by an image of the molten liquid alloy. Playback speed is 10~frames~per~second (fps). [gif video file of 69 frames is 11.2MB].}
\label{V_S5}  
\end{Video} 

\end{document}